\documentclass[a4paper,11pt]{report}
\usepackage{graphicx,amssymb,amstext,amsmath}
\usepackage{subfigure}
\usepackage[retainorgcmds]{IEEEtrantools}
\usepackage[section] {placeins}
\usepackage[cm]{fullpage}
\usepackage{booktabs}
\usepackage{multirow}
\usepackage{hyperref}  
\hypersetup{bookmarks=true, pdfauthor={Pei Zhang},colorlinks=true,linkcolor=black,citecolor=blue,breaklinks=true}
\usepackage{amsmath}
\usepackage{multirow}
\usepackage{pdflscape} 
\newcommand{\nocontentsline}[3]{}
\newcommand{\tocless}[2]{\bgroup\let\addcontentsline=\nocontentsline#1{#2}\egroup}
\begin{document}
\title{\textbf{A Study of Beam Position Diagnostics with Beam-excited Dipole Higher Order Modes using a Downconverter Test Electronics in Third Harmonic 3.9~GHz Superconducting Accelerating Cavities at FLASH}}

\author{P.~Zhang$^{\dagger \ddagger \ast}$, N.~Baboi$^\ddagger$, N.~Eddy$^\#$, B.~Fellenz$^\#$, R.M.~Jones$^{\dagger \ast}$, \\B.~Lorbeer$^\ddagger$, T.~Wamsat$^\ddagger$, M.~Wendt$^\#$\\
\mbox{$^\dagger$The University of Manchester, Manchester, U.K.}\\
\mbox{$^\ddagger$DESY, Hamburg, Germany}\\
\mbox{$^\ast$The Cockcroft Institute, Daresbury, U.K.}\\
\mbox{$^\#$Fermilab, Batavia, Illinois 60510, U.S.A.}}

\date{\today}

\maketitle
\begin{abstract}
Beam-excited higher order modes (HOM) in accelerating cavities contain transverse beam position information. Previous studies have narrowed down three modal options for beam position diagnostics in the third harmonic 3.9~GHz cavities at FLASH. Localized modes in the beam pipes at approximately 4.1~GHz and in the f{}if{}th cavity dipole band at approximately 9~GHz were found, that can provide a local measurement of the beam position. In contrast, propagating modes in the f{}irst and second dipole bands between 4.2 and 5.5~GHz can reach a better resolution. All the options were assessed with a specially designed test electronics built by Fermilab. The aim is to def{}ine a mode or spectral region suitable for the HOM electronics. Two data analysis techniques are used and compared in extracting beam position information from the dipole HOMs: direct linear regression and singular value decomposition. Current experiments suggest a resolution of 50~$\mu m$ accuracy in predicting local beam position using modes in the f{}ifth dipole band, and a global resolution of 20~$\mu m$ over the complete module. Based on these results we decided to build a HOM electronics for the second dipole band and the f{}if{}th dipole band, so that we will have both high resolution measurements for the whole module, and localized measurements for individual cavity. The prototype electronics is being built by Fermilab and planned to be tested in FLASH by the end of 2012.
\end{abstract}

\renewcommand{\abstractname}{Acknowledgements}

\tableofcontents
\chapter{Introduction}

\section{FLASH and the third harmonic cavities}\label{sec:flash}
\begin{figure}[h]\center
\includegraphics[width=0.8\textwidth]{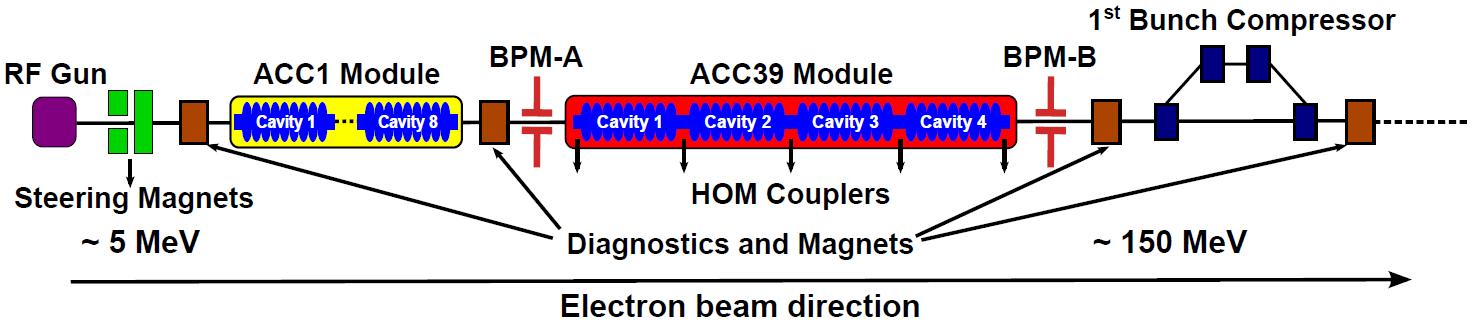}
\caption{Schematic of injector section of the FLASH facility (not to scale, cavities in ACC1 are approximately three times larger than those in ACC39).}
\label{injector}
\end{figure}
FLASH \cite{flash-1} is a free-electron laser (FEL) user facility at DESY in Hamburg providing ultrashort radiation pulses with an unprecedented brilliance. A schematic of the injector section of FLASH is shown in Fig.~\ref{injector}. A laser-driven RF gun generates a high quality electron beam. Seven superconducting 1.3~GHz TESLA type \cite{tesla-1} modules accelerate the electron beam to a maximum energy of 1.2~GeV. Then the electron beam goes through an undulator section to produce laser like coherent FEL radiation. 

\begin{figure}[h]\center
\subfigure[The ACC39 module]{
\includegraphics[width=0.45\textwidth]{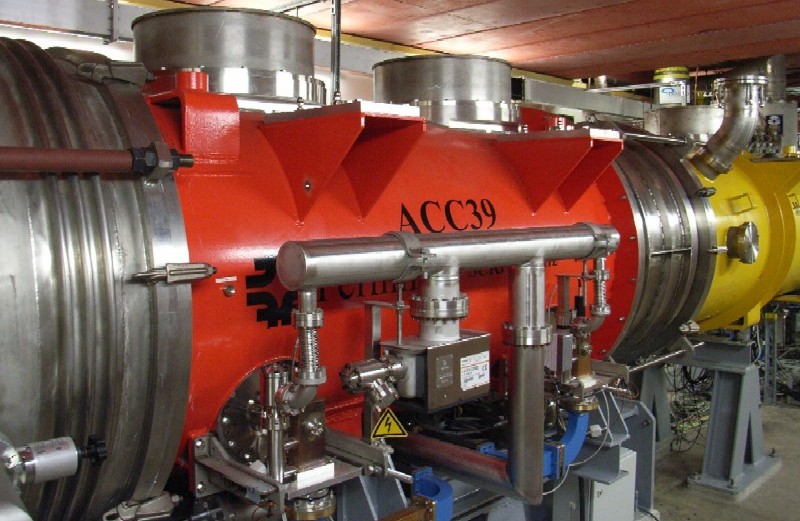}
\label{acc39-module}
}
\quad\quad
\subfigure[Cavities]{
\includegraphics[width=0.45\textwidth]{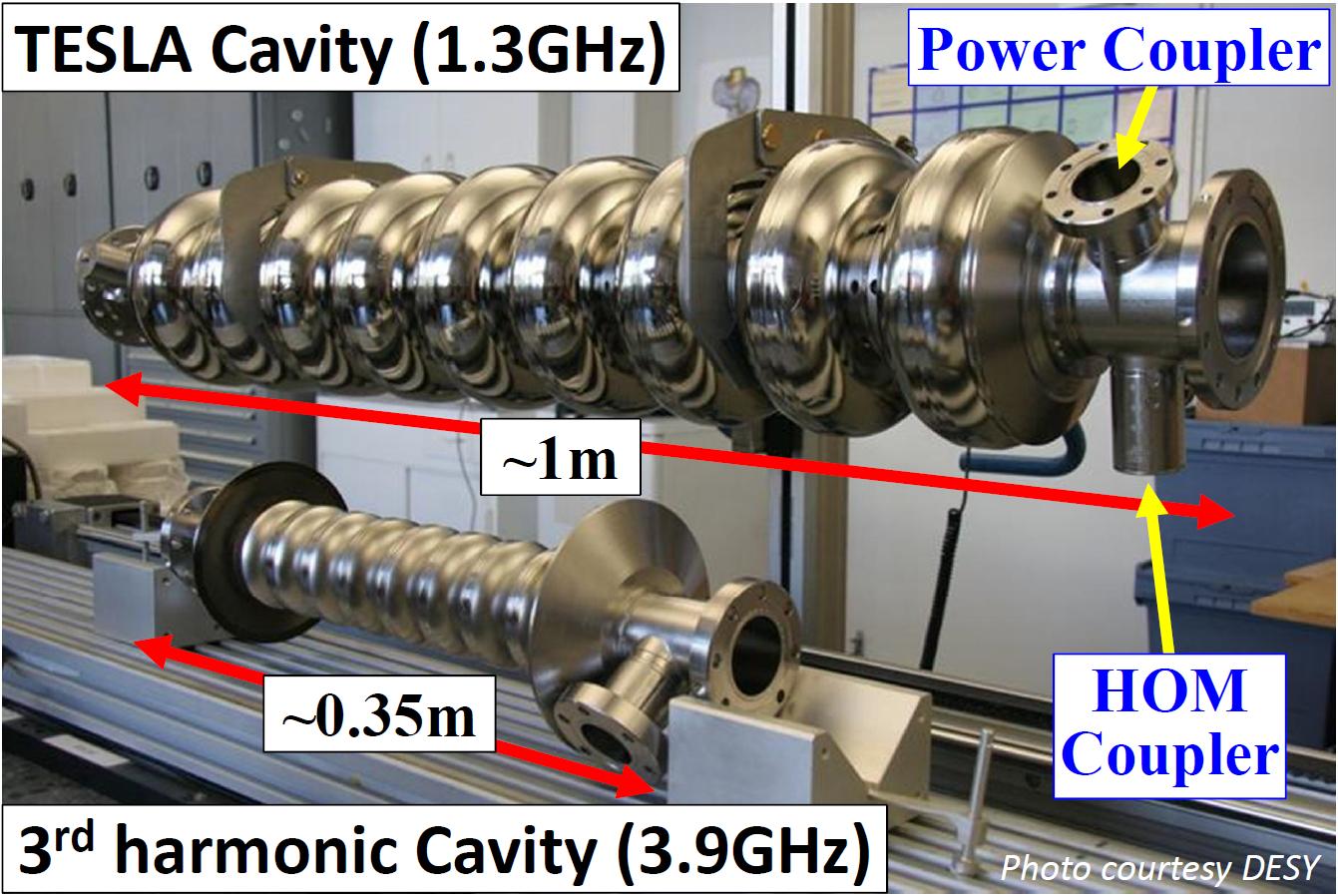}
\label{acc39-acc1}
}
\caption{(a) The ACC39 module in FLASH beam line. The module in yellow, on the right, is ACC1. The direction of travel of the multi-bunch beam, is from right to left. (b) A TESLA style cavity operating at 1.3~GHz (top) and the corresponding third harmonic cavity (bottom).}
\label{acc39-module-cavity}
\end{figure}

A third harmonic module containing four 3.9~GHz superconducting cavities is installed in FLASH (Fig.~\ref{acc39-module}) in order to linearize the longitudinal phase space of the electron bunch \cite{acc39-1}. Due to a special design \cite{acc39-2}, most higher order modes (HOMs) are above the cutof{}f frequency of the connecting beam pipes, therefore are able to propagate among cavities. This leads to a signif{}icantly complex HOM spectrum of the 3.9~GHz cavity. Moreover, wakef{}ields are much stronger in the 3.9~GHz cavities than those in the TESLA 1.3~GHz cavities. This is due to the smaller iris of the 3.9~GHz cavity which consequently scales up the wakef{}ield with the second power for the longitudinal components and the third power for the transverse components \cite{wake-1}. Fig.~\ref{acc39-acc1} shows a comparison of the 1.3~GHz TESLA cavity and the 3.9~GHz cavity. If left unchecked, these HOMs will add adverse ef{}fects on the beam resulting a dilution of the beam quality. On the other hand, dipole higher order modes have linear dependence on the transverse beam position \cite{wake-1}, and this enables the beam position to be determined remotely by monitoring these modes \cite{tesla-hombpm-1,tesla-hombpm-2,tesla-hombpm-3}. 

Higher order mode beam position monitors (HOMBPM) have the capability of measuring the beam position inside the accelerating cavity without additional vacuum components, while the traditional beam position monitors (BPMs) fall short of providing this information directly. However, a careful study on the cavity spectrum is required in order to secure an appropriate dipole mode. Since the cavity is not designed to be used as a diagnostics tool, the position resolution might be limited by the characterizations of the cavities.

Prior to designing any electronics for HOM-based beam diagnostics, extensive studies on the third harmonic cavities have been conducted. Concerning simulations, at f{}irst, eigenmode simulations have been performed on an ideal third harmonic cavity without couplers using three dif{}ferent simulation codes: MAFIA\textregistered \cite{mafia-0,mafia}, HFSS\textregistered \cite{hfss,hfss-1} and CST Microwave Studio\textregistered \cite{cst,cst-1}. Then the entire third harmonic module consisting of four 3.9~GHz cavities have been simulated with three dif{}ferent schemes: Generalized Scattering Matrix (GSM) \cite{acc39-hombpm-2,acc39-hombpm-4}, Coupled S-Parameter Calculation (CSC) \cite{acc39-hombpm-2,acc39-hombpm-10} and direct calculation using the ACE3P code suite \cite{acc39-hombpm-9,acc39-hombpm-11}. Regarding the measurement, a series of transmission measurements without beam excitations were conducted \cite{acc39-hombpm-2,acc39-hombpm-12}, then beam-excited HOM spectra were measured \cite{acc39-hombpm-3,acc39-hombpm-12} and the dipole dependence on the beam of{}fset were studied \cite{acc39-hombpm-12,acc39-hombpm-5,acc39-hombpm-6,acc39-hombpm-8}. Three modal options are therefore narrowed down \cite{acc39-hombpm-7} and the test electronics are subsequently designed and built by FNAL in order to study these modal options using both amplitude and phase of each mode for beam position diagnostics. Then a dedicated study has been conducted with the test electronics and the results are presented in this report.  

In this report, the principle of the test electronics is introduced in Chapter~\ref{ca:hom-meas} along with example output signals and various modal options tested in this study. In Chapter~\ref{ca:ana}, two analysis methods used to extract beam position information from the HOM signals are explained and compared: direct linear regression (DLR) and singular value decomposition (SVD). Then the position resolution using one 20~MHz band of signals from each modal option is described in Chapter~\ref{ca:res}. Both time-domain and frequency-domain signals are studied and compared. Extensive tables of position resolutions for all modal options studied are listed in Appendix~\ref{app:res}, while the number of SVD modes used to achieve these resolutions is listed in Appendix~\ref{app:nsvd}. 

\chapter{HOM Measurements with the Test Electronics}\label{ca:hom-meas}
Previous studies suggest three promising modal options for beam position diagnostics \cite{acc39-hombpm-7}. These are dipole beam-pipe modes at approximately 4~GHz, the f{}irst two dipole cavity bands at 4--5.5~GHz and the f{}ifth dipole cavity band at approximately 9~GHz. A test electronics has been built to study these options in FLASH as explained in this chapter.

\section{The Principle of the Test Electronics}
\begin{figure}[h]\center
\includegraphics[width=0.8\textwidth]{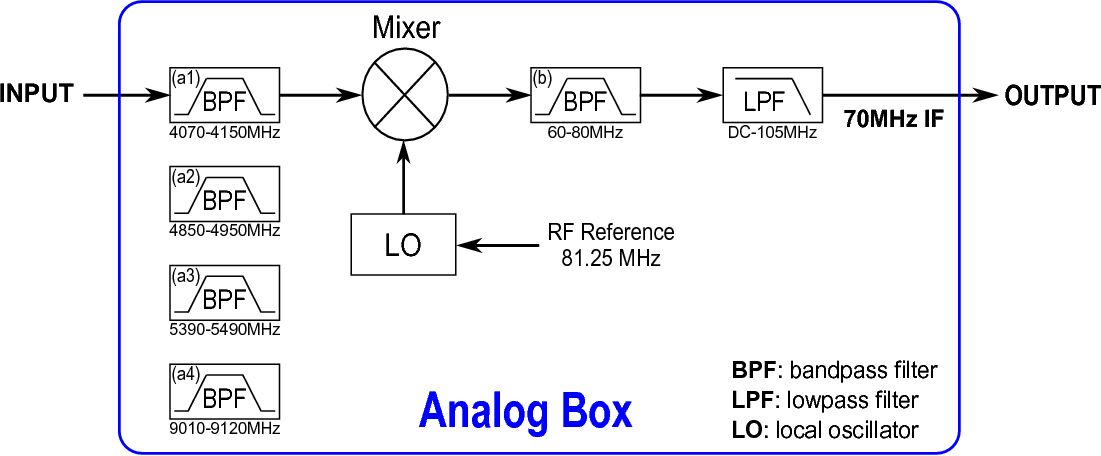}
\caption{Schematics of the analog box. Only one of the four BPFs (a1-a4) is connected in the chain during each measurement.}
\label{analog-box-chart}
\end{figure}
\begin{figure}[h]\center
\includegraphics[width=0.6\textwidth]{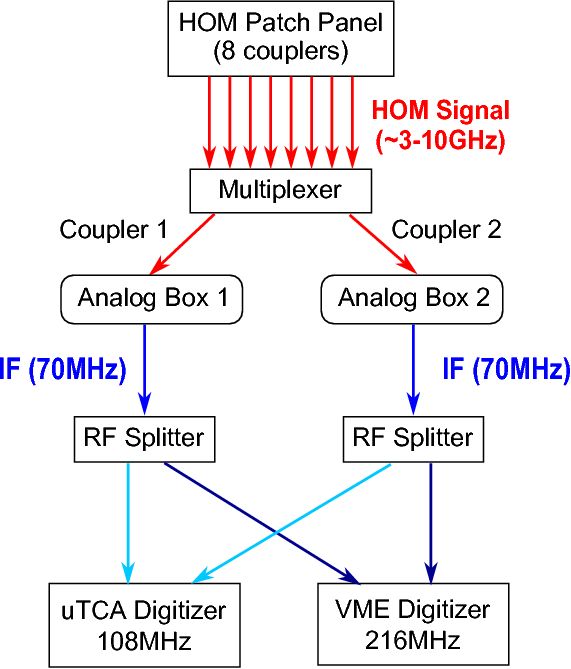}
\caption{Schematic device setup in the injector barrack outside the FLASH tunnel.}
\label{device-setup-chart}
\end{figure}

The analog electronics was designed to have the f{}lexibility to study various modal options of interest as well as accommodate the large mode bandwidths (BW) of these options. Its simplif{}ied block diagram is shown in Fig.~\ref{analog-box-chart}. One of four dif{}ferent analog bandpass f{}ilters (BPF) can be connected into the chain to study the localized dipole beam-pipe modes at $\sim$4.1~GHz, strong multi-cavity modes in the f{}irst dipole band at $\sim$4.9~GHz and in the second dipole band at $\sim$5.4~GHz, and the trapped cavity modes in the f{}ifth dipole band at $\sim$9~GHz. Af{}ter f{}iltering, the signal is mixed with a selectable local oscillator (LO) to an intermediate frequency (IF) of approximately 70~MHz. Then the 70~MHz IF signal is further f{}iltered with a 20~MHz analog BPF to select the specif{}ic band of modes. In order to ensure that the possible remaining high frequency components of the IF signal generated during the mixing step are well suppressed, a lowpass f{}ilter (LPF) is applied to preserve only the frequency below 105~MHz with a dominant component from 60~MHz to 80~MHz. In Fig.~\ref{analog-box-chart}, the ``INPUT'' is connected to a RF multiplexer to allow for a fast switching among HOM couplers as shown in Fig.~\ref{device-setup-chart}. The 70~MHz IF out of the analog box is further split into two signals. One is digitized by a VME digitizer operating at 216~MS/s with 14~bit resolution\footnote{The output value from the VME digitizer has 16~bit with the last two digits both set to 0. It was for easy data processing with the FPGA. The real resolution is still 14~bit.} and 2~Volt peak to peak signal range along with a programmable FPGA for signal processing. The digitizer is triggered by a 10~Hz beam trigger, which is the RF-pulse repetition rate of FLASH. Both the selectable LO and the digitizer clock of 216~MHz are locked to the FLASH accelerator by using RF signals delivered from the FLASH master oscillator as a reference. This locking allows correct phase information of the digitized signal. The other IF signal is processed by a $\mu$TCA \cite{uTCA} digitizer with 16~bit resolution and 2~Volt peak to peak signal range operating at 108~MS/s. Since the IF is approximately 70~MHz, the $\mu$TCA ADC is undersampled. Both the trigger and the clock of the $\mu$TCA digitizer are directly delivered from the FLASH master oscillator, which is automatically locked to the FLASH accelerator. All the devices are set up in the FLASH injector barrack outside the accelerator tunnel. The detailed circuit drawing of the analog electronics is shown in Appendix~\ref{app:draw} along with photos of the device setup, the analog box and the digitizers. Digitized data is collected from the VME digitizer with an EPICS \cite{epics} software tool, while FLASH control system DOOCS \cite{doocs} takes care of the data collection for the $\mu$TCA digitizer. The beam charge, steerer current and BPM readouts are recorded synchronously from DOOCS. The data processing for position diagnostics is conducted of{}f{}line using MATLAB \cite{matlab}.

Inside the analog box, individual attenuation and amplif{}ication was used for dif{}ferent frequency band. This is for the optimization of the signal level while accommodating the variation of HOM power amongst couplers. These are shown in Appendix~\ref{app:atten}.

\FloatBarrier
\section{Downconverted HOM Signals}
\begin{figure}[h]\center
\includegraphics[width=0.8\textwidth]{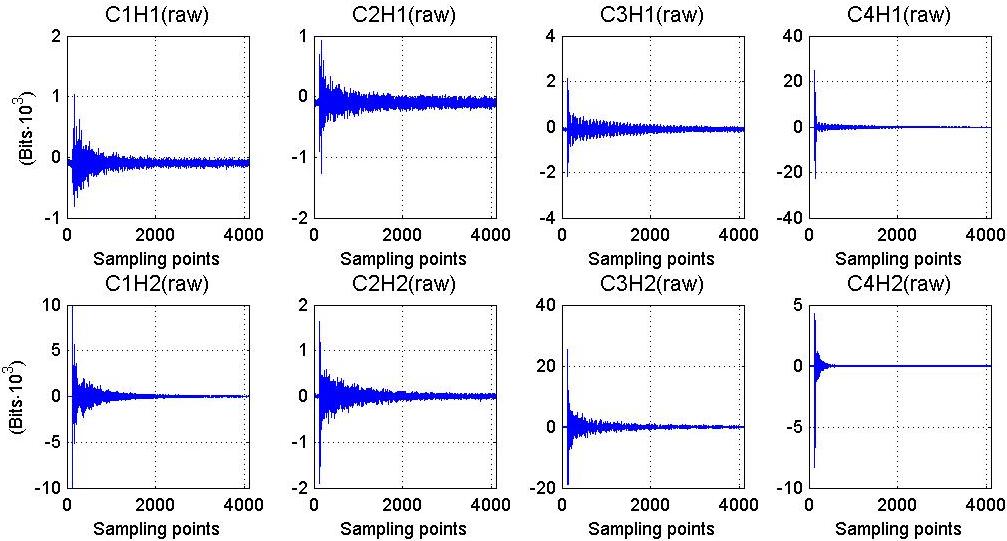}
\caption{Raw waveforms measured by the VME digitizer for each HOM coupler. The center frequency is 9066~MHz with a 20~MHz BPF. Dif{}ferent vertical scale is used for each waveform.}
\label{raw-8wfm-9066MHz-D503}
\end{figure}

\begin{figure}[h]\center
\includegraphics[width=0.8\textwidth]{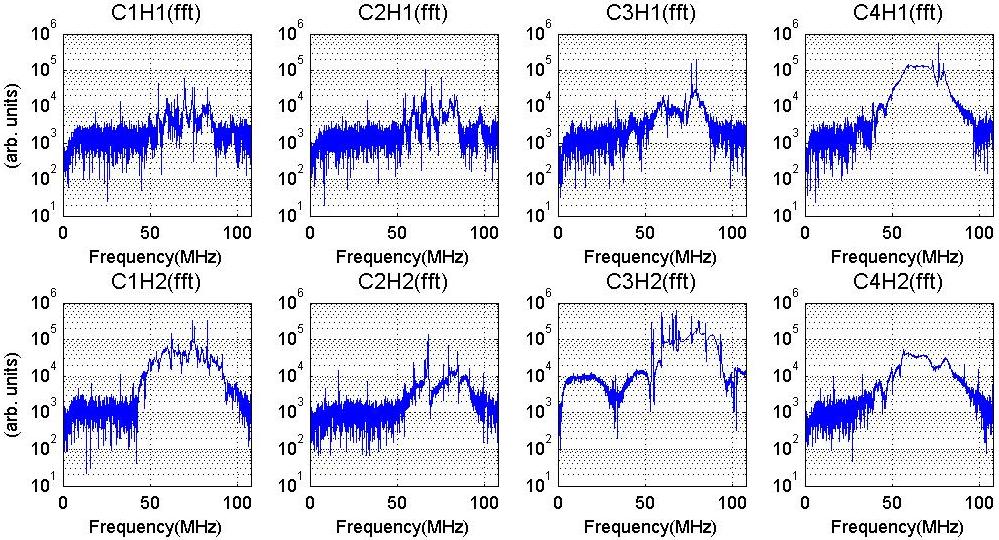}
\caption{FFT amplitude of the raw waveform measured by the VME digitizer for each HOM coupler. The center frequency is 9066~MHz with a 20~MHz BPF.}
\label{raw-8fft-9066MHz-D503}
\end{figure}

The downconverted HOM signals after splitting are processed by both the VME and $\mu$TCA digitizer. Fig.~\ref{raw-8wfm-9066MHz-D503} shows the raw waveform of each HOM coupler for the 9066~MHz (BW: 20~MHz) digitized by VME. This frequency range is within the f{}ifth dipole band. The strength of the signal varies dramatically among couplers. This might be due to the dif{}ferent coupling of the HOM couplers, which can also be contributed by the fact that modes within the 20~MHz band are trapped inside each cavity. 
Therefore dissimilar cavities attributed to fabrication tolerances give rise to the dif{}ferences presented in the waveforms. Applying a fast Fourier transform (f{}ft) on the waveforms in Fig.~\ref{raw-8wfm-9066MHz-D503}, the amplitude of each f{}ft output is shown in Fig.~\ref{raw-8fft-9066MHz-D503}. The spectra dif{}fer among cavities and HOM couplers.   

\begin{figure}[h]\center
\includegraphics[width=0.8\textwidth]{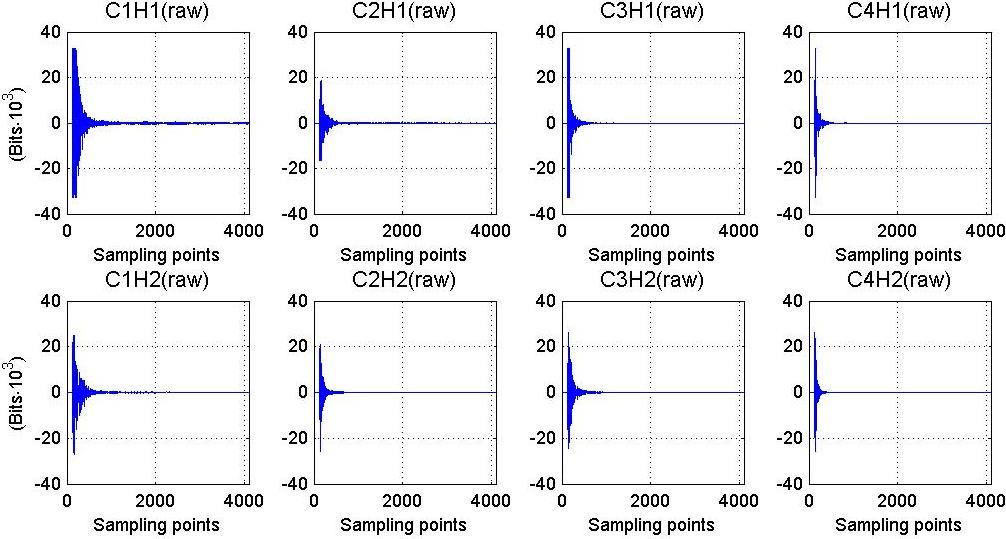}
\caption{Raw waveforms measured by the VME digitizer for each HOM coupler. The center frequency is 5437~MHz with a 20~MHz BPF.}
\label{raw-8wfm-5437MHz-D208}
\end{figure}

\begin{figure}[h]\center
\includegraphics[width=0.8\textwidth]{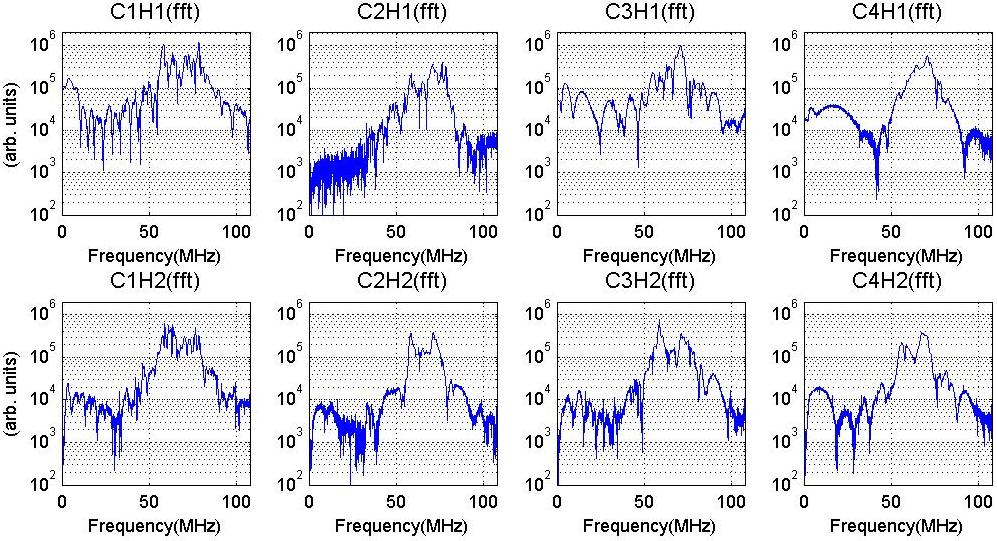}
\caption{FFT amplitude of the raw waveform measured by the VME digitizer for each HOM coupler. The center frequency is 5437~MHz with a 20~MHz BPF.}
\label{raw-8fft-5437MHz-D208}
\end{figure}

\FloatBarrier
Fig.~\ref{raw-8wfm-5437MHz-D208} shows the waveforms for the 5437~MHz (BW: 20~MHz) digitized by VME. This frequency range is within the second dipole band. The signal strength is similar for each coupler. Since modes in this band propagate through the module, couplers pick up similar signals. The dif{}ferences come from the couplers' dissimilar coupling and their locations. Fig.~\ref{raw-8fft-5437MHz-D208} shows the f{}f{}t amplitudes of the waveforms.

%

\FloatBarrier
\section{Measurement Scheme}
\begin{figure}[h]\center
\includegraphics[width=0.8\textwidth]{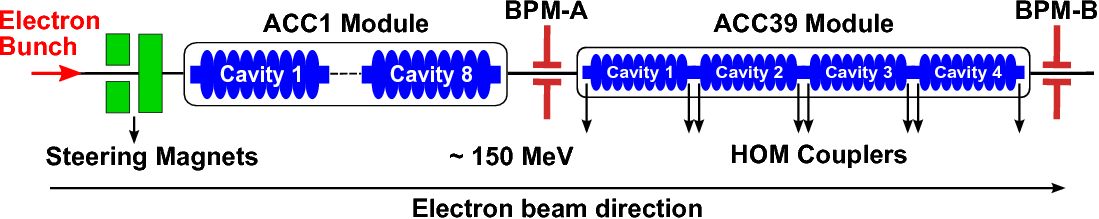}
\caption{Schematic of measurement setup for HOM study (not to scale, cavities in ACC1 are approximately three times larger than those in ACC39).}
\label{hom-setup}
\end{figure}

The schematics of the measurement setup is shown in Fig.~\ref{hom-setup}. An electron bunch of approximately 0.5~nC is accelerated on-crest through ACC1 before entering the ACC39 module. Two steering magnets located upstream of ACC1 are used to produce horizontal and vertical of{}fsets of the electron bunch in ACC39. Two beam position monitors (BPM-A and BPM-B) are used to record transverse beam positions before and af{}ter ACC39. Switching of{}f the accelerating f{}ield in ACC39 and all quadrupoles close to ACC39, a straight line trajectory of the electron bunch is produced between those two BPMs. Therefore, the transverse of{}fset of the electron bunch in the module can be determined by interpolating the readouts of the two BPMs.

\begin{figure}[h]\center
\includegraphics[width=0.8\textwidth]{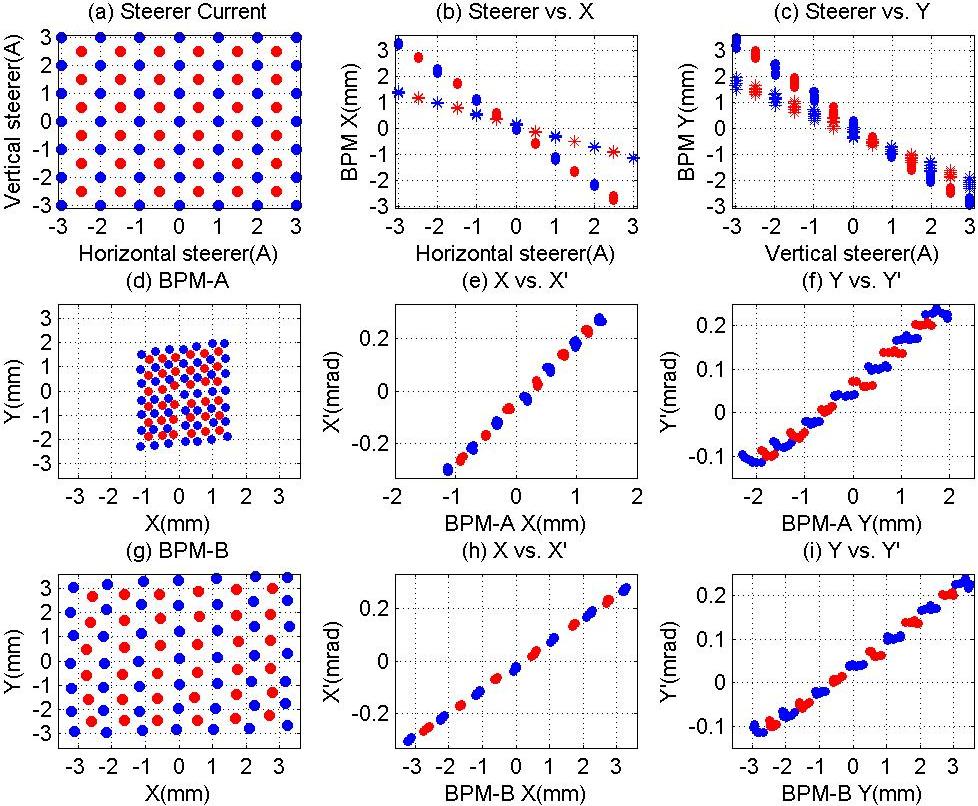}
\caption{2D grid-like beam movement. The calibration samples are in blue while the validation samples are in red. In (b) and (c), dots represent the readings from BPM-A while asterisks for BPM-B.}
\label{bpm-reading}
\end{figure}

To study the position dependencies of HOMs, we moved the beam in a 2D grid manner. Fig.~\ref{bpm-reading}(a) shows the steerer current during the movement. The samples in blue are used for calibration, and then validation\footnote{A set of coef{}f{}icients is f{}irst generated during calibration. Then the newly measured HOM signals are applied with these coef{}f{}icients to make predictions of beam positions. This is called validation.} samples were taken as shown in red. The usage of the samples are explained in Chapter~\ref{ca:ana}. Fig.~\ref{bpm-reading}(b) and Fig.~\ref{bpm-reading}(c) show the relations of the steerer current to the readings of BPM-A (dots) and BPM-B (asterisks). Fig.~\ref{bpm-reading}(d) and Fig.~\ref{bpm-reading}(g) show the readings of BPM-A and BPM-B during the beam movement. The angle of the beam trajectory is represented by $x'$ and $y'$, and its relations to the bpm readings are shown in Fig.~\ref{bpm-reading}(e)(f)(h)(i). When the readings of BPM-A were close to zero for both $x$ and $y$ plane, the readouts jumped between positive and negative unpredictably. In order to cure this, we use the currents of steering magnets to correct the problematic BPM readings. From the measurement results, the jumping in $x$ plane is not severe, therefore only the readings of $y$ have been corrected for all measurements shown in this report.   

\begin{figure}[h]\center
\includegraphics[width=0.6\textwidth]{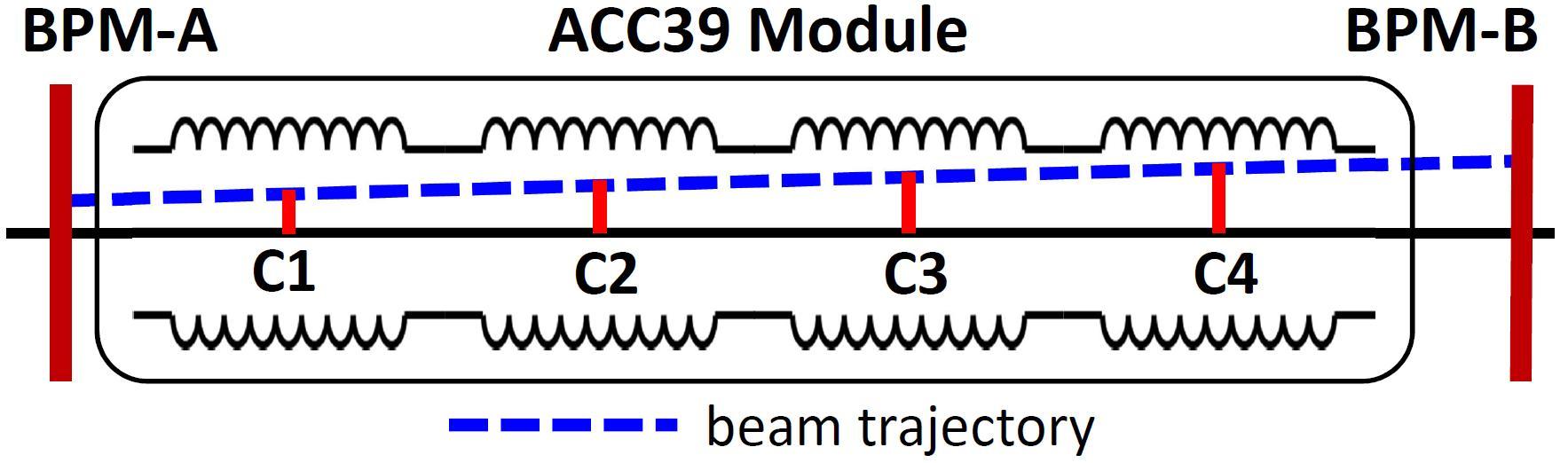}
\caption{Position interpolation into the center of each cavity for localized modes.}
\label{interp-cav}
\end{figure}

\begin{figure}[h]\center
\subfigure[Interpolation into each cavity]{
\includegraphics[width=0.4\textwidth]{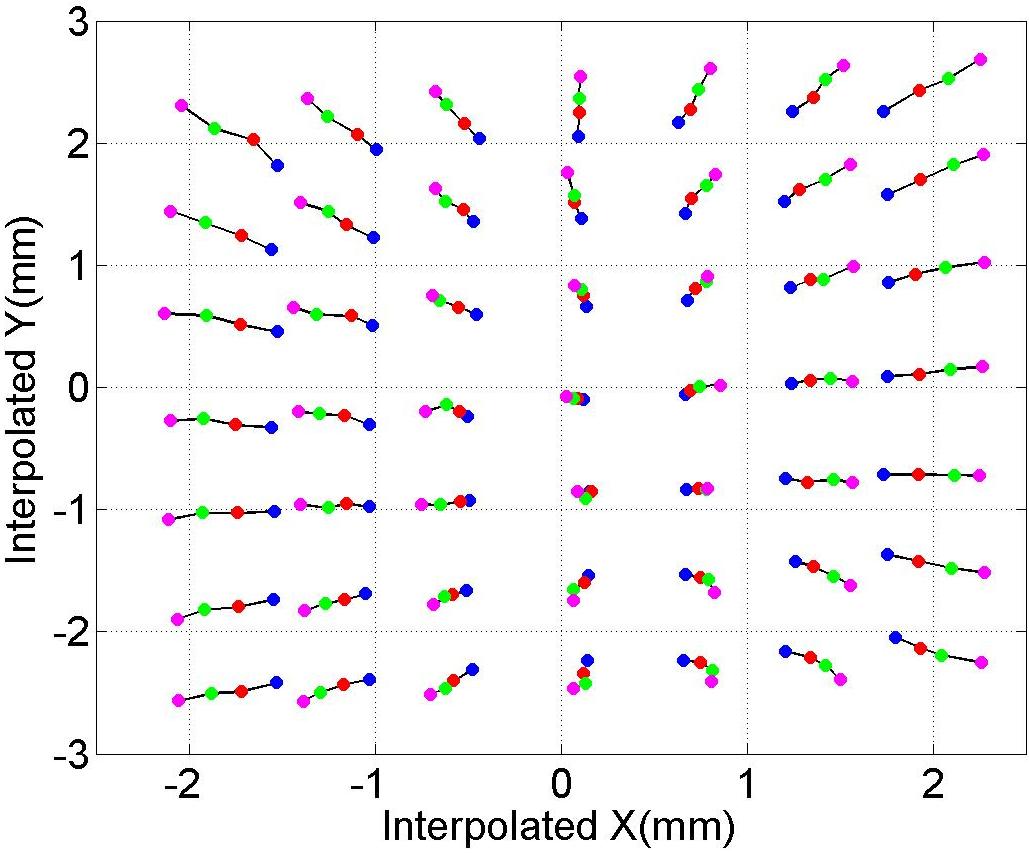}
\label{interp-pos-cav}
}
\quad\quad
\subfigure[Interpolation into module center]{
\includegraphics[width=0.4\textwidth]{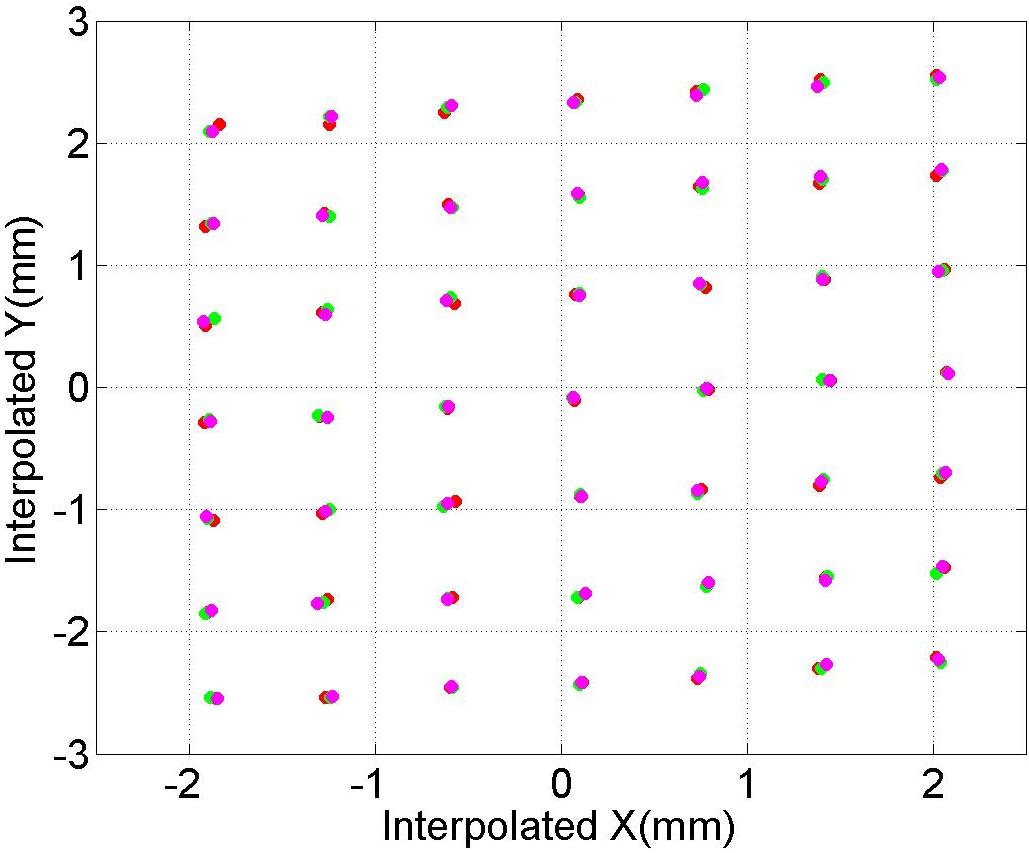}
\label{interp-pos-module}
}
\caption{(a) Interpolated position at each cavity center colored as C1 (blue), C2 (red), C3 (green) and C4 (magenta). Points connected with black lines belong to the same beam position. (b) Interpolated position at the module center when measuring C1 (blue), C2 (red), C3 (green) and C4 (magenta). Only the calibration samples are shown.}
\label{interp-pos}
\end{figure}

\begin{figure}[h]\center
\includegraphics[width=0.6\textwidth]{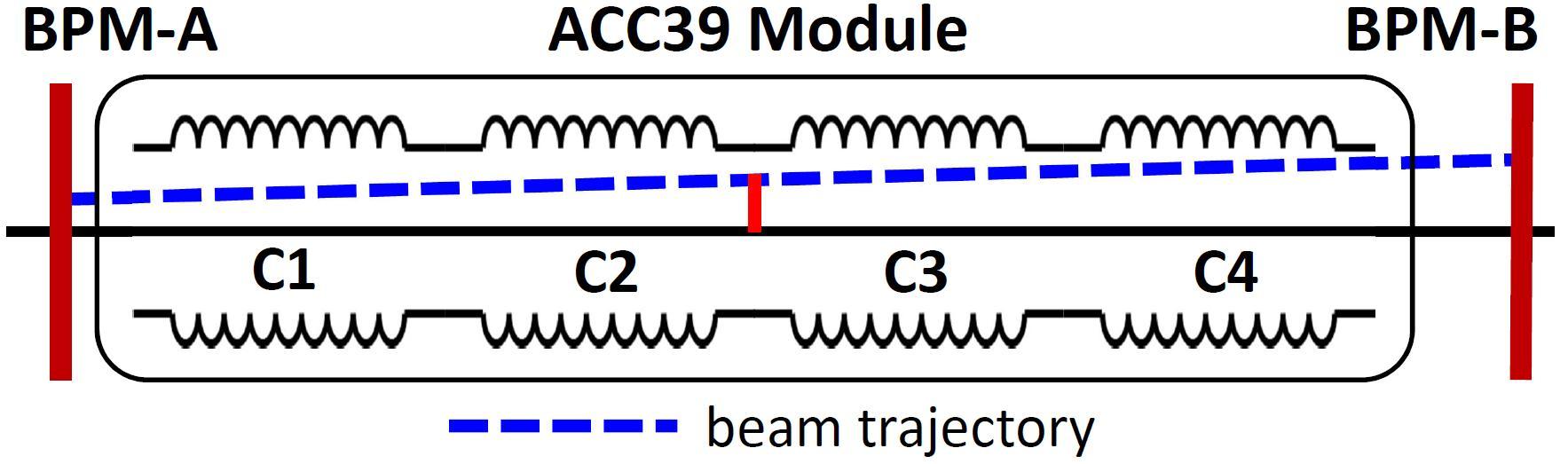}
\caption{Position interpolation into the center of the whole module for coupled modes.}
\label{interp-module}
\end{figure}

\FloatBarrier
As we are dealing with trapped and coupled modes, two schemes of position interpolations are used. The transverse beam position is interpolated into each cavity center for trapped modes as shown in Fig.~\ref{interp-cav}. Fig.~\ref{interp-pos-cav} shows the interpolated position of each cavity center. This enables the local positions in each cavity to be determined by the non-propagating modes extracted from each cavity. Since the coupled modes are propagating through the four-cavity module, we decide to interpolate the beam position into the center of the whole module considering the four cavities in its entirety. The interpolation is shown in Fig.~\ref{interp-module} and the interpolated position of the module center is shown in Fig.~\ref{interp-pos-module}. One notices some non-straight lines in Fig.~\ref{interp-pos-cav} on the bottom right, and it might be due to the beam jitter as only one cavity was measured at each time.   

\FloatBarrier
In this report, three modal options have been studied. The center frequencies and the BWs for these options that we have set on the test electronics during the measurement are listed in Table~\ref{table-modal-options}. For the f{}irst and second dipole band, we also studied the diagnostics with a broadband signal by removing the 20~MHz BPF after the mixer. In this case, the BW of the IF signal was 100~MHz determined by the BPF in front of the mixer and the LPF.  

\begin{table}[h]\center
\caption{Modal options studied with the test electronics.}
\label{table-modal-options}
\begin{tabular}{c|c|c|c|c|c|c|c}
\hline
\multicolumn{2}{c|}{Beam-pipe modes} & \multicolumn{2}{c|}{1$^{st}$ dipole band} & \multicolumn{2}{c|}{2$^{nd}$ dipole band} & \multicolumn{2}{c}{5$^{th}$ dipole band} \\
\hline
Center & BW & Center & BW & Center & BW & Center & BW \\
\hline
4082~MHz & 20~MHz & 4859~MHz & 20~MHz & 5437~MHz & 20~MHz & 9048~MHz & 20~MHz \\
\hline
4118~MHz & 20~MHz & 4904~MHz & 20~MHz & 5464~MHz & 20~MHz  & 9066~MHz & 20~MHz \\
\hline
& & 4940~MHz & 20~MHz & 5482~MHz & 20~MHz  & & \\
\hline
& & 4900~MHz & 100~MHz & 5450~MHz & 100~MHz & & \\
\hline
\end{tabular}
\end{table}

\FloatBarrier
\chapter{Analysis Methods for Beam Position Diagnostics}\label{ca:ana}
Two dif{}ferent analysis methods are used to extract beam position information from the HOM signals: direct linear regression and singular value decomposition. These are explained and compared in this chapter.

\section{Data Samples}
\begin{figure}[h]\center
\includegraphics[width=0.8\textwidth]{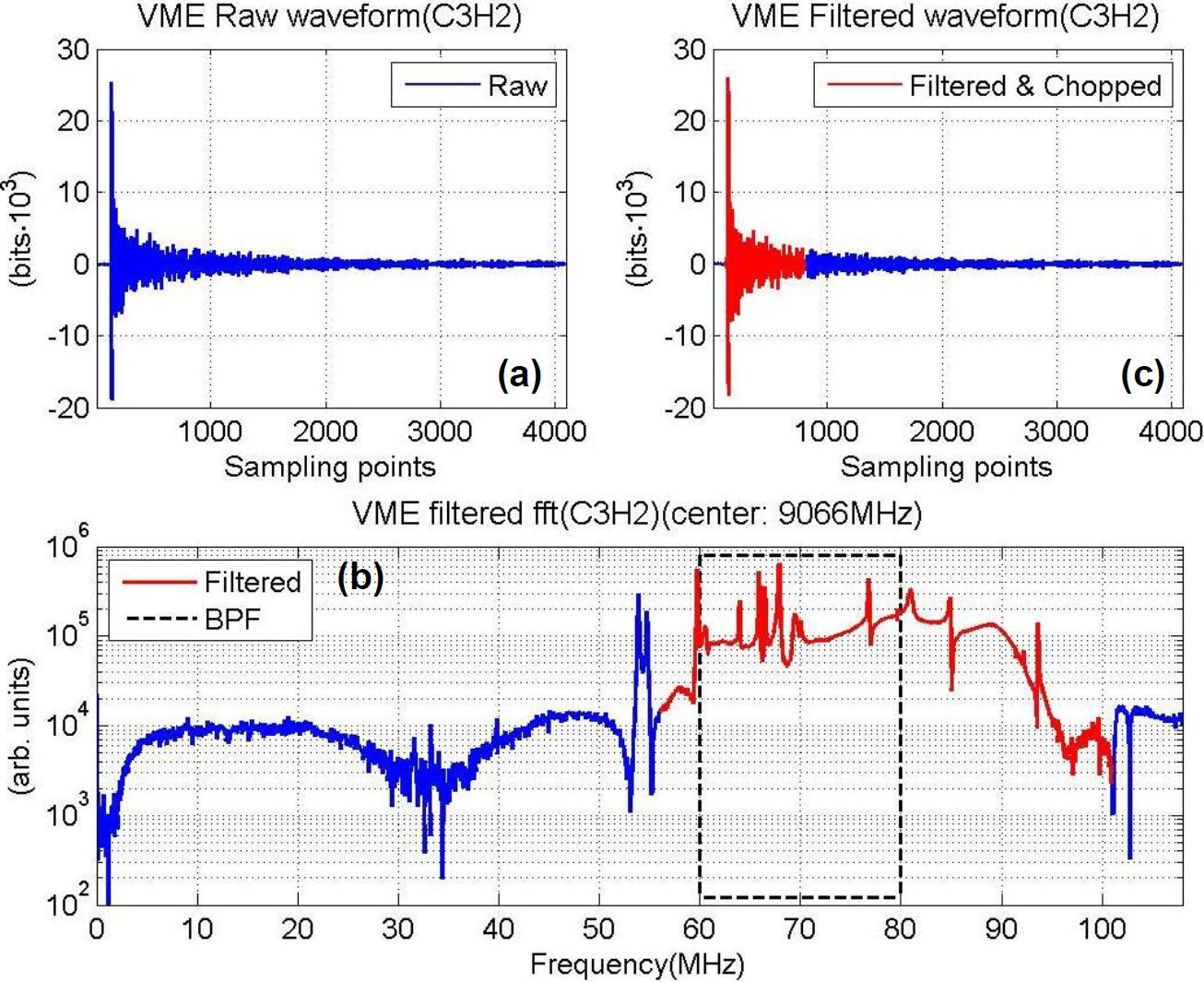}
\caption{Waveform measured with VME-based digitizer (center frequency: 9066~MHz, BW: 20~MHz).}
\label{vme-wfm-C3H2}
\end{figure}

\begin{figure}[h]\center
\subfigure[One waveform]{
\includegraphics[width=0.46\textwidth]{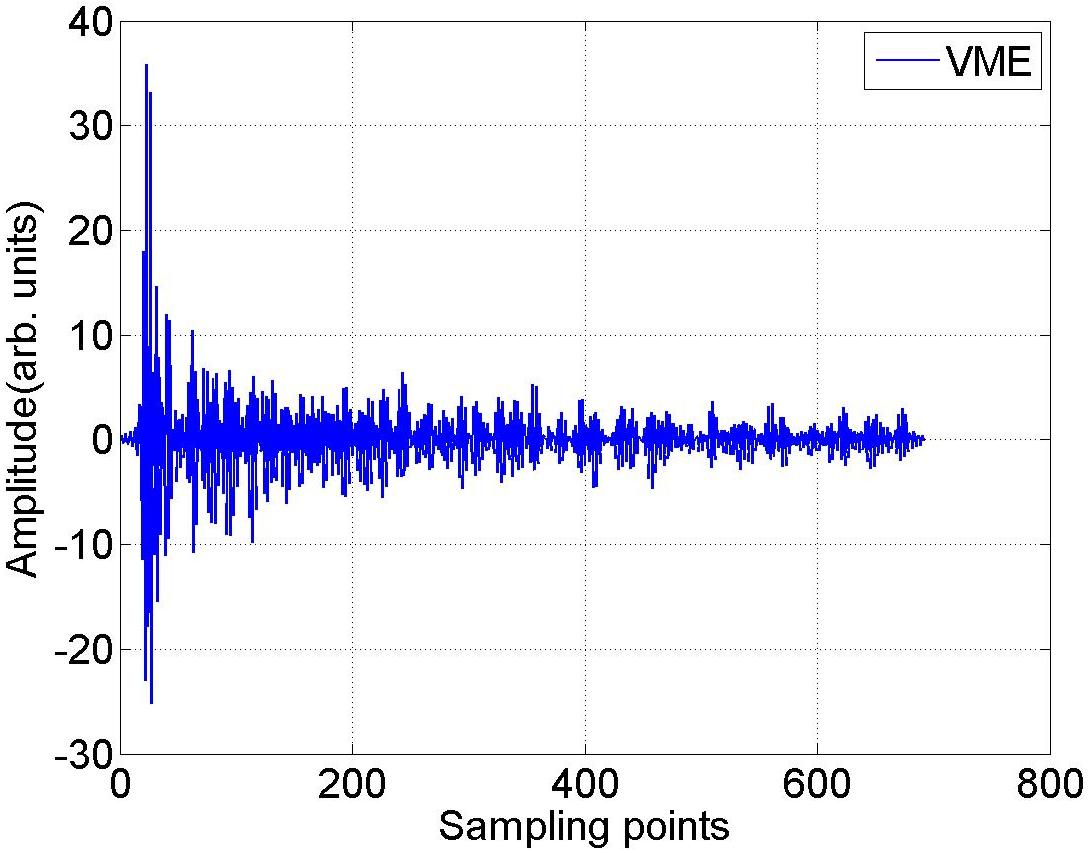}
\label{1wfm-C3H2}
}
\quad
\subfigure[Beam positions]{
\includegraphics[width=0.45\textwidth]{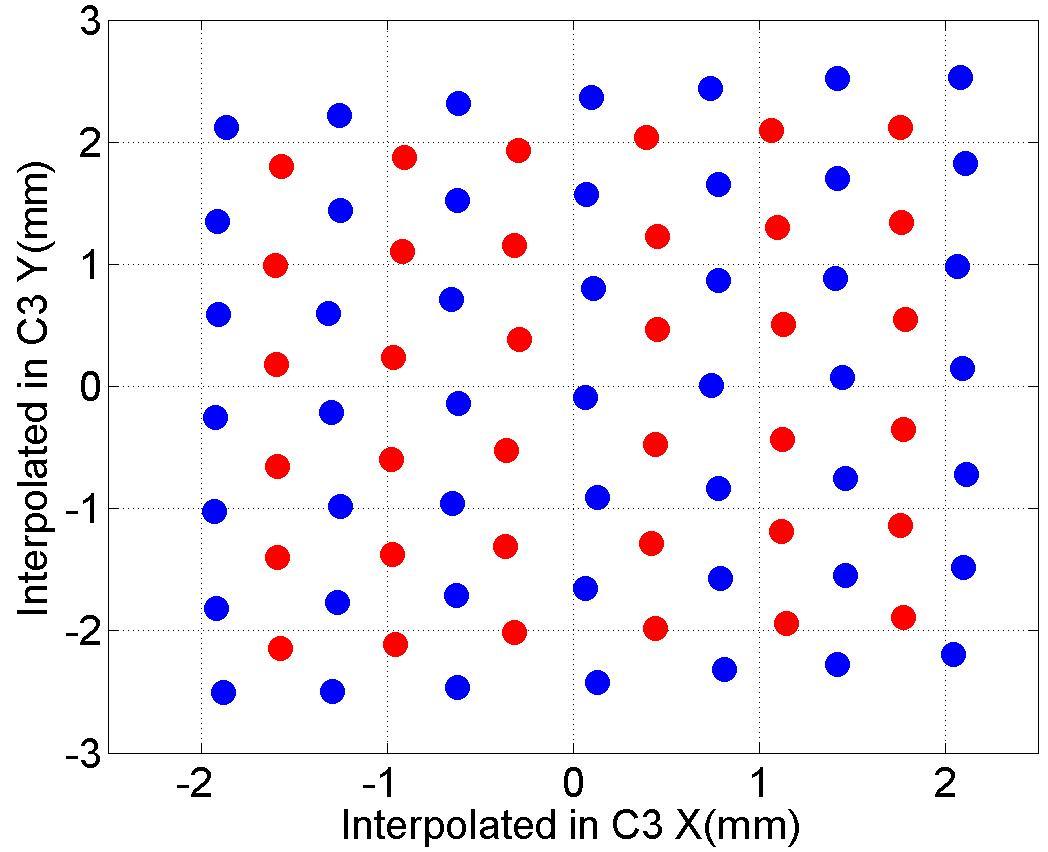}
\label{2D-grid-C3}
}
\caption{(a) One waveform measured from C3H2. (b) Grid-like beam movement (calibration samples are in blue and validation samples are in red). The positions are interpolated in C3. }
\label{2D-grid-1wfm-C3H2}
\end{figure}

In this chapter, waveforms measured from HOM coupler C3H2 with the VME digitizer are used to describe two data analysis techniques. The center frequency has been set to 9066~MHz with a 20~MHz BW. One example of the downconverted waveforms is shown in Fig.~\ref{vme-wfm-C3H2}(a). The downconverted waveforms are applied with a mathematical ideal f{}ilter to keep the range from 56~MHz to 101~MHz in order to cut away the coupling modes at approximately 55~MHz. Fig.~\ref{vme-wfm-C3H2}(b) shows one example of the spectrum after a f{}ft was applied on the waveform shown in Fig.~\ref{vme-wfm-C3H2}(a). The f{}iltered waveform is shown in Fig.~\ref{vme-wfm-C3H2}(c) in blue as one example. Since the signal decays fast, a time window is further applied on the f{}iltered signal which is shown as the red waveform in Fig.~\ref{vme-wfm-C3H2}(c). All waveforms have been normalized to 1~nC bunch charge, and one of the them is shown in Fig.~\ref{1wfm-C3H2}. The beam has been moved in a grid manner. As shown in Fig.~\ref{2D-grid-C3}, a $7\times 7$ grid is used for calibration (49 samples, blue dots) and a $6\times 6$ grid is used as validation samples (36 samples, red dots) in order to obtain full coverage in 2D space for both samples within the linear range of BPM-B (-3~mm to +3~mm).

\section{Direct Linear Regression}
To extract beam position information from the dipole modes, a straightforward method is direct linear regression (DLR),
\begin{equation}
A\cdot M=B.
\label{eq:DLR}
\end{equation}
In this equation, matrix $A$ contains the waveforms, $B$ is the matrix of transverse beam position $x$ and $y$ interpolated from the two BPM readouts. The angle of the beam trajectory $x'$ and $y'$ is not the subject of this study. $M$ is the matrix of regression coef{}f{}icients. The arrangement of matrix $A$ and $B$ are
\begin{equation}
A =
\begin{pmatrix}
\mathit{waveform}_1\\
\mathit{waveform}_2\\
\vdots \\
\mathit{waveform}_m
\end{pmatrix}
= 
\begin{pmatrix}
a_{11} & a_{12} & \ldots & a_{1n} \\
a_{21} & a_{22} & \ldots & a_{2n} \\
\vdots & \vdots & \ddots & \vdots \\
a_{m1} & a_{m2} & \ldots & a_{mn} \\
\end{pmatrix}
= 
\begin{pmatrix}
r_1^T \\
r_2^T \\
\vdots \\
r_m^T
\end{pmatrix}
= 
\begin{pmatrix}
c_1^T \\
c_2^T \\
\vdots \\
c_n^T
\end{pmatrix}
^T \in \mathbb{R}^{m\times n},
\label{eq:DLR-A}
\end{equation}
\begin{equation}
B = 
\begin{pmatrix}
B_1 \\
B_2 \\
\vdots \\
B_m
\end{pmatrix}
=
\begin{pmatrix}
x_1 & y_1 \\
x_2 & y_2 \\
\vdots & \vdots \\
x_m & y_m 
\end{pmatrix}
\in \mathbb{R}^{m\times 2}.
\label{eq:DLR-B}
\end{equation}
$A$ is a $m\times n$ matrix. $a_{ij}$ represents the value of $j^{th}$ sampling point in $\mathit{waveform}_i$. The superscript $T$ denotes the vector or matrix transpose. $r_i^T\in \mathbb{R}^n$ is the $i^{th}$ row of $A$, which is the $waveform_i$ for a particular beam position. $c_i\in \mathbb{R}^m$ is the $i^{th}$ column of $A$, representing the $i^{th}$ regressor or variable. $B_i \in \mathbb{R}^2$ is the $i^{th}$ row of matrix $B \in \mathbb{R}^{m\times 2}$. It represents the interpolated horizontal coordinate $x_i$ and vertical coordinate $y_i$ of the $i^{th}$ beam position corresponding to the measurement of $waveform_i$. In order to characterizes the DC components in the waveforms, a intercept term is added into $A$ and $M$ by adding one column of 1 in $A$ and one row of coef{}f{}icients in $M$ as 
\begin{equation}
A^\ast = (I, A) \in \mathbb{R}^{m\times (n+1)},
\label{eq:DLR-A-2}
\end{equation}
\begin{equation}
M^\ast =
\begin{pmatrix}
M_0\\
M
\end{pmatrix} \in \mathbb{R}^{(n+1)\times 2},
\label{eq:DLR-M-2}
\end{equation}
where $I\in\mathbb{R}^{m\times 1}$ and $M_0\in\mathbb{R}^{1\times 2}$. Matrix $M$ is obtained by solving linear system composed by $A^{\ast}$ and $B$ for calibration samples.

\begin{figure}[h]\center
\includegraphics[width=0.8\textwidth]{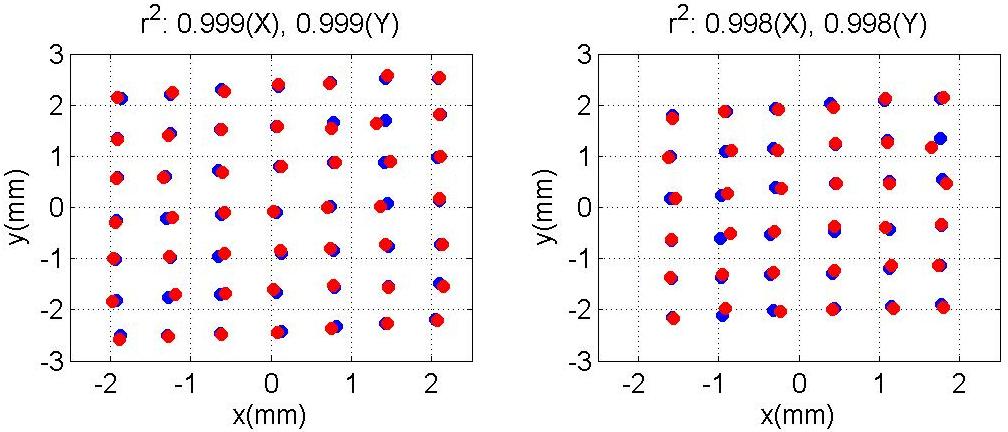}
\caption{Measurement (blue) and prediction (red) of the transverse beam position from calibration (left) and validation (right) samples. The applied method is DLR.}
\label{DLR-B-Bp}
\end{figure}

\begin{figure}[h]\center
\includegraphics[width=0.65\textwidth]{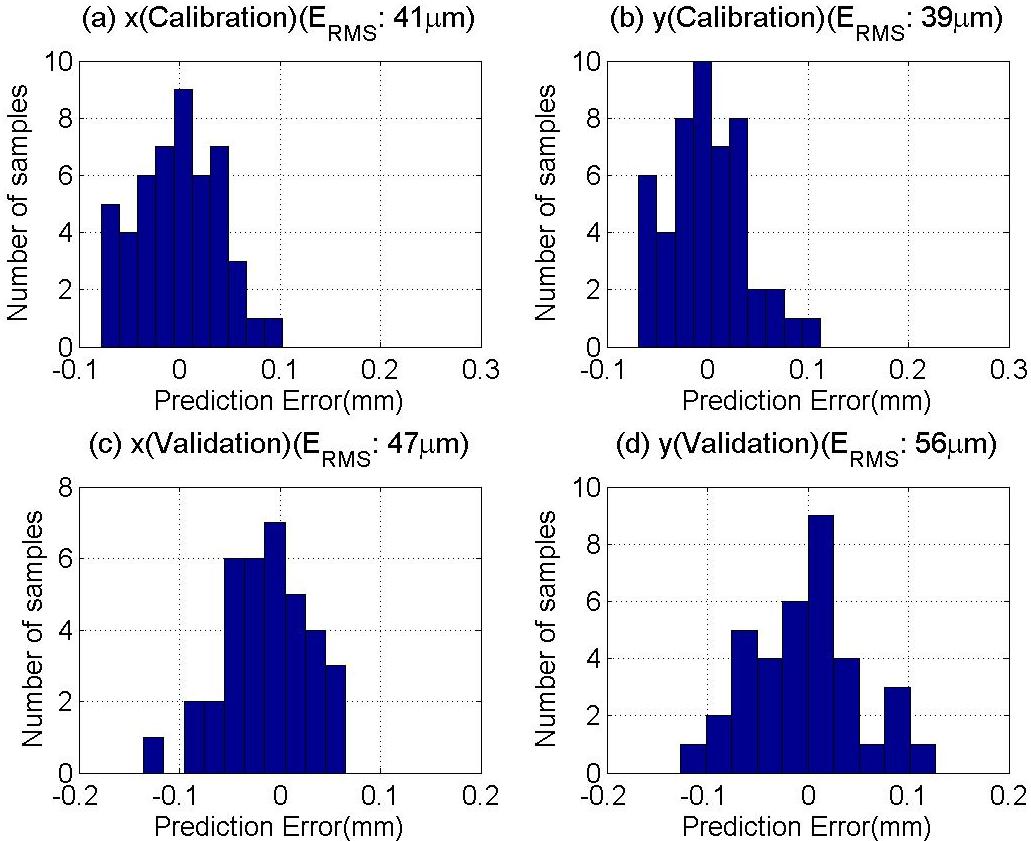}
\caption{Dif\mbox{}ference of measured and predicted transverse beam position from calibration and validation samples. The applied method is DLR.}
\label{DLR-diff}
\end{figure}

The predictions of $B$ can be obtained by
\begin{equation}
B_{pre}=A^{\ast}\cdot M^{\ast}
\label{eq:DLR-pre}
\end{equation}
for both calibration and validation samples. Applying DLR on the waveforms measured from HOM coupler C3H2 split as Fig.~\ref{2D-grid-C3}, we obtain the results for both samples. In Fig.~\ref{DLR-B-Bp}(a), the prediction and the measurement almost overlap, which proves the linear dependence of the waveform on the transverse beam of{}fset. Fig.~\ref{DLR-B-Bp}(b) shows that the $M$ obtained from calibration has good prediction power. To measure the consistency, the coef{}f{}icient of determination $r^2$ is calculated ($r^2=1$, perfect f{}it; $r^2=0$, poor f{}it) \cite{stat-1}. The dif{}ference between the predicted position $B_{pre}$ and the measured position $B$ is def{}ined as the \textit{prediction error},
\begin{equation}
\mathit{prediction\mbox{ }error}=|B_{pre}-B|.
\label{eq:pre-err}
\end{equation}
Fig.~\ref{DLR-diff} shows the prediction error for calibration and validation samples by using DLR, together with the rms value of the prediction error, $E_{RMS}$. Then the position resolution is given by $E_{RMS}$ of the validation samples. Therefore, the resolution achieved by DLR is 47~$\mu m$ for $x$ and 56~$\mu m$ for $y$.

\begin{figure}[h]\center
\includegraphics[width=0.7\textwidth]{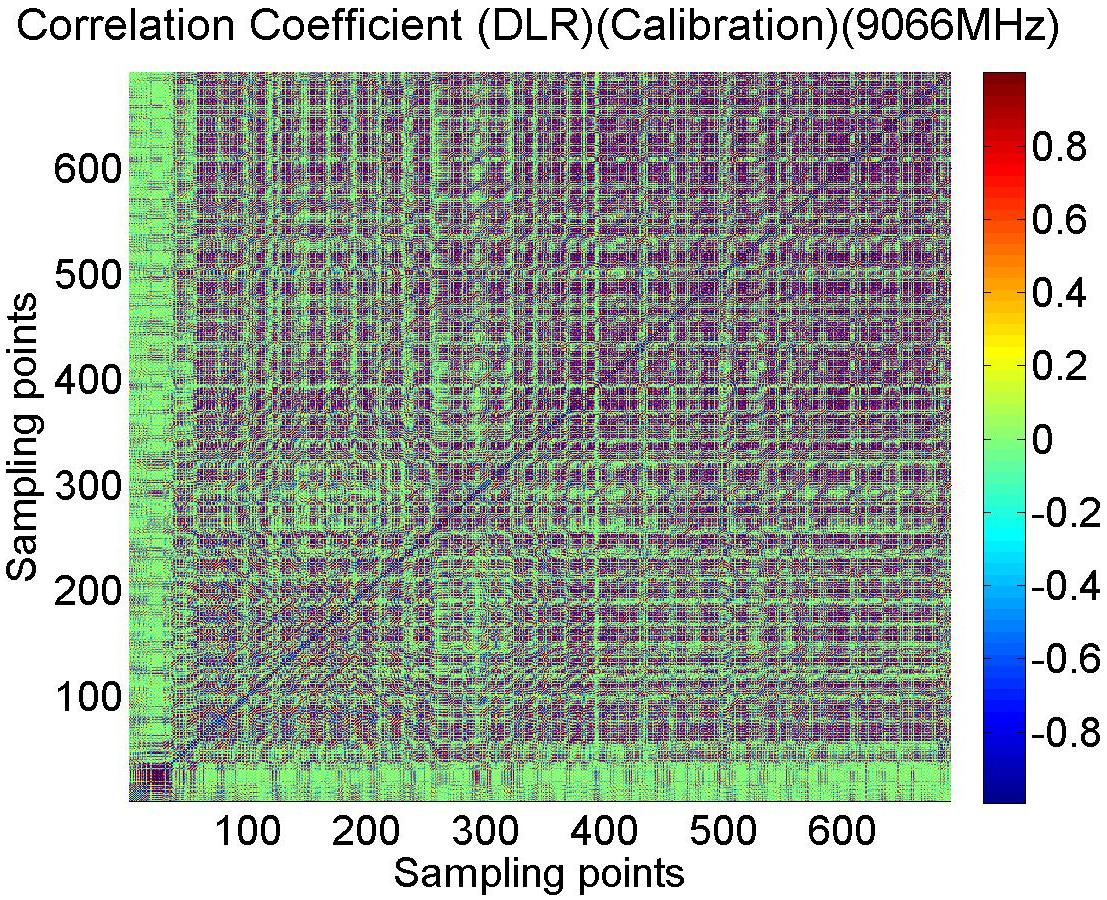}
\caption{Correlation coef{}f{}icients of the calibration samples.}
\label{DLR-corr}
\end{figure}

The algorithm used in solving the linear system (in the form of Eq.~\ref{eq:DLR}) is least squares, which is a standard method to solve linear regression problems. The method relies on minimizing the dif{}ference of prediction ($B_{pre}$) and measurement ($B$) whilst modifying the elements of the matrix $M^{\ast}$. In our case, the size of matrix $M^{\ast}$ is related to the number of sampling points, therefore a considerable number of unknown variables (in this particular case, 701) needs to be determined. This is computationally expensive. As all sampling points are used in the regression, each column of matrix $A$ is a regressor. The correlation coef{}f{}icients $R \in \mathbb{R}^{n\times n}$ of these regressors are calculated as
\begin{equation}
R(i,j) = \frac{\mathrm{Cov}(c_i,c_j)}{\sqrt{\mathrm{Cov}(c_i,c_i)\cdot \mathrm{Cov}(c_j,c_j)}}, 
\label{eq:corr}
\end{equation}
where $c_i$ and $c_j$ are $i^{th}$ and $j^{th}$ column of matrix $A$ respectively def{}ined in Eq.~\ref{eq:DLR-A}, $\mathrm{Cov}(c_i,c_j)$ is the covariance between two regressors $c_i$ and $c_j$. In this def{}inition, $R$ = $\pm 1$ means strong correlation while $R$=$0$ means no correlation. Fig.~\ref{DLR-corr} shows the correlation coef{}f{}icients of the calibration samples. Most regressors are strongly correlated and this makes the linear system sensitive to the f{}luctuations of the calibration samples. In our case, the system is vulnerable to noise. Moreover, even though it is not the case here, overf{}itting is always a potential problem unless the calibration samples are fairly large. To avoid these risks, it is better to reduce the system size from multi-hundred to several tens, and then apply linear regression on the reduced system. The method we used for this purpose is described in the following section.

\section{Singular Value Decomposition}
In order to f{}ind a small number of prominent components from the signal matrix $A$, a method known as \textit{Singular Value Decomposition} (SVD) is used \cite{stat-4}. In general, SVD looks for patterns of a matrix in terms of SVD modes without relating them to any physical parameters (like dipole mode frequency, quality factor, beam position, etc.). Those SVD modes are natural groupings of the signal matrix, which are not clearly visible or explicitly def{}ined in the matrix itself.

\begin{figure}[h]\center
\subfigure[]{
\includegraphics[width=0.35\textwidth]{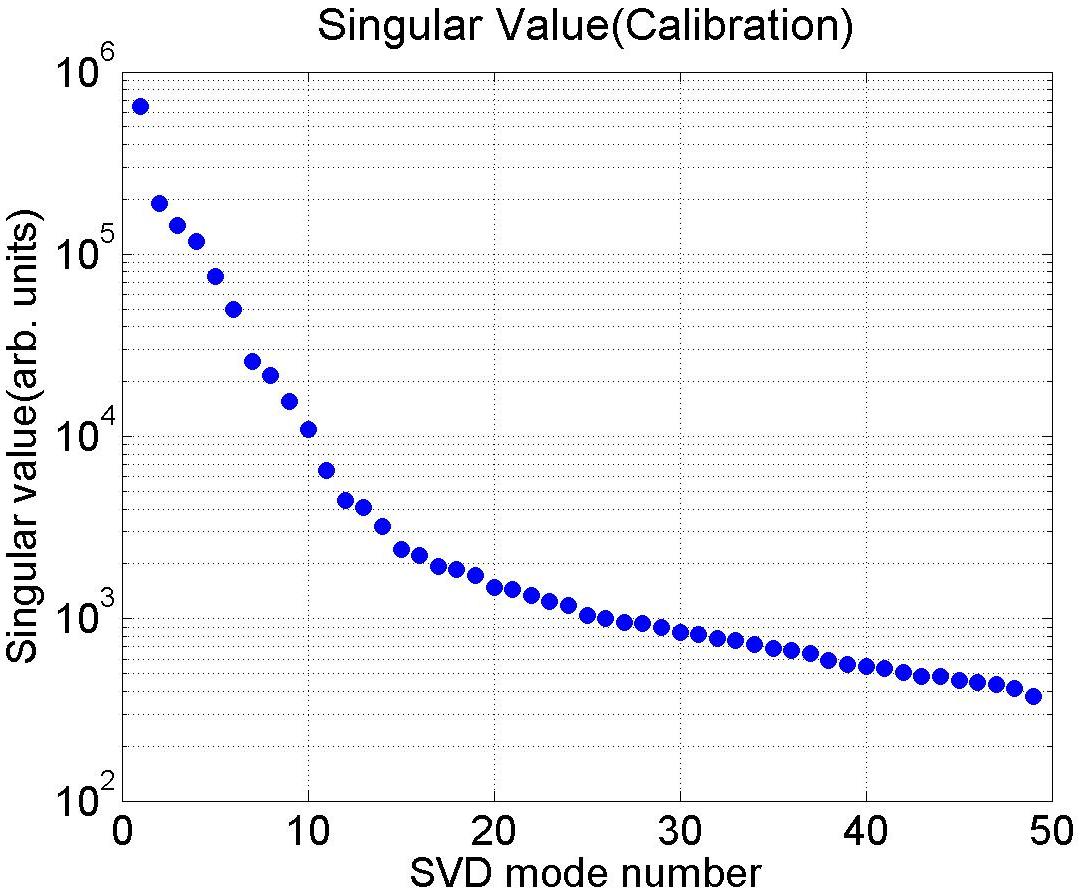}
\label{SVD-sv}
}
\quad\quad
\subfigure[]{
\includegraphics[width=0.35\textwidth]{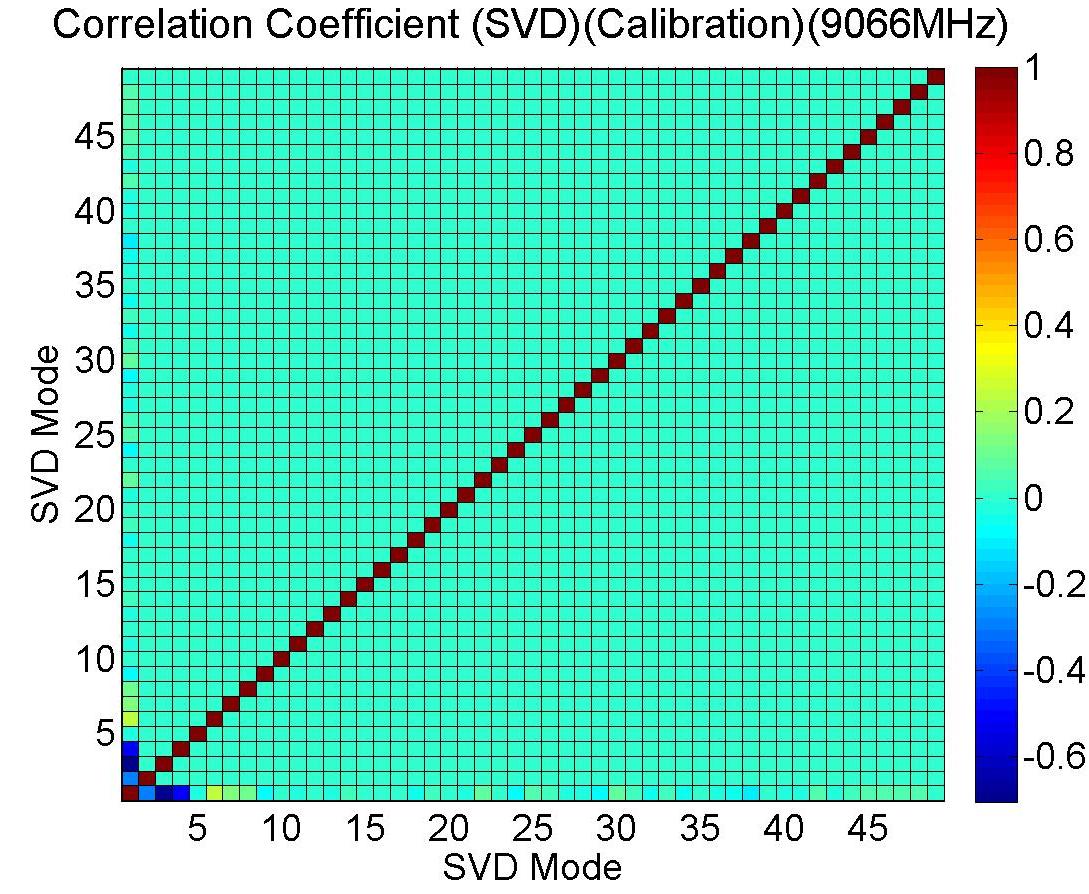}
\label{SVD-corr}
}
\caption{(a) Singular values of SVD modes; (b) Correlation coef{}f{}icients of SVD modes decomposed from the calibration samples.}
\label{SVD-sv-corr}
\end{figure}

\begin{figure}[h]\center
\subfigure[]{
\includegraphics[width=0.3\textwidth]{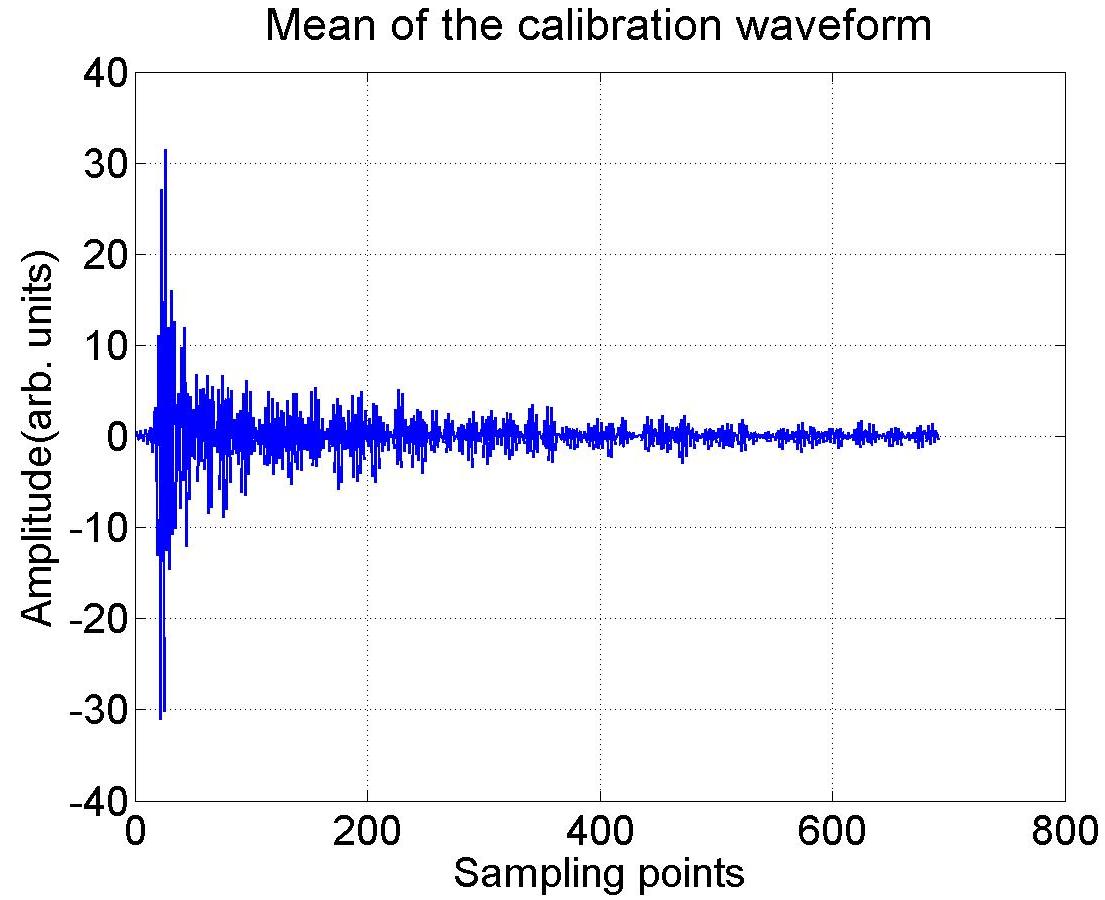}
\label{SVD-mean-wfm}
}
\subfigure[]{
\includegraphics[width=0.3\textwidth]{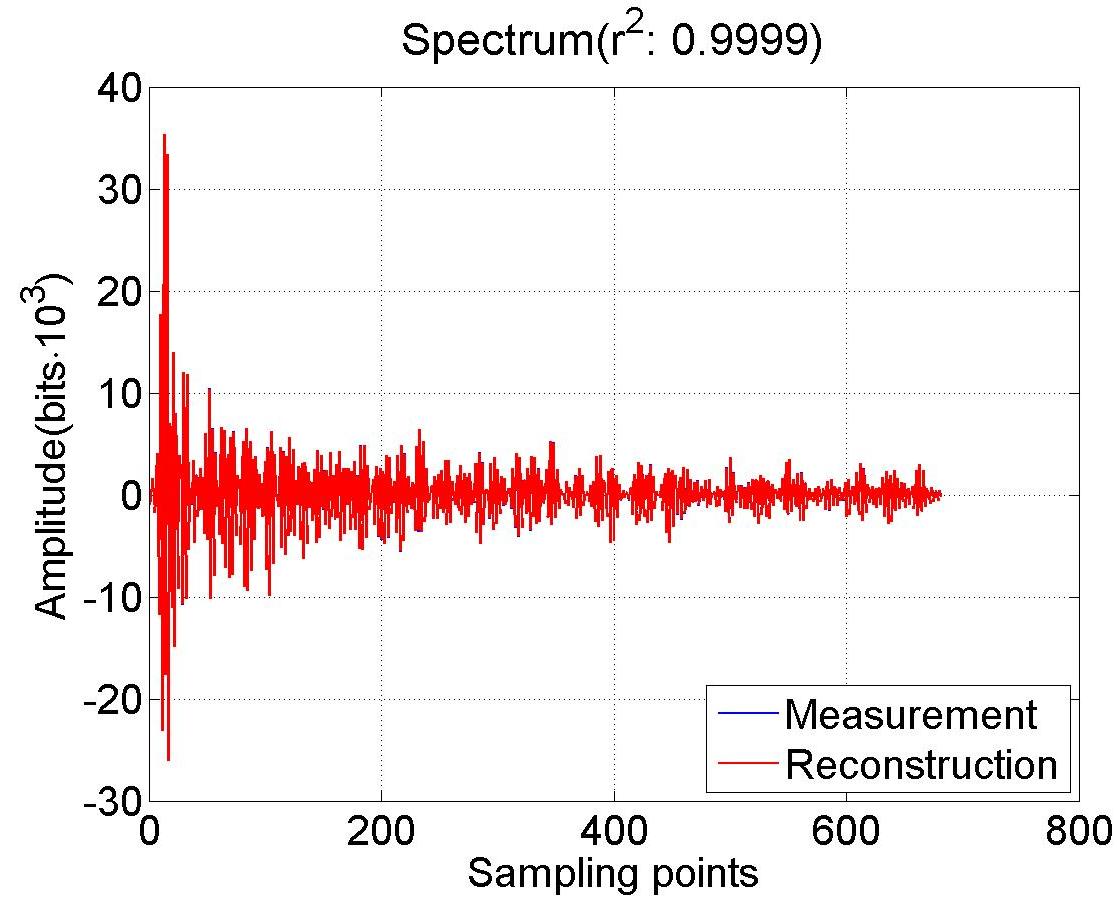}
\label{SVD-reco}
}
\subfigure[]{
\includegraphics[width=0.3\textwidth]{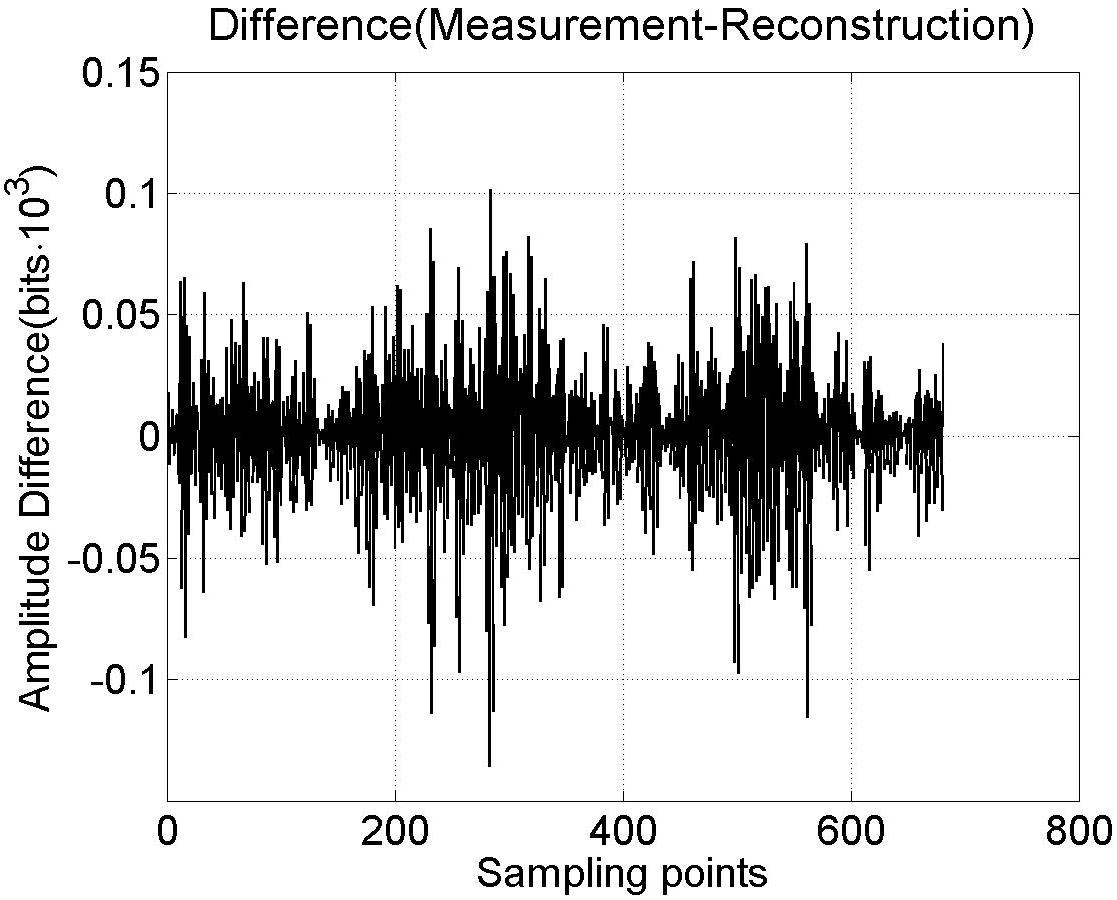}
\label{SVD-reco-res}
}
\caption{(a) Mean of the calibration waveforms; (b) Reconstructed waveform (red) from the f{}irst twelve SVD modes compared with original waveform (blue). (c) Dif{}ference of the original and reconstructed waveform.}
\label{SVD-wfm-reco}
\end{figure}

Applying SVD, the matrix $A$ is decomposed into the product of three matrices,
\begin{equation}
A=U\cdot S\cdot V^T,
\label{eq:SVD-1}
\end{equation}
where $U\in\mathbb{R}^{m\times m}$ and $V\in\mathbb{R}^{n\times m}$. The columns of $U$ and $V^T$ are singular vectors of $A$, which form the bases of the decomposition\footnote{In general case, $U\in\mathbb{R}^{m\times m}$, $S\in\mathbb{R}^{m\times n}$ and $V\in\mathbb{R}^{n\times n}$. However, in the case of $m<n$, only the f{}irst $m$ SVD modes have non-zero singular values.}. $S$ is a diagonal matrix whose non-zero elements are singular values. Applying SVD on the calibration samples (blue dots in Fig.~\ref{2D-grid-C3}), the singular value of each SVD mode is obtained and shown in Fig.~\ref{SVD-sv}. It can be seen that the f{}irst few SVD modes have relatively large singular values, in other words, they are the dominant patterns of matrix $A$. The correlation coef{}f{}icients of all SVD modes decomposed from the calibration samples are calculated as shown in Fig.~\ref{SVD-corr}. The f{}irst SVD mode has relatively strong correlations with the third and the fourth SVD mode. This is because the mean of the 49 calibration waveforms is not zero as shown in Fig.~\ref{SVD-mean-wfm}.

\begin{figure}[h]\center
\includegraphics[width=0.9\textwidth]{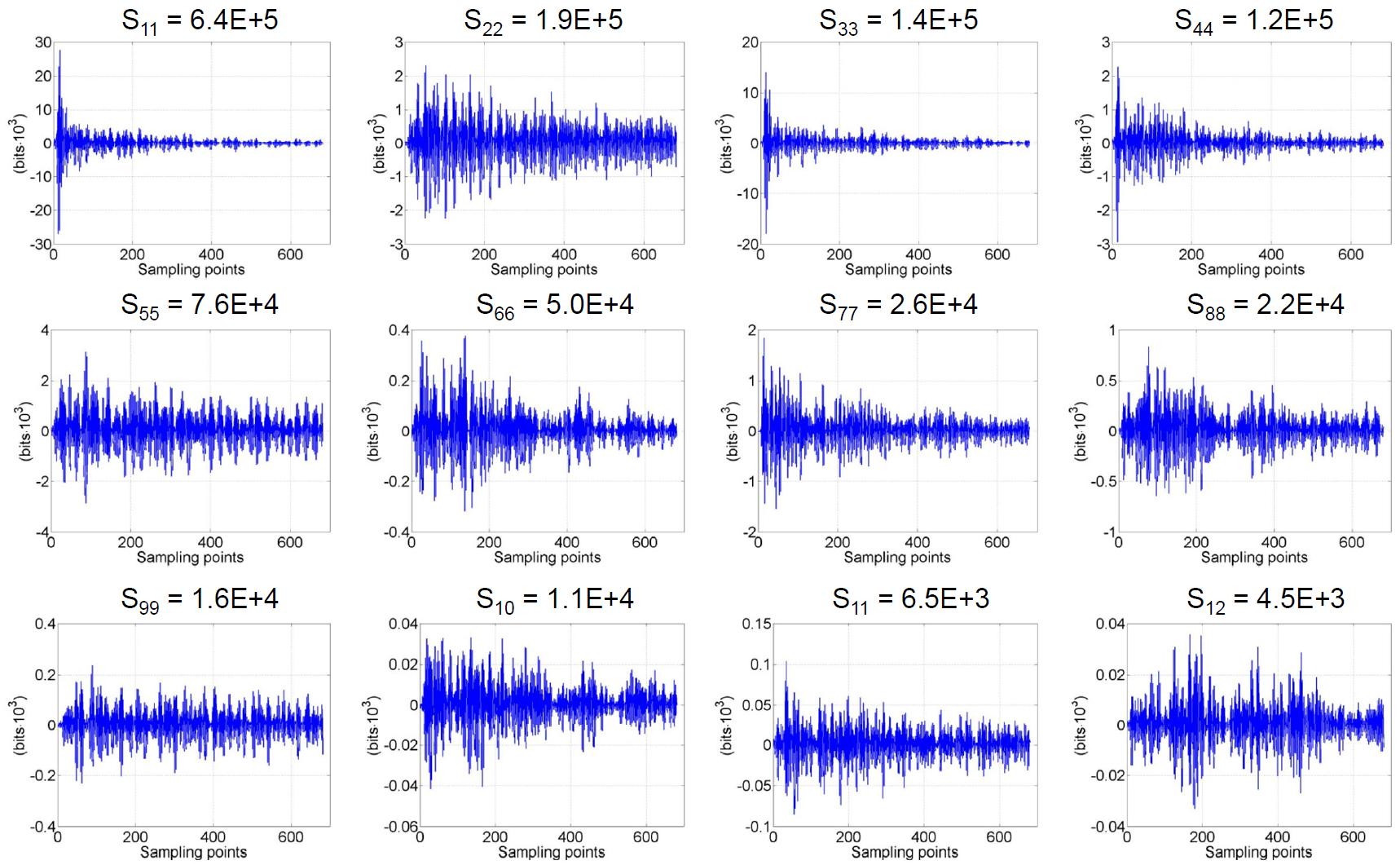}
\caption{Spectra reconstructed from each of the f{}irst twelve SVD modes.}
\label{SVD-12reco}
\end{figure}

Each of the SVD modes can be used to produce a spectra matrix by
\begin{equation}
A_i=U_i\cdot S_{ii}\cdot V^T_i,
\label{eq:SVD-reco-1}
\end{equation}	
where $U_i$ is the $i^{th}$ column of $U$, and $V^T_i$ is the $i^{th}$ row of $V^T$. $S_{ii}$ denotes the $i^{th}$ diagonal element of $S$. $A_i$ has the same size as the original spectra matrix $A$. Fig.~\ref{SVD-12reco} shows one spectrum for each of the f{}irst twelve $A_i$'s. The approximation $A_{reco}$ can be calculate by simply summing over the related $A_i$ by
\begin{equation}
A_{reco}=\sum_{i=1}^{p} A_i.
\label{eq:SVD-reco-2}
\end{equation}
By combining the f{}irst twelve SVD modes ($p$ = 12), $A_{reco}$ is plotted together with the original spectra matrix $A$ in Fig.~\ref{SVD-reco} and the dif{}ference of $A$ and $A_{reco}$ is shown in Fig.~\ref{SVD-reco-res}. The spectrum can be well represented by using only the f{}irst twelve SVD modes.

Having the SVD base vectors from the decompositions of spectra matrix $A$, the amplitude of all SVD modes, $A_{svd, full}$, can be obtained for each beam movement by
\begin{equation}
A_{svd, full}=A \cdot V
\label{eq:SVD-2}
\end{equation}
The f{}irst twelve columns of the matrix $A_{svd, full}$ constitute the matrix $A_{svd}$ along with one column of $1$ representing the intercept term. The size of the matrix $A_{svd}$ is signif{}icantly smaller (13 columns, representing 12 SVD modes and one intercept term) compared to the original matrix $A$ (701 columns, representing 700 sampling points and one intercept term). Replacing $A$ by $A_{svd}$ in the regression (Eq.~\ref{eq:DLR}), the linear system composed by $A_{svd}$ and $B$ is now over-determined, which has a well-known best solution in least-square sense. The position prediction $B_{pre}$ can then be obtained using Eq.~\ref{eq:DLR-pre} by replacing $A$ with $A_{svd}$ for calibration samples. The prediction for the validation samples has two steps: f{}irst, project $A$ onto the base vectors obtained from calibration samples to get the amplitude of the f{}irst twelve SVD modes using Eq.~\ref{eq:SVD-2}; second, substitute the SVD amplitude matrix in Eq.~\ref{eq:DLR-pre} to get the position prediction.

\begin{figure}[h]\center
\includegraphics[width=0.55\textwidth]{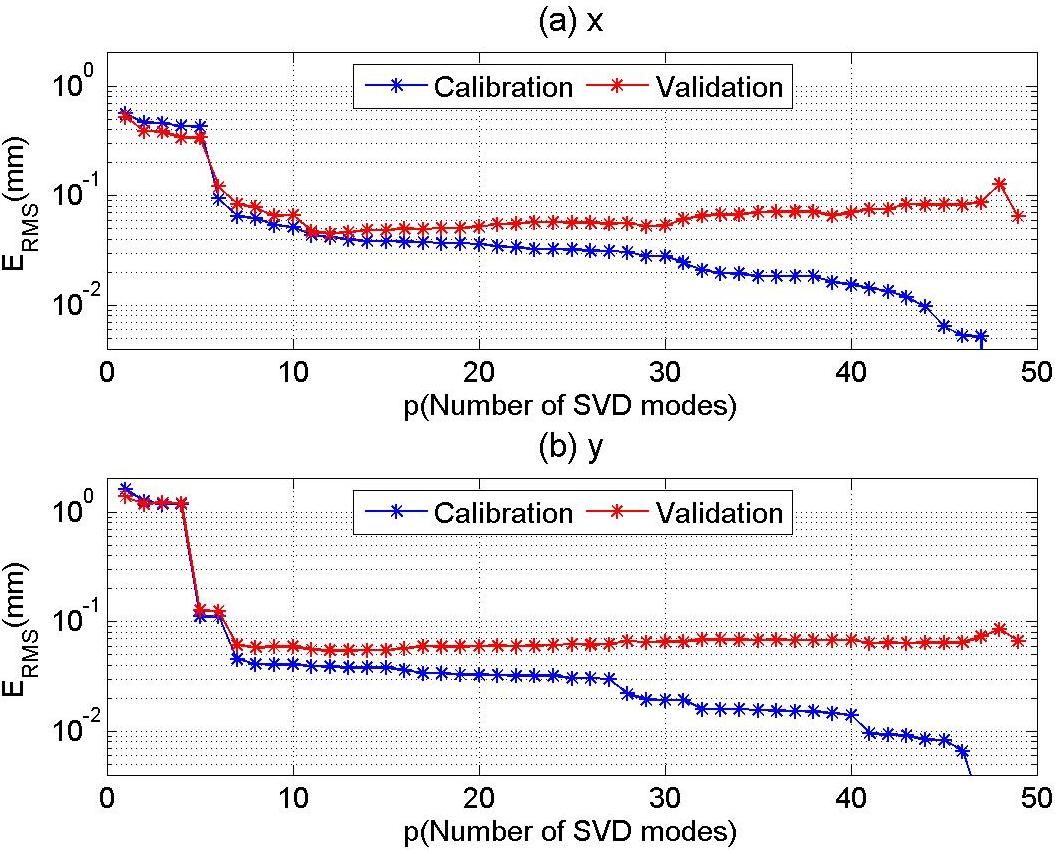}
\caption{Contribution of the f{}irst $p$ SVD modes to determine the transverse beam position $x$ and $y$ measured by the $E_{RMS}$.}
\label{SVD-sv-rms}
\end{figure}

The contribution of combining the f{}irst $p$ SVD modes to determine the transverse beam position $x$ and $y$ is measured by rms prediction error as shown in Fig.~\ref{SVD-sv-rms}. Using the f{}irst twelve SVD modes is seen to give an optimal performance for the $x$ plane. It is also a reasonable choice for the $y$ plane (Fig.~\ref{SVD-sv-rms}(b)) even though less SVD modes would be suf{}f{}icient. This conf{}irms, with Fig.~\ref{SVD-sv} and Fig.~\ref{SVD-reco}, that the f{}irst few modes contain most of the information. By combining the f{}irst twelve SVD modes, one can get the majority of beam position information.

\begin{figure}[h]\center
\includegraphics[width=0.7\textwidth]{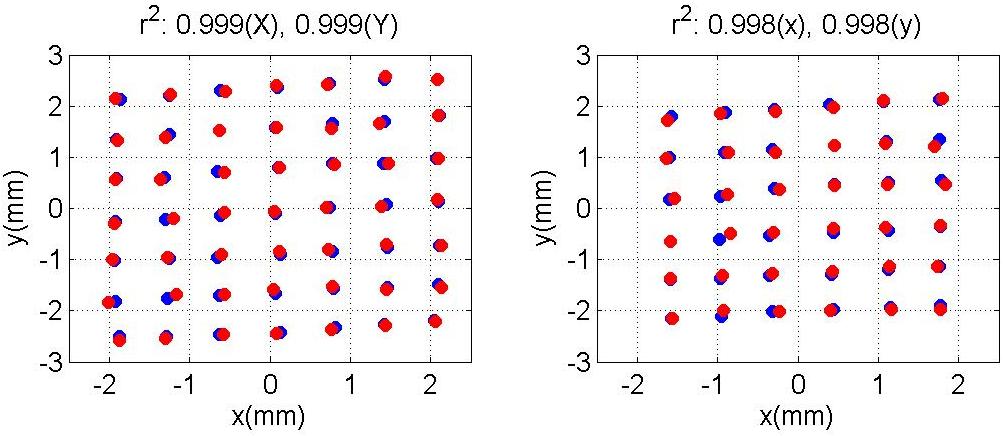}
\caption{Measurement (blue) and prediction (red) of the transverse beam position from calibration (left) and validation (right) samples. The applied method is SVD with the f{}irst twelve SVD modes.}
\label{SVD-B-Bp}
\end{figure}

\begin{figure}[h]\center
\includegraphics[width=0.6\textwidth]{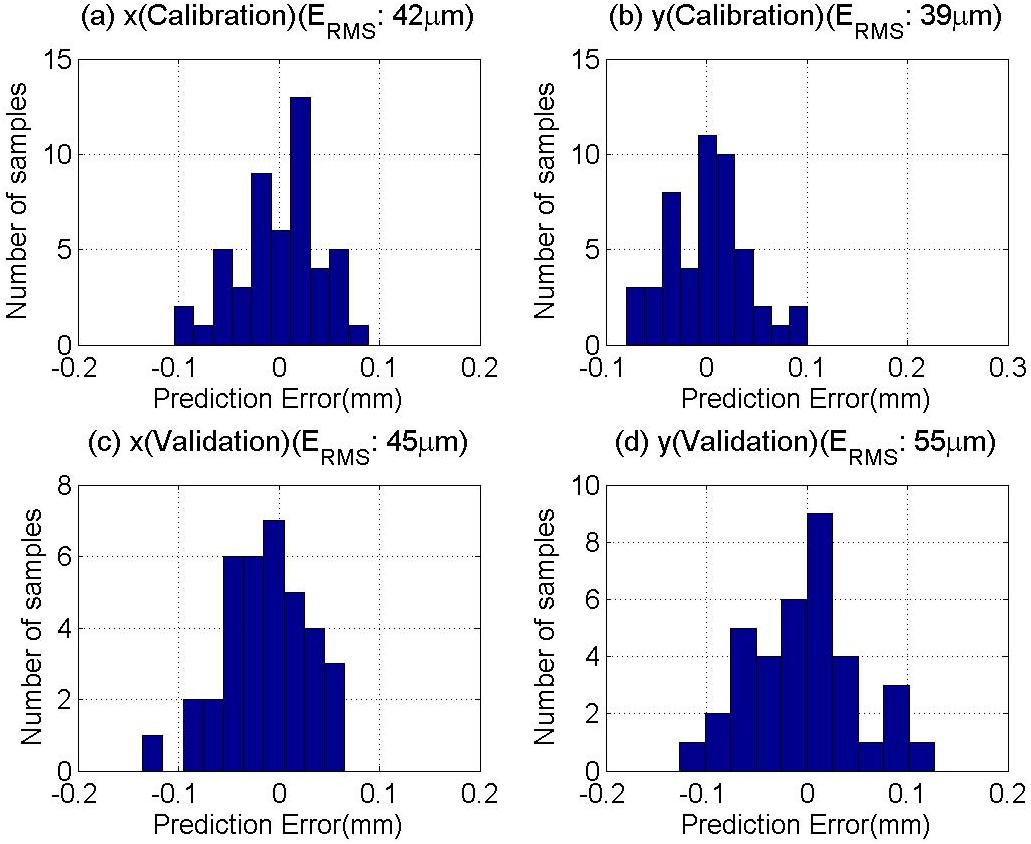}
\caption{Dif{}ference of measured and predicted transverse beam position from calibration and validation samples. The applied method is SVD with the f{}irst twelve SVD modes. The position resolution is denoted as rms value on each plot.}
\label{SVD-diff}
\end{figure}

Fig.~\ref{SVD-B-Bp} shows the HOM response from calibration and validation samples by using the f{}irst twelve SVD modes to determine the transverse beam position $x$ and $y$. The prediction errors remain small and comparable for calibration and validation samples as shown in Fig.~\ref{SVD-diff}. The position resolution is 45~$\mu m$ for $x$ and 55~$\mu m$ for $y$.

\FloatBarrier
\section{Comparison of DLR and SVD}
\begin{figure}[h]\center
\includegraphics[width=0.7\textwidth]{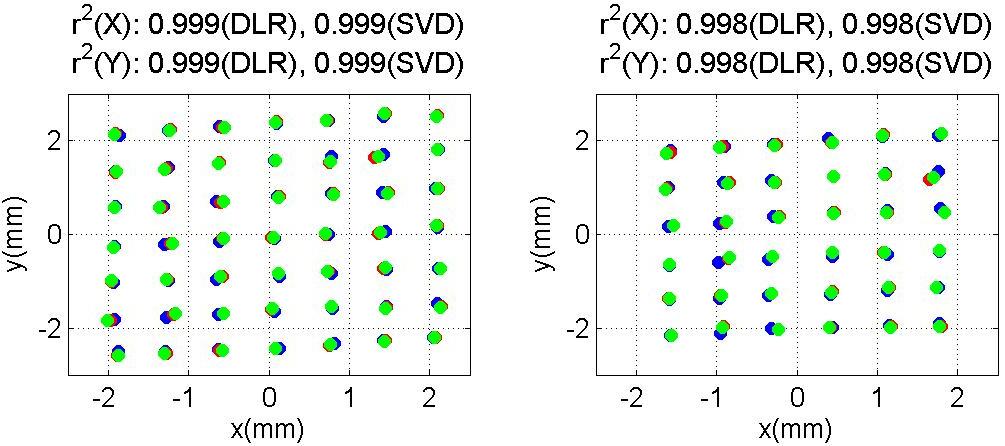}
\caption{Measurement (blue) and prediction of the transverse beam position from calibration (left) and validation (right) samples. The applied method is DLR (red) and SVD with the f{}irst twelve SVD modes (green).}
\label{SVD-DLR-B-Bp}
\end{figure}

A direct comparison of DLR and SVD on the same samples (split as Fig.~\ref{2D-grid-C3}) is shown in Fig.~\ref{SVD-DLR-B-Bp}. The results obtained by the two dif{}ferent methods are similar. The comparisons of the rms prediction errors are listed in Table.~\ref{table-SVD-DLR-err}. 

\begin{table}[h]\center
\caption{Comparison of rms prediction errors of DLR and SVD corresponding to Fig.~\ref{DLR-diff} and Fig.~\ref{SVD-diff}.}
\label{table-SVD-DLR-err}
\begin{tabular}{c|c|c}
\hline
& \textit{\textbf{x}} ($\mu m$) & \textit{\textbf{y}} ($\mu m$)\\
\hline
Calib(DLR) & 41& 39 \\
\hline
Calib(SVD) & 42& 39 \\
\hline\hline
Valid(DLR) & 47& 56 \\
\hline
Valid(SVD) & 45& 55 \\
\hline
\end{tabular}
\end{table}

\begin{figure}[h]\center
\subfigure[One example]{
\includegraphics[width=0.3\textwidth]{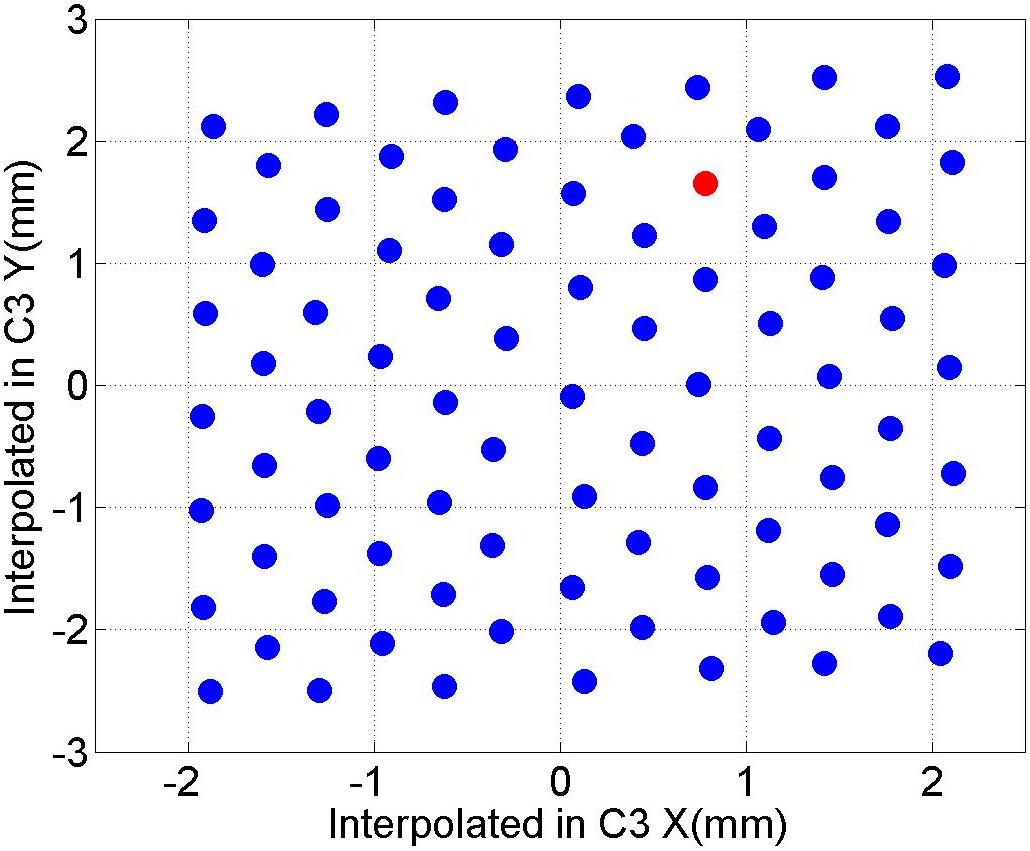}
\label{kfold-B}
}
\quad\quad
\subfigure[LOOCV]{
\includegraphics[width=0.3\textwidth]{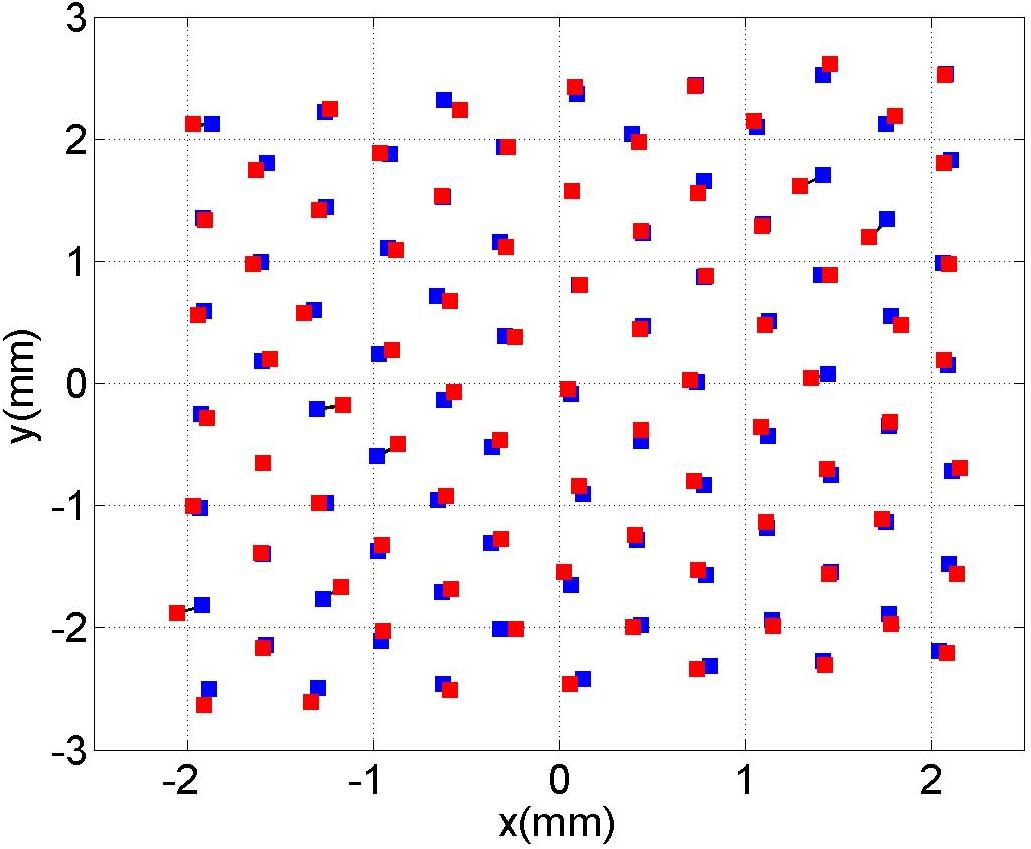}
\label{kfold-B-Bp}
}
\caption{(a) Calibration (blue) and validation (red) samples from one of the 85 sample splits during the cross validation. (b) Measurement (blue) and prediction (red) of the transverse beam position from each of the 85 sample splits using the cross validation.}
\label{kfold-B-B-Bp}
\end{figure}

\begin{figure}[h]\center
\includegraphics[width=0.5\textwidth]{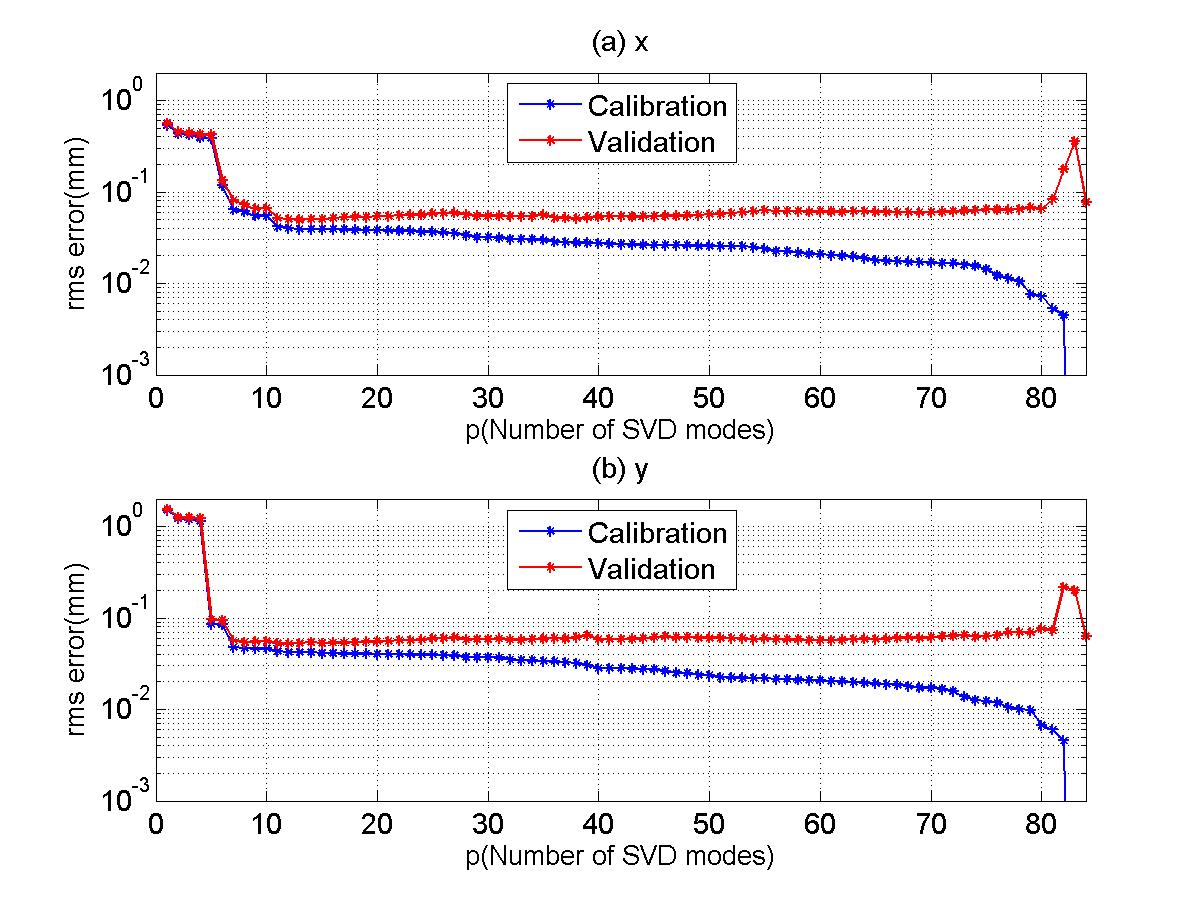}
\caption{Contribution of the f{}irst $p$ SVD modes to determine the transverse beam position $x$ and $y$ measured by the sample-independent rms prediction error.}
\label{kfold-sv-rms}
\end{figure}

Until now, our analysis is based on a specif{}ic sample split (Fig.~\ref{2D-grid-C3}). To remove sample dependence, a technique namely \textit{cross-validation} is used \cite{stat-8}. As in the discussed example the total sample size is only 85 (49 calibration positions plus 36 validation positions), one can apply the \textit{leave-one-out cross-validation} (LOOCV) \cite{stat-8} technique. LOOCV uses only one from the total samples for the validation and the remainings as calibrations as shown in Fig.~\ref{kfold-B} for one example. This is repeated for every sample (85 times in the considered case). Fig.~\ref{kfold-B-Bp} shows the measurement and the prediction of each validation sample. The method we used is SVD with the f{}irst 12 SVD modes. The rms prediction error is then calculated on the prediction error for the validation sample from all 85 dif{}ferent sample splits. The sample-independent rms prediction error is 50~$\mu m$ for $x$ and 52~$\mu m$ for $y$. Fig.~\ref{kfold-sv-rms} shows the sample-independent rms error using the f{}irst $p$ SVD modes. Similar to the case of sample split shown in Fig.~\ref{2D-grid-C3}, the f{}irst 12 SVD modes give an optimal prediction accuracy for $x$ and even less SVD modes (the f{}irst 7) are needed for $y$. Since the calibration samples for each of the 85 splits are similar, the sample-independent rms error for the validations is a good estimation of the position resolution. The results are similar compared with the f{}ixed sample split case shown in Table~\ref{table-SVD-DLR-err}.

\chapter{Position Resolutions}\label{ca:res}
Resolutions for the position determination using various modes are discussed in this chapter. Only one set of modes for each modal option (Table~\ref{table-modal-options}) is described. The data obtained by the VME digitizer are used for the f{}irst two cavity dipole bands and the f{}ifth cavity dipole band, while the $\mu$TCA digitizer is used for the beam-pipe modes. The selection is based on data availability for this set of measurements\footnote{The VME digitizer was unavailable when measuring the beam-pipe modes due to technical problems.}. The data analysis technique used in this chapter is SVD as explained in Chapter~\ref{ca:ana}. The number of SVD modes for regression is determined based on a stable and good position resolution. The results for the f{}irst (C1H2) and the last (C4H2) HOM coupler in the four-cavity string are excluded in this chapter. HOM coupler C1H2 picks up the modes transmitted from the f{}irst accelerating module, while C4H2 has weak signals. Both the time-domain waveforms and frequency-domain f{}ft amplitudes are studied. The frequency domain is used for comparisons with previous studies using a real-time spectrum analyzer. In general, time-domain signals give better position resolution than frequency-domain f{}f{}t amplitudes. This proves that the phase information is crucial for the beam position determination. A full list of resolutions for all modes studied with the test electronics can be found in Appendix~\ref{app:res} along with the number of SVD modes used for a stable and good position determination presented in Appendix~\ref{app:nsvd}. 

\section{The Localized Beam-pipe Dipole Modes}
\begin{figure}[h]\center
\includegraphics[width=0.8\textwidth]{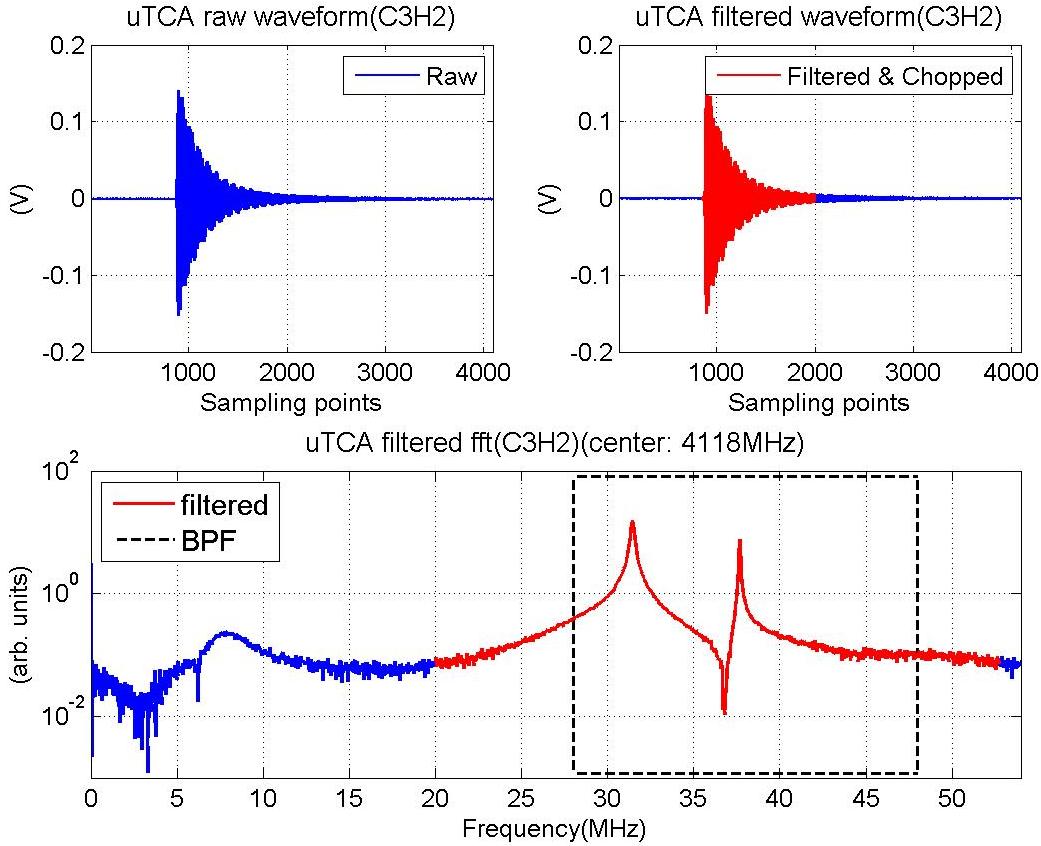}
\caption{HOM signal. The center frequency is 4118~MHz with a 20~MHz BPF.}
\label{wfm-fft-4118MHz-BP03}
\end{figure}

The LO was set to downconvert 4118~MHz to 70~MHz, and a 20~MHz BPF was then applied to the down-converted signal. Fig.~\ref{wfm-fft-4118MHz-BP03} shows the down-converted signal after it was processed with the $\mu$TCA digitizer, in both the time- and frequency domain (after a f{}f{}t). As mentioned in the previous chapter, the mathematical ideal f{}ilter is applied to the down-converted time-domain waveform for cutting away the other modes after the down-conversion. The f{}inal f{}iltered signal is shown as red in Fig.~\ref{wfm-fft-4118MHz-BP03}. Compared with the spectrum analyzer signal obtained in previous studies \cite{acc39-hombpm-12}, the displayed signal f{}iltered from 4100~MHz to 4133~MHz has proved to contain the dipole modes localized in the beam pipes. As the BPF was set to 70~MHz with a 10~MHz bandwidth on each side, the main component of the f{}iltered signal is from 4108~MHz to 4128~MHz. The IF signal is undersampled with the 108~MS/s $\mu$TCA digitizer.  

\begin{figure}[h]\center
\subfigure[time domain]{
\includegraphics[width=0.7\textwidth]{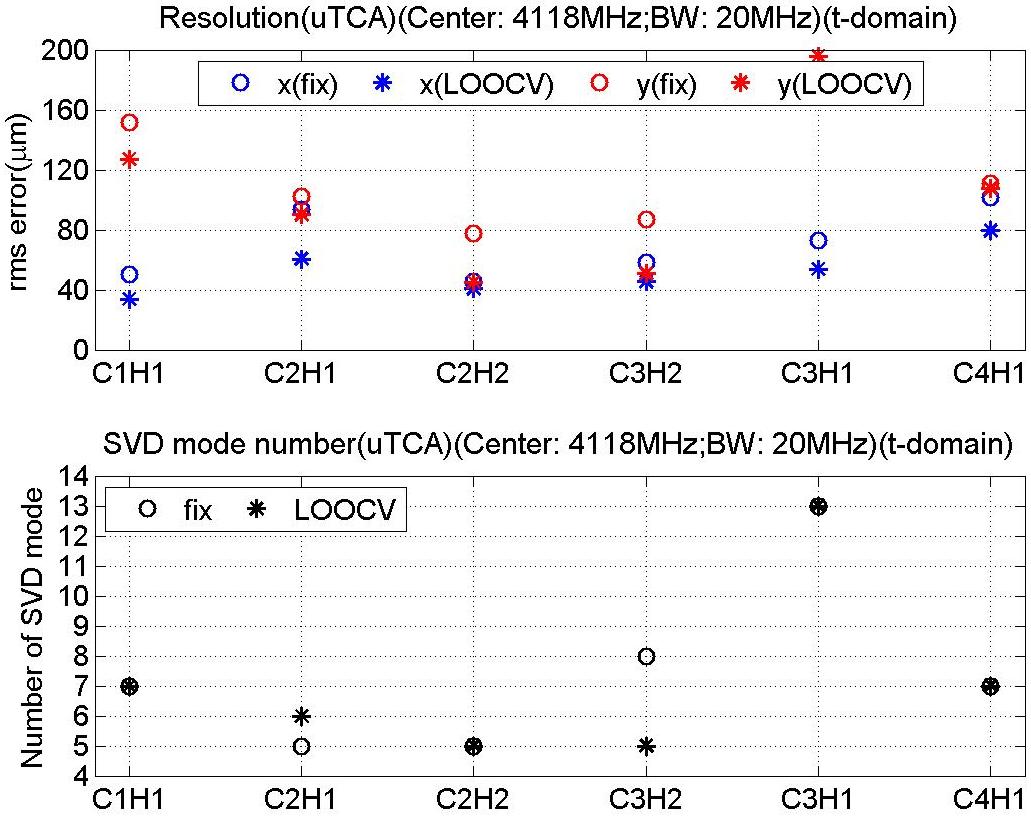}
\label{rms-8HOM-4118MHz-BP03-uTCA-wfm}
}
\subfigure[frequency domain]{
\includegraphics[width=0.7\textwidth]{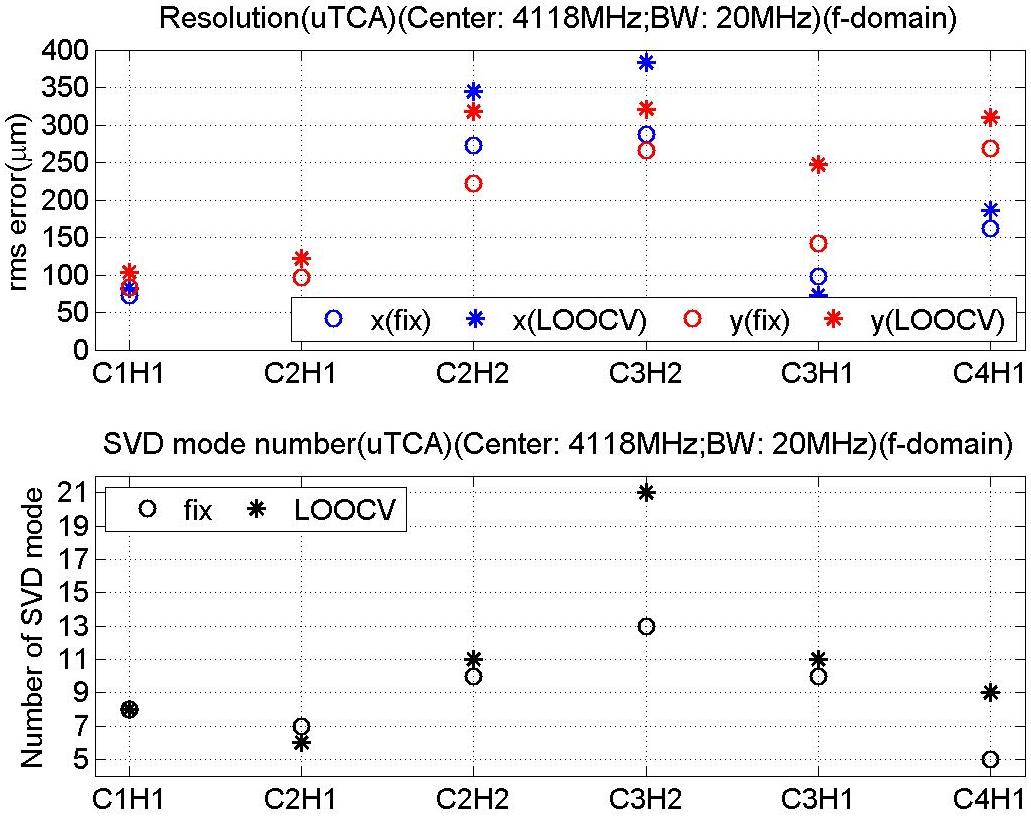}
\label{rms-8HOM-4118MHz-BP03-uTCA-fft}
}
\caption{Position resolution and SVD mode number used for position determination for both time-domain and frequency-domain. The center frequency is 4118~MHz with a 20~MHz BPF.}
\label{rms-8HOM-4118MHz-BP03-uTCA}
\end{figure}

To determine the beam position, we considered both the time-domain and the frequency-domain (the f{}f{}t amplitude, discard f{}f{}t phase) signal. They are shown as red waveform and red spectrum in Fig.~\ref{wfm-fft-4118MHz-BP03} for one example. The samples have been split in a f{}ixed manner as shown in Fig.~\ref{bpm-reading}(a) along with a LOOCV case. The position resolution for these modes is shown in Fig.~\ref{rms-8HOM-4118MHz-BP03-uTCA-wfm} for time-domain waveform and Fig.~\ref{rms-8HOM-4118MHz-BP03-uTCA-fft} for frequency-domain f{}f{}t amplitude along with the number of SVD modes used in the respective regression. One may notice that the number of SVD modes is similar for coupler C2H2 and C3H2. Both of these two couplers see clearly orthogonal polarizations of these two peaks from previous studies \cite{acc39-hombpm-12,acc39-hombpm-5}. 

\begin{figure}[h]\center
\includegraphics[width=0.85\textwidth]{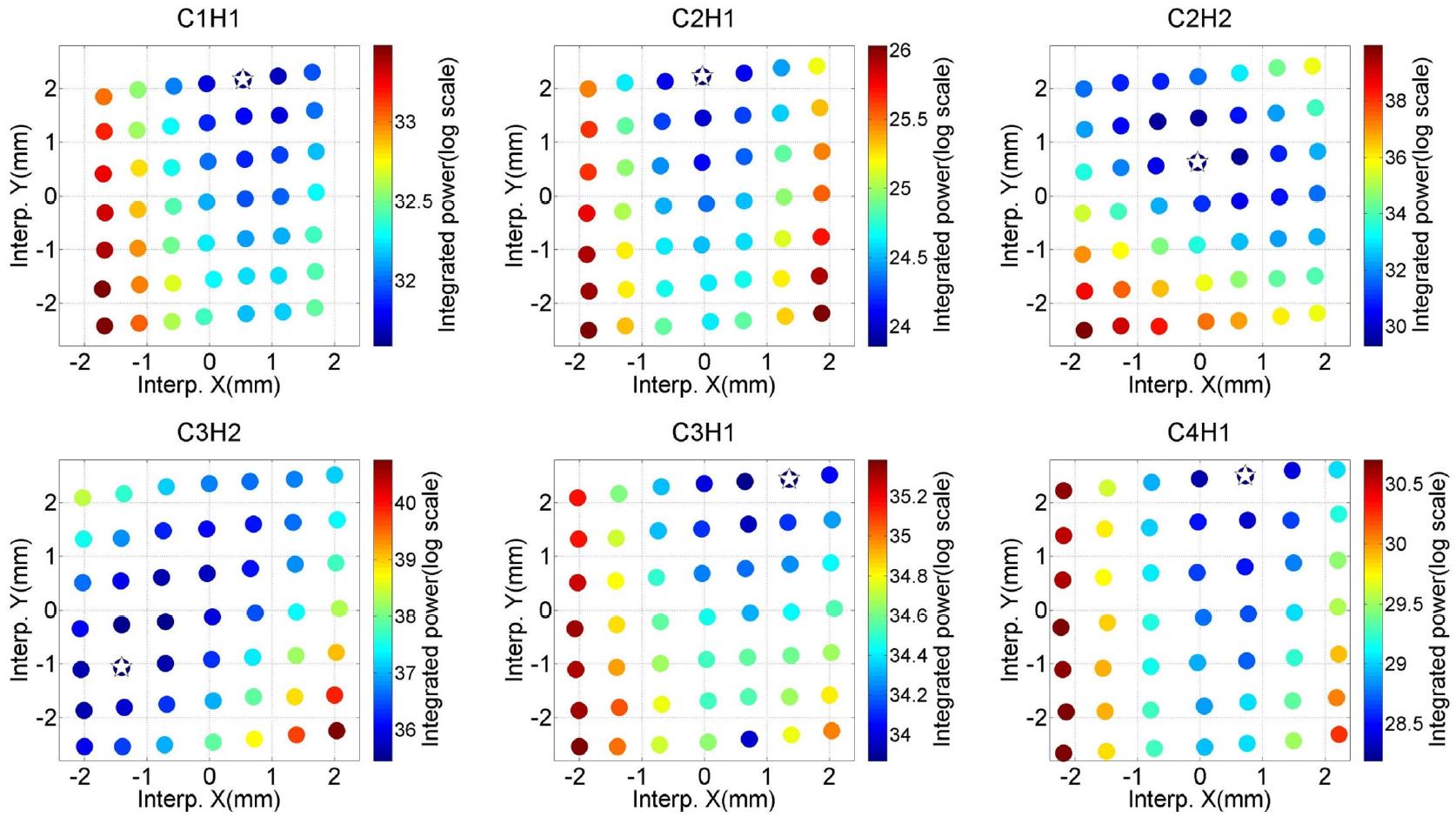}
\caption{Integrated power as a function of transverse position interpolated in each cavity. The log scale magnitude of power is denoted by dif{}ferent color. The minimum power is marked with white pentagon. The center frequency is 4118~MHz with a 20~MHz BPF.}
\label{power-well-B-4118MHz-BP03}
\end{figure}

\begin{figure}[h]\center
\includegraphics[width=0.98\textwidth]{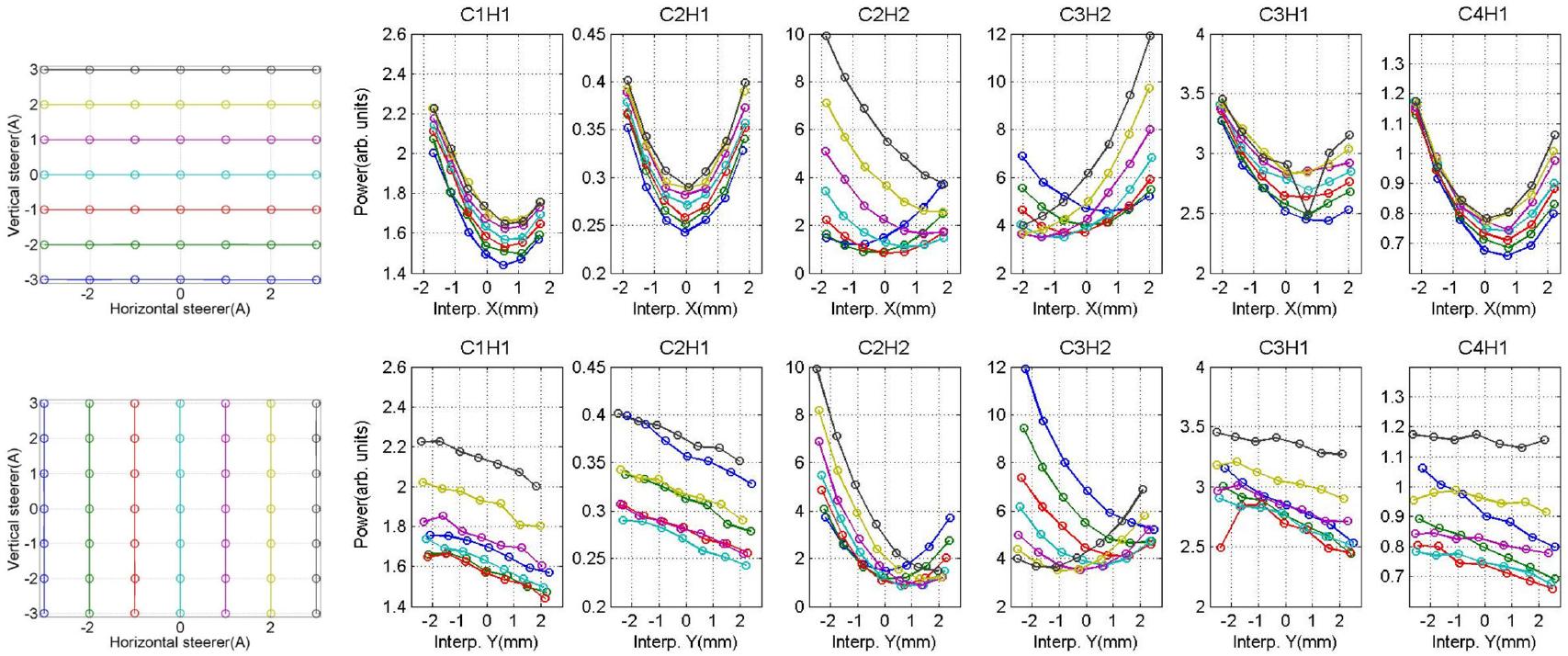}
\caption{Integrated power as a function of transverse position interpolated in each cavity. The center frequency is 4118~MHz with a 20~MHz BPF. }
\label{power-scatter-B-4118MHz-BP03}
\end{figure}

\FloatBarrier
The integrated power over the frequency range shown in Fig.~\ref{wfm-fft-4118MHz-BP03} (red spectrum) is calculated for each beam position. Fig.~\ref{power-well-B-4118MHz-BP03} shows the integrated power distribution for each HOM coupler. The position, which has minimum integrated power, is marked with white pentagon. Neither the power distribution nor the power minimum is similar among couplers. This might be a proof that modes are localized inside each beam pipe presented in the frequency region for power integration. It might also be an implication of dif{}ferent electrical axes for individual cavities. This dif{}ference can be attributed to asymmetric structure due to couplers and fabrication tolerance of the cavities. Fig.~\ref{power-scatter-B-4118MHz-BP03} decomposes the power distribution plot into horizontal and vertical moves. 

%

\FloatBarrier
\section{Coupled Cavity Modes - The First Dipole Band}
\begin{figure}[h]\center
\includegraphics[width=0.8\textwidth]{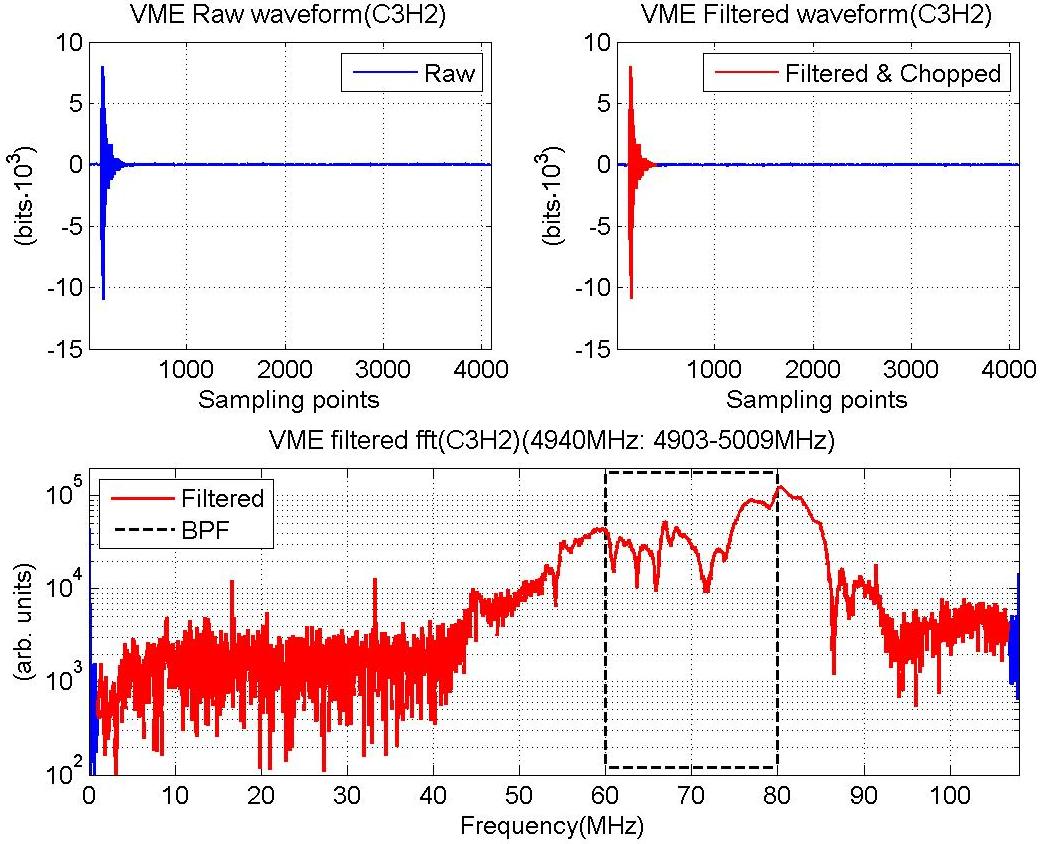}
\caption{HOM signal. The center frequency is 4940~MHz with a 20~MHz BPF.}
\label{wfm-fft-4940MHz-D109}
\end{figure}

The LO was set to downconvert 4940~MHz to 70~MHz, and a 20~MHz BPF was then applied to the down-converted signal. Fig.~\ref{wfm-fft-4940MHz-D109} shows the down-converted signal after it was processed by the VME digitizer, in both time- and frequency domain (after a f{}f{}t). Modes in this band are coupled, therefore, the mathematical ideal f{}ilter is used only to preserve good signal quality as shown in Fig.~\ref{wfm-fft-4940MHz-D109}. Compared with the spectrum analyzer signal obtained in previous studies \cite{acc39-hombpm-12}, the displayed signal f{}iltered from 4903~MHz to 5009~MHz has proved to contain dipole modes propagating amongst cavities. As the BPF was set to 70~MHz with a 10~MHz bandwidth on each side, the main component of the f{}iltered signal is from 4930~MHz to 4950~MHz.    

\begin{figure}[h]\center
\subfigure[time domain]{
\includegraphics[width=0.7\textwidth]{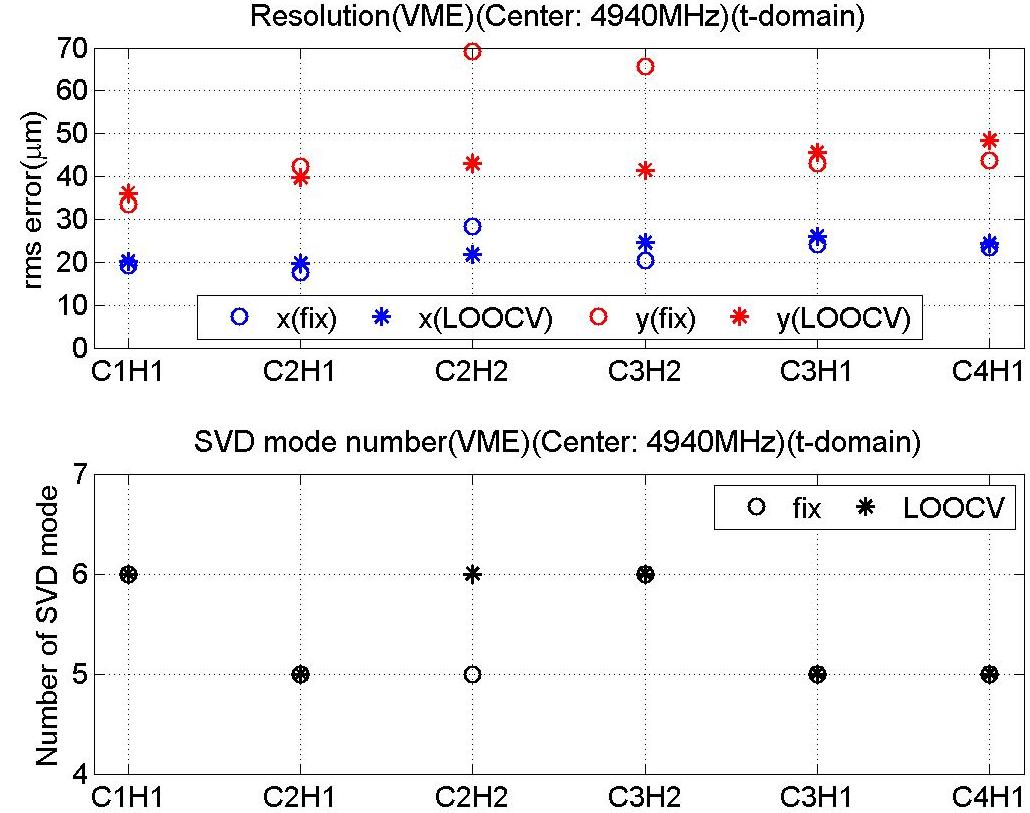}
\label{rms-8HOM-4940MHz-D109-VME-wfm}
}
\subfigure[frequency domain]{
\includegraphics[width=0.7\textwidth]{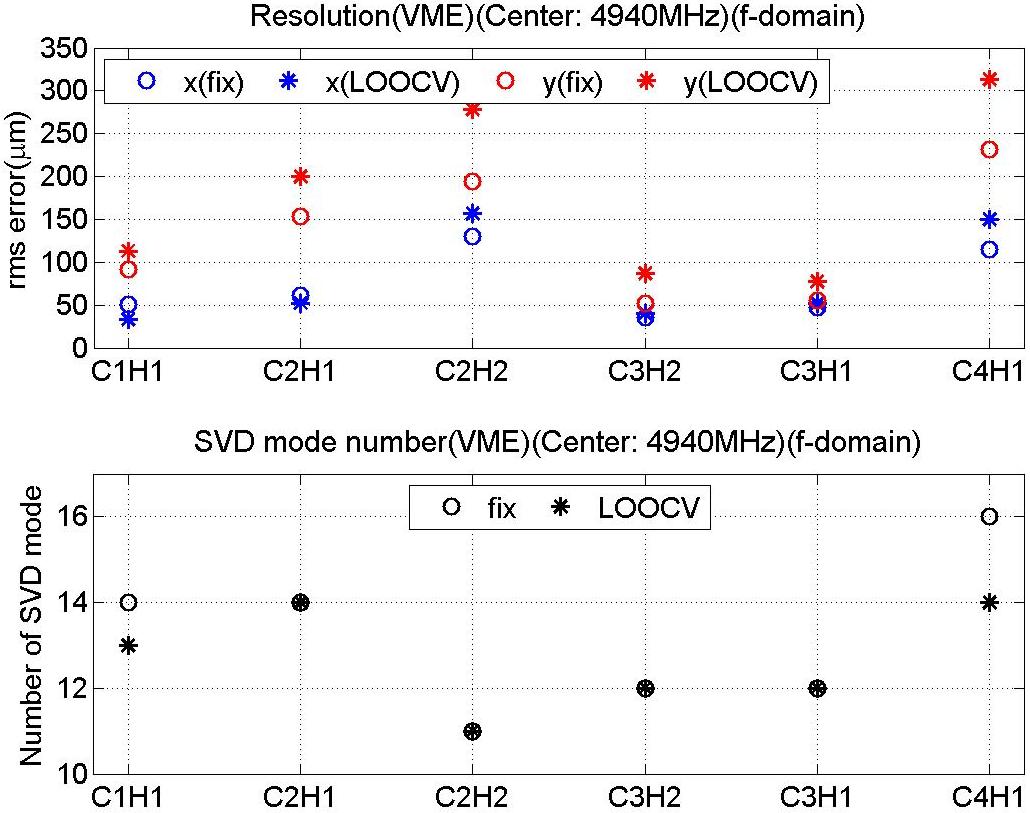}
\label{rms-8HOM-4940MHz-D109-VME-fft}
}
\caption{Position resolution and SVD mode number used for position determination for both time-domain and frequency-domain. The center frequency is 4940~MHz with a 20~MHz BPF.}
\label{rms-8HOM-4940MHz-D109-VME}
\end{figure}

The position resolution of this band is shown in Fig.~\ref{rms-8HOM-4940MHz-D109-VME-wfm} for time-domain waveform and Fig.~\ref{rms-8HOM-4940MHz-D109-VME-fft} for frequency-domain f{}f{}t amplitude along with the number of SVD modes used in the respective regression. The resolution of $y$ is approximately two times worse than that of $x$ for time-domain, which is not understood.

\begin{figure}[h]\center
\includegraphics[width=0.85\textwidth]{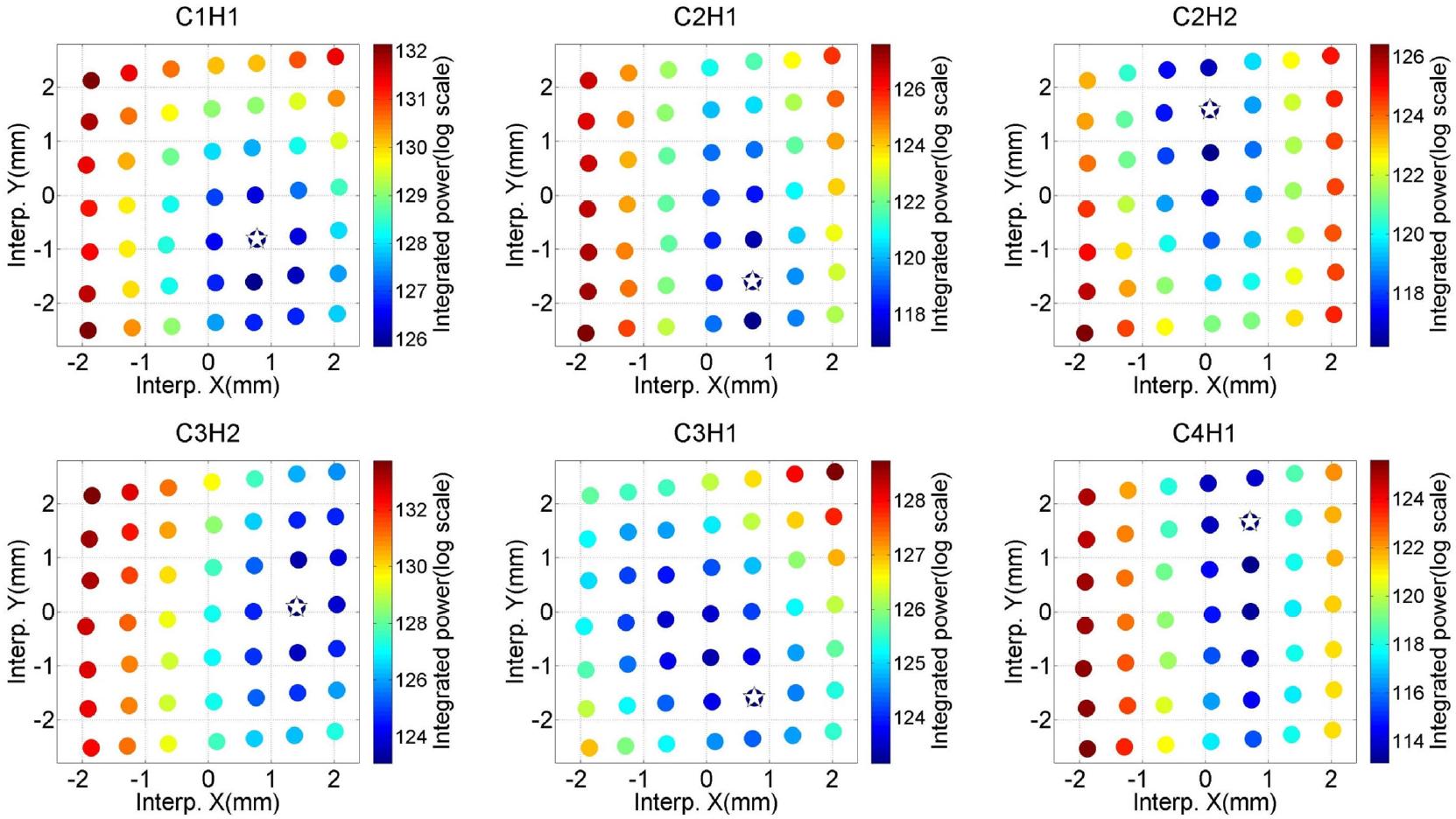}
\caption{Integrated power as a function of transverse position interpolated in each cavity. The log scale magnitude of power is denoted by dif{}ferent color. The minimum power is marked with white pentagon. The center frequency is 4940~MHz with a 20~MHz BPF.}
\label{power-well-B-4940MHz-D109}
\end{figure}

\begin{figure}[h]\center
\includegraphics[width=0.98\textwidth]{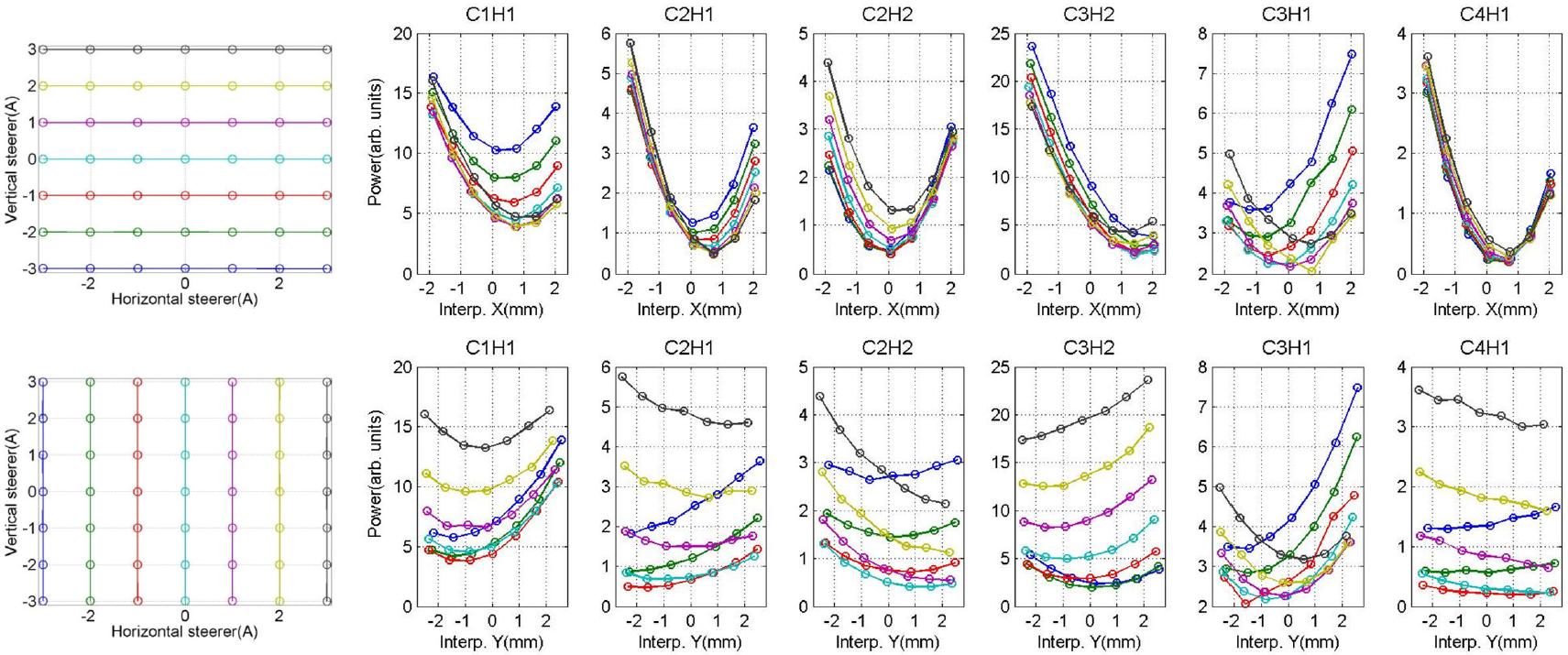}
\caption{Integrated power as a function of transverse position interpolated in each cavity. The center frequency is 4940~MHz with a 20~MHz BPF.}
\label{power-scatter-B-4940MHz-D109}
\end{figure}

\FloatBarrier
The integrated power over the frequency range shown in Fig.~\ref{wfm-fft-4940MHz-D109} (red spectrum) is calculated for each beam position. Fig.~\ref{power-well-B-4940MHz-D109} shows the integrated power distribution for each HOM coupler. The position, which has minimum integrated power, is marked with white pentagon. A dissimilar position of the power minimum is found for each of the six HOM couplers. This might be due to dif{}ferent coupling of individual coupler and the longitudinal f{}ield distribution of the various modes. Fig.~\ref{power-scatter-B-4940MHz-D109} decomposes the power distribution plot into horizontal and vertical moves.

\FloatBarrier
\section{Coupled Cavity Modes - The Second Dipole Band}
\begin{figure}[h]\center
\includegraphics[width=0.8\textwidth]{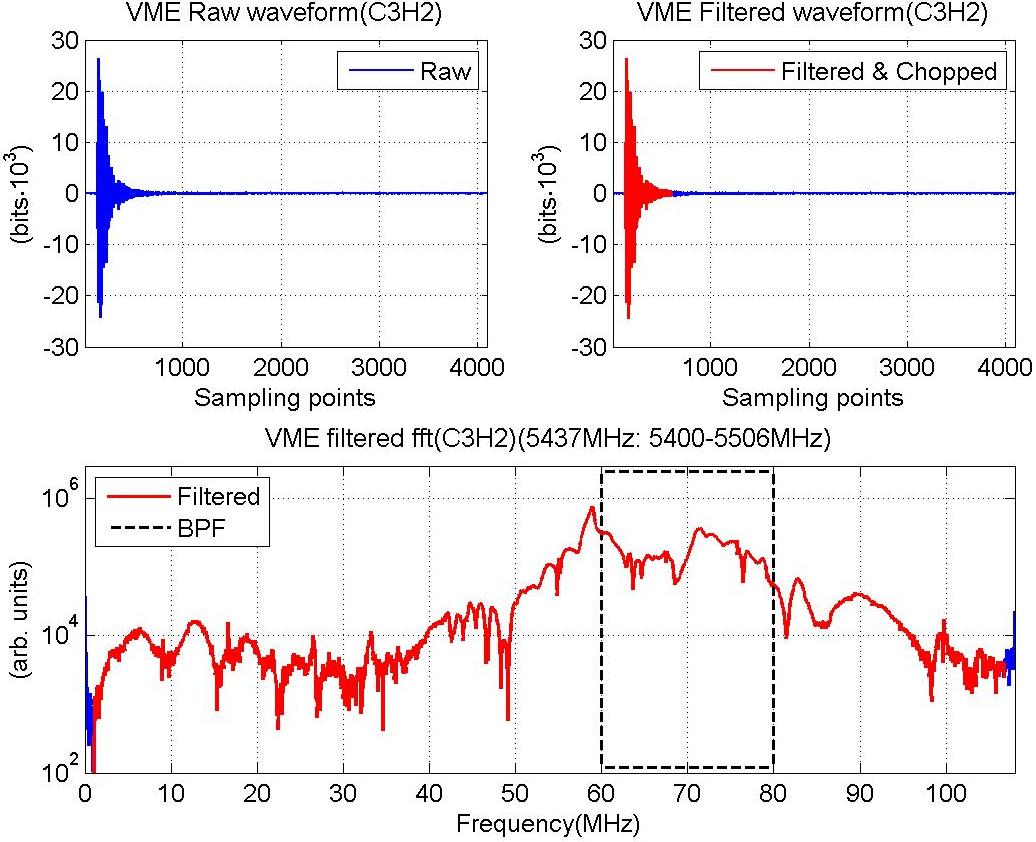}
\caption{HOM signal. The center frequency is 9048~MHz with a 20~MHz BPF.}
\label{wfm-fft-5437MHz-D208}
\end{figure}

The LO was set to downconvert 5437~MHz to 70~MHz, and a 20~MHz BPF was then applied to the down-converted signal. Fig.~\ref{wfm-fft-5437MHz-D208} shows the down-converted signal processed by the VME digitizer, in both time- and frequency domain (after a f{}f{}t). Modes in this band are coupled, therefore, the mathematical ideal f{}ilter is only to preserve good signal quality as shown in Fig.~\ref{wfm-fft-5437MHz-D208}. Compared with the spectrum analyzer signal obtained in previous studies \cite{acc39-hombpm-12}, the signal f{}iltered from 5400~MHz to 5506~MHz has proved to contain dipole modes propagating amongst cavities. As the BPF was set to 70~MHz with a 10~MHz bandwidth on each side, the main component of the f{}iltered signal is from 5427~MHz to 5447~MHz.    

\begin{figure}[h]\center
\subfigure[time domain]{
\includegraphics[width=0.7\textwidth]{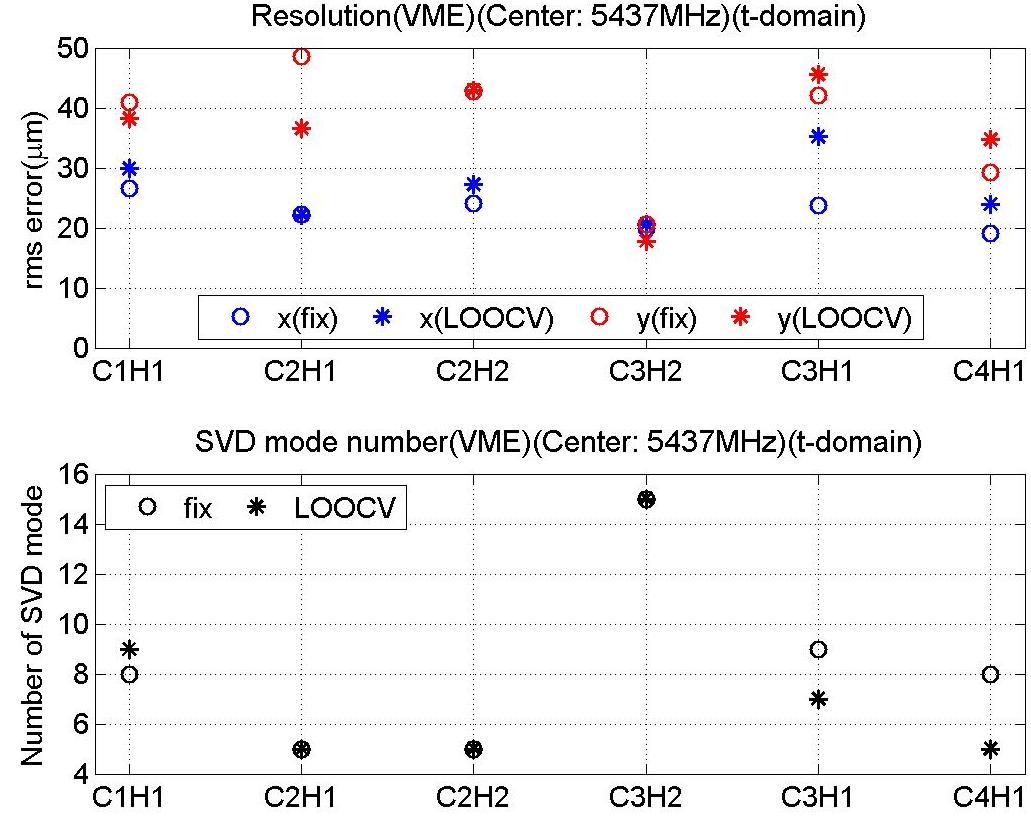}
\label{rms-8HOM-5437MHz-D208-VME-wfm}
}
\subfigure[frequency domain]{
\includegraphics[width=0.7\textwidth]{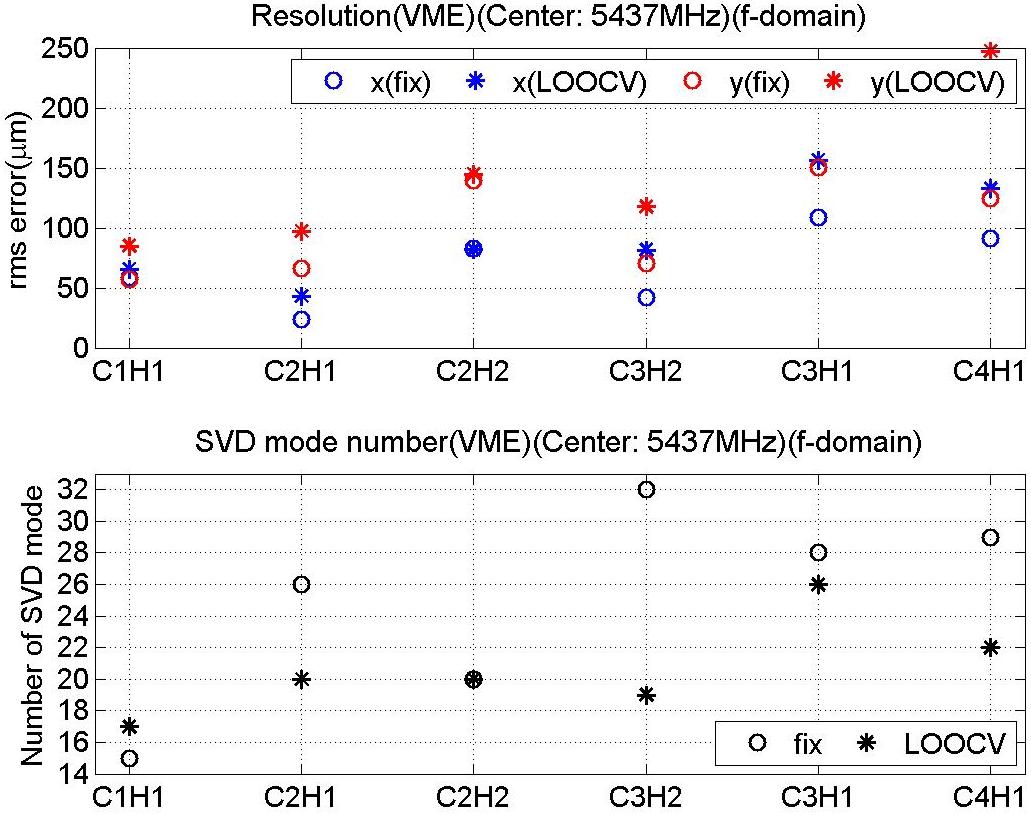}
\label{rms-8HOM-5437MHz-D208-VME-fft}
}
\caption{Position resolution and SVD mode number used for position determination for both time-domain and frequency-domain. The center frequency is 5437~MHz with a 20~MHz BPF.}
\label{rms-8HOM-5437MHz-D208-VME}
\end{figure}

The position resolution of this band is shown in Fig.~\ref{rms-8HOM-5437MHz-D208-VME-wfm} for time-domain waveform and Fig.~\ref{rms-8HOM-5437MHz-D208-VME-fft} for frequency-domain f{}f{}t amplitude along with the number of SVD modes used in the respective regression. The resolution of $y$ is approximately two times worse than that of $x$ for time-domain. Although the modes are propagating, the HOM signal extracted from each coupler is still dif{}ferent due to the individual coupling ef{}fect of each coupler. This is ref{}lected by the dissimilar number of SVD modes used to determine beam position.

\begin{figure}[h]\center
\includegraphics[width=0.85\textwidth]{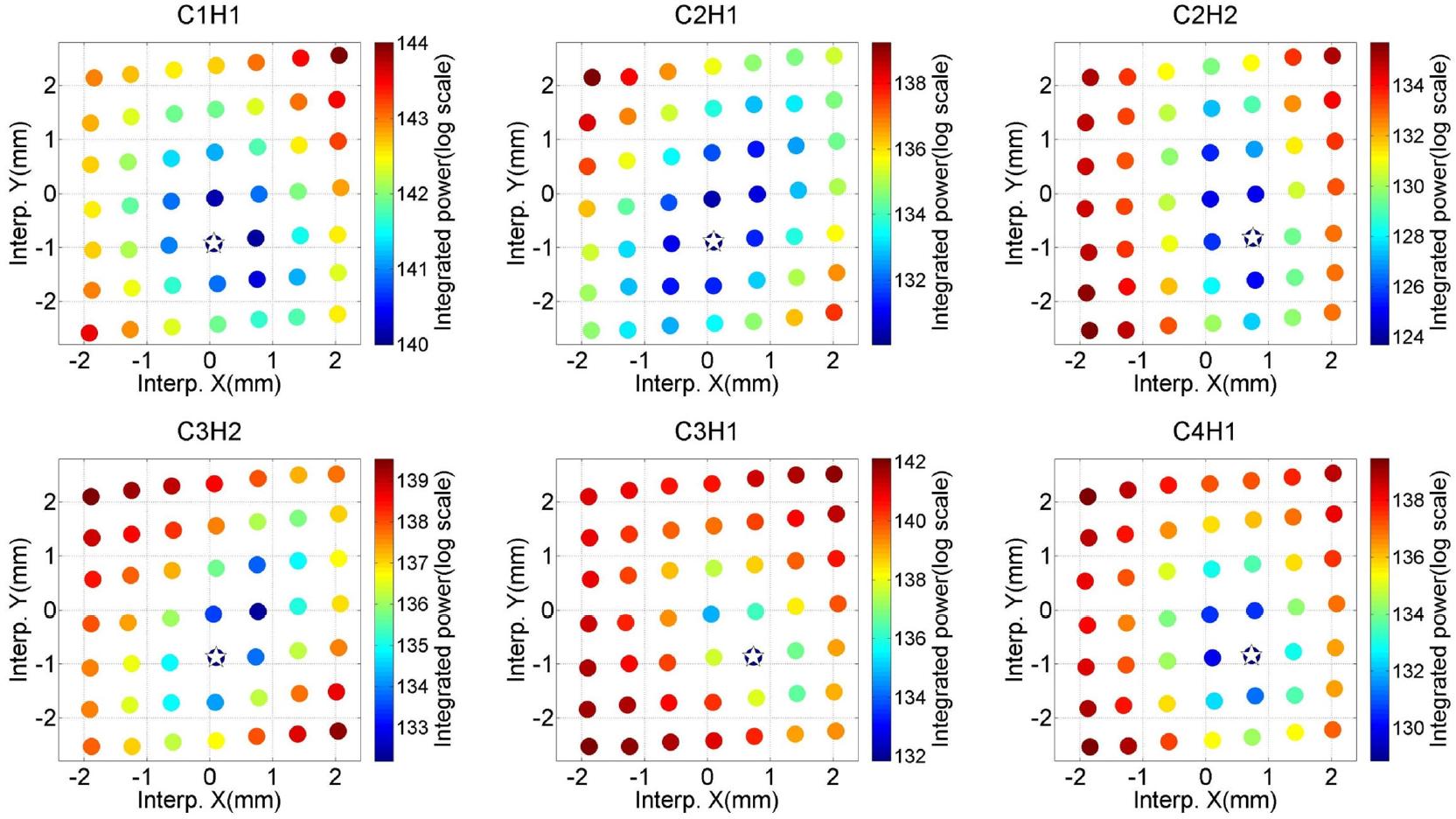}
\caption{Integrated power as a function of transverse position interpolated in each cavity. The log scale magnitude of power is denoted by dif{}ferent color. The minimum power is marked with white pentagon. The center frequency is 5437~MHz with a 20~MHz BPF.}
\label{power-well-B-5437MHz-D208}
\end{figure}

\begin{figure}[h]\center
\includegraphics[width=0.98\textwidth]{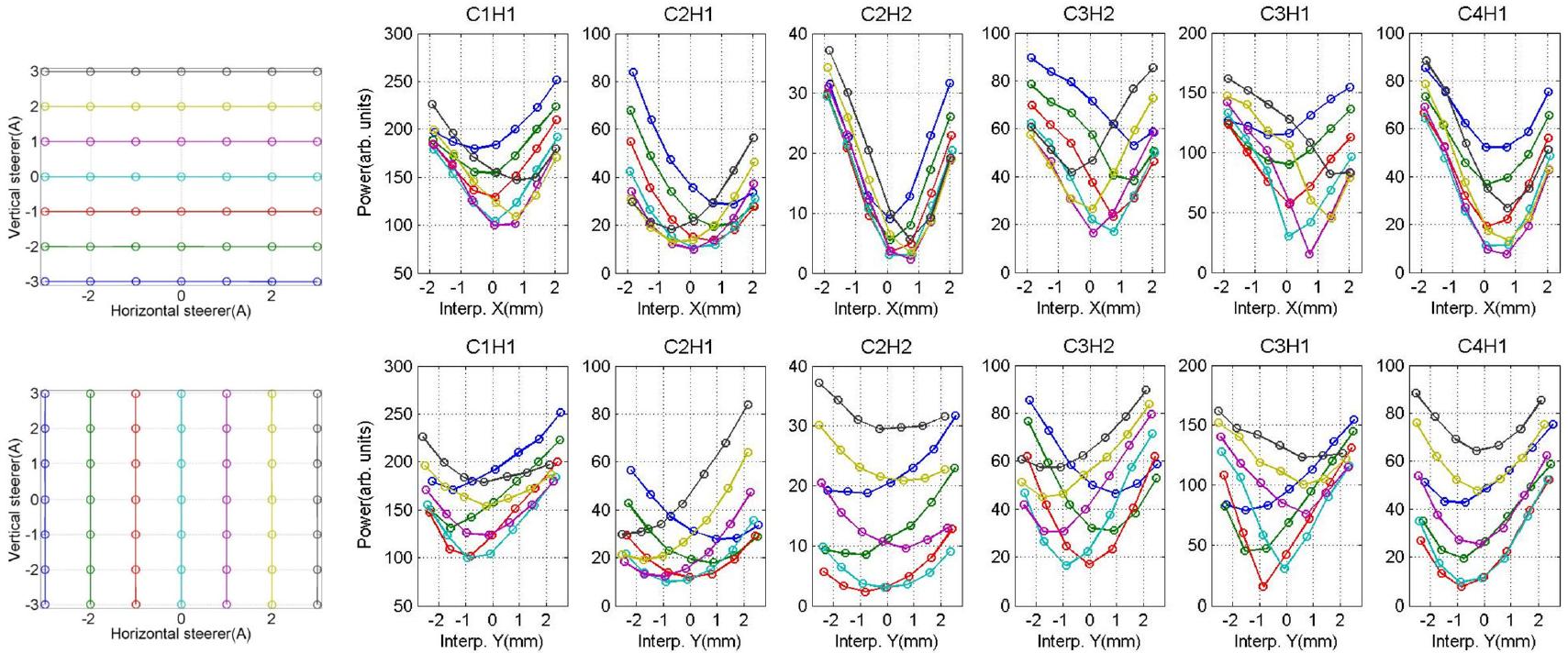}
\caption{Integrated power as a function of transverse position interpolated in each cavity. The center frequency is 5437~MHz with a 20~MHz BPF.}
\label{power-scatter-B-5437MHz-D208}
\end{figure}

\begin{figure}[h]\center
\includegraphics[width=0.6\textwidth]{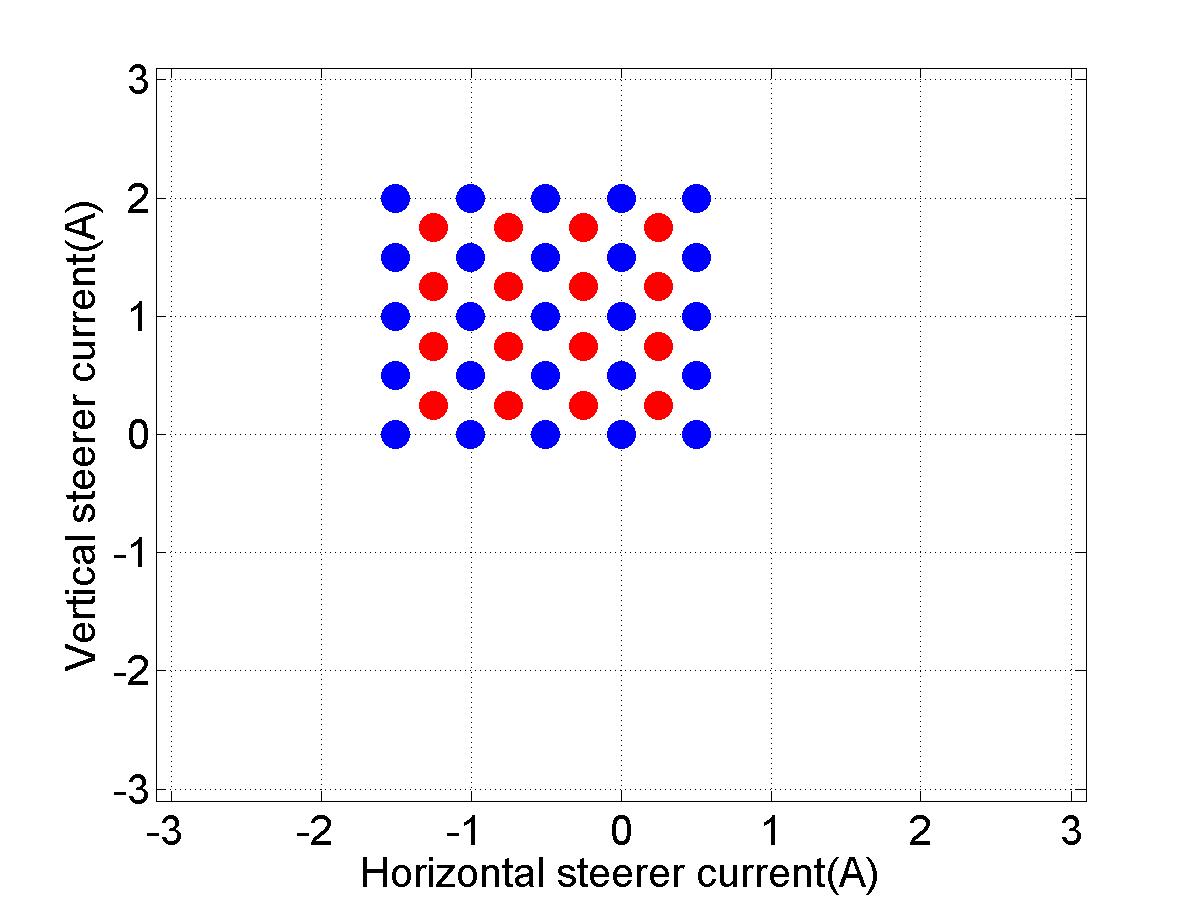}
\caption{Smaller scan around the power minimum. The center frequency is 5437~MHz with a 20~MHz BPF.}
\label{steerer-5437MHz-D209}
\end{figure}

\begin{figure}[h]\center
\subfigure[time domain]{
\includegraphics[width=0.7\textwidth]{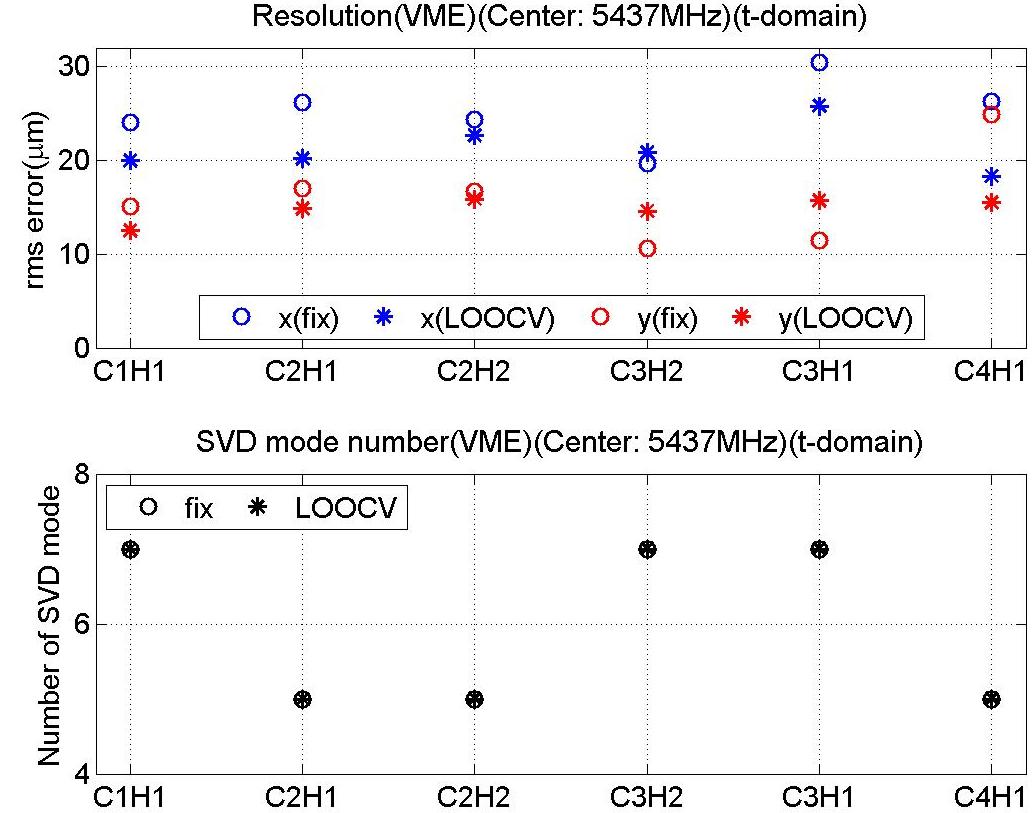}
\label{rms-8HOM-5437MHz-D209-VME-wfm}
}
\subfigure[frequency domain]{
\includegraphics[width=0.7\textwidth]{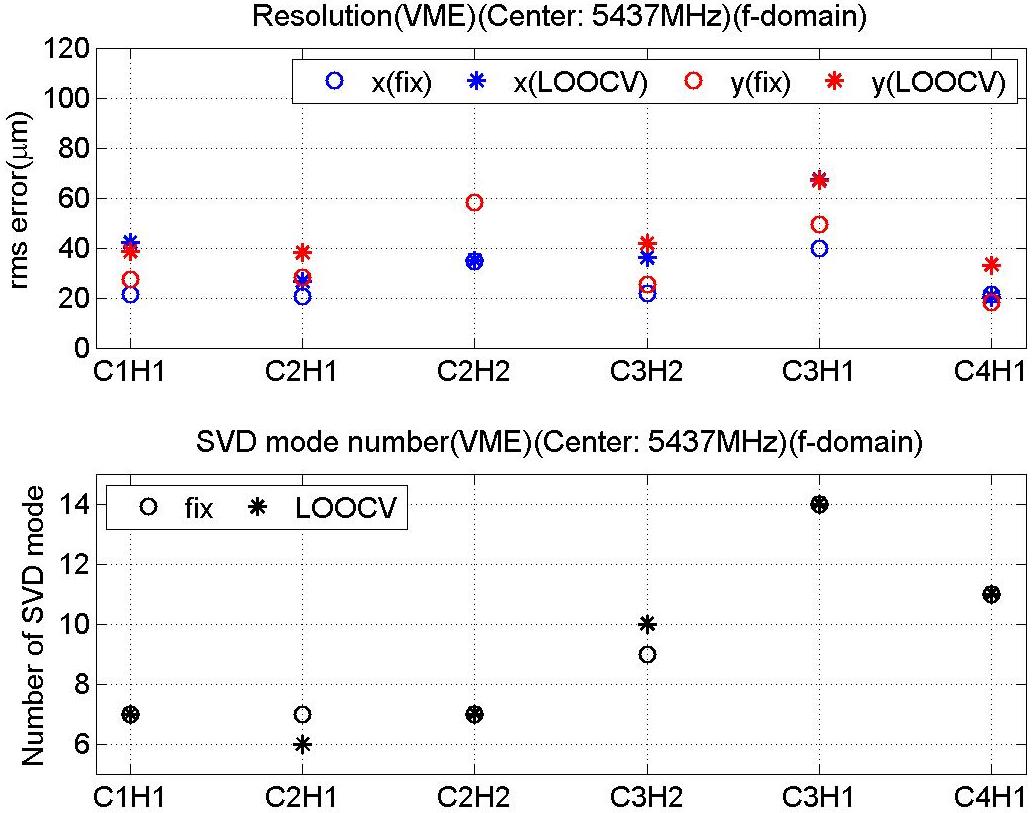}
\label{rms-8HOM-5437MHz-D209-VME-fft}
}
\caption{Position resolution and SVD mode number used for position determination for both time-domain and frequency-domain. The center frequency is 5437~MHz with a 20~MHz BPF.}
\label{rms-8HOM-5437MHz-D209-VME}
\end{figure}

\FloatBarrier
The integrated power over the frequency range shown in Fig.~\ref{wfm-fft-5437MHz-D208} (red spectrum) is calculated for each beam position. Fig.~\ref{power-well-B-5437MHz-D208} shows the integrated power distribution for each HOM coupler. The position, which has minimum integrated power, is marked with white pentagon. A similar position of the power minimum can be found for all six HOM couplers. This might be a proof that modes are propagating through cavities presented in the frequency range for power integration. Fig.~\ref{power-scatter-B-5437MHz-D208} decomposes the power distribution plot into horizontal and vertical moves. Around this ``common'' power-minimum position, we did a smaller scan as shown in Fig.~\ref{steerer-5437MHz-D209}. The position resolutions for this smaller scan are shown in Fig.~\ref{rms-8HOM-5437MHz-D209-VME}. The position resolution of 30~$\mu m$ for $x$ and 20~$\mu m$ for $y$ can be achieved using the signal from each of the six HOM couplers.


\FloatBarrier
\section{Trapped Cavity Modes - The F{}ifth Dipole Band}
\begin{figure}[h]\center
\includegraphics[width=0.8\textwidth]{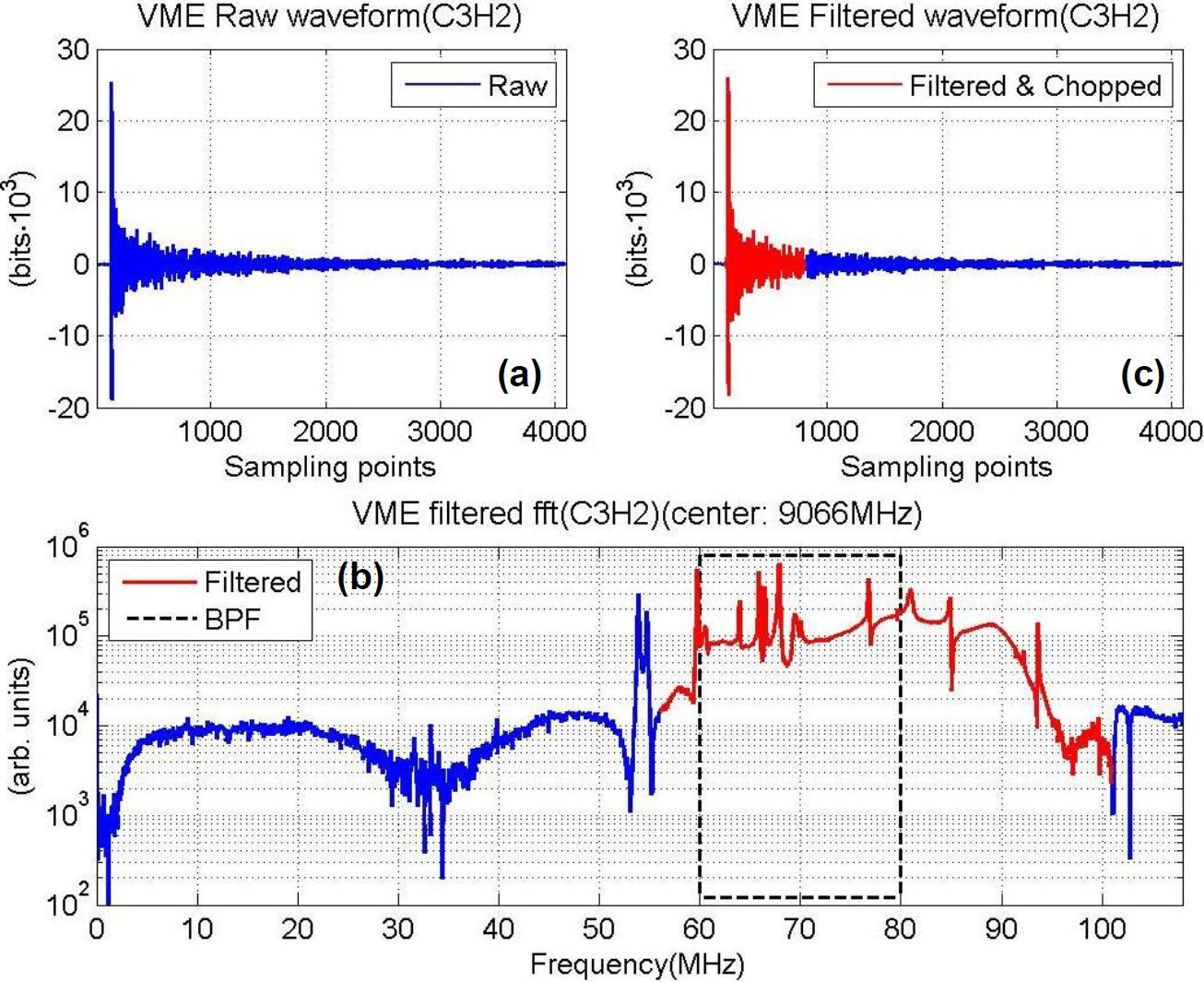}
\caption{HOM signal. The center frequency is 9066~MHz with a 20~MHz BPF.}
\label{wfm-fft-9066MHz-D503}
\end{figure}

The LO was set to downconvert 9066~MHz to 70~MHz, and a 20~MHz BPF was then applied to the down-converted signal. Fig.~\ref{wfm-fft-9066MHz-D503} shows the down-converted signal processed by the VME digitizer, in both time- and frequency domain (after a f{}f{}t). According to the previous study \cite{acc39-hombpm-8}, the double-peak at approximately 55~MHz are coupled modes. Therefore, we applied a mathematical ideal f{}ilter to the down-converted time-domain waveform and the f{}iltered signal is shown as red in Fig.~\ref{wfm-fft-9066MHz-D503}. Compared with the spectrum analyzer signal obtained in previous studies \cite{acc39-hombpm-12}, the signal f{}iltered from 9035~MHz to 9080~MHz has proved to contain dipole modes localized inside each cavity. As the BPF was set to 70~MHz with a 10~MHz bandwidth on each side, the main component of the f{}iltered signal is from 9056~MHz to 9076~MHz.   

\begin{figure}[h]\center
\subfigure[time domain]{
\includegraphics[width=0.7\textwidth]{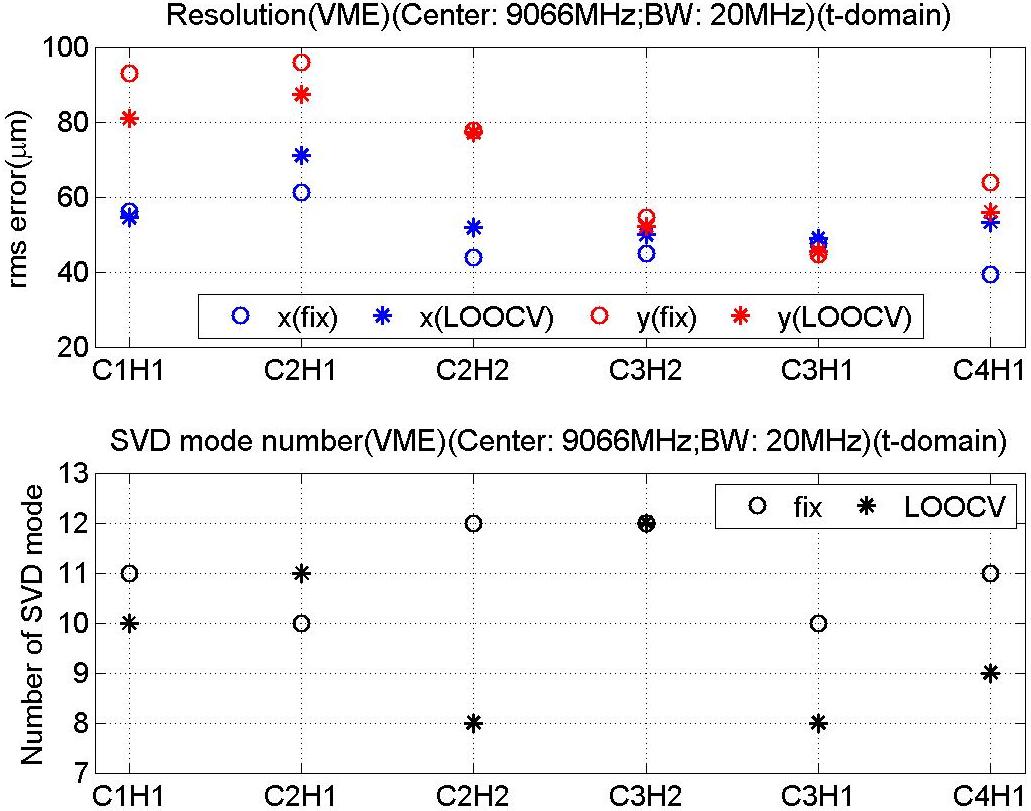}
\label{rms-8HOM-9066MHz-D503-VME-wfm}
}
\subfigure[frequency domain]{
\includegraphics[width=0.7\textwidth]{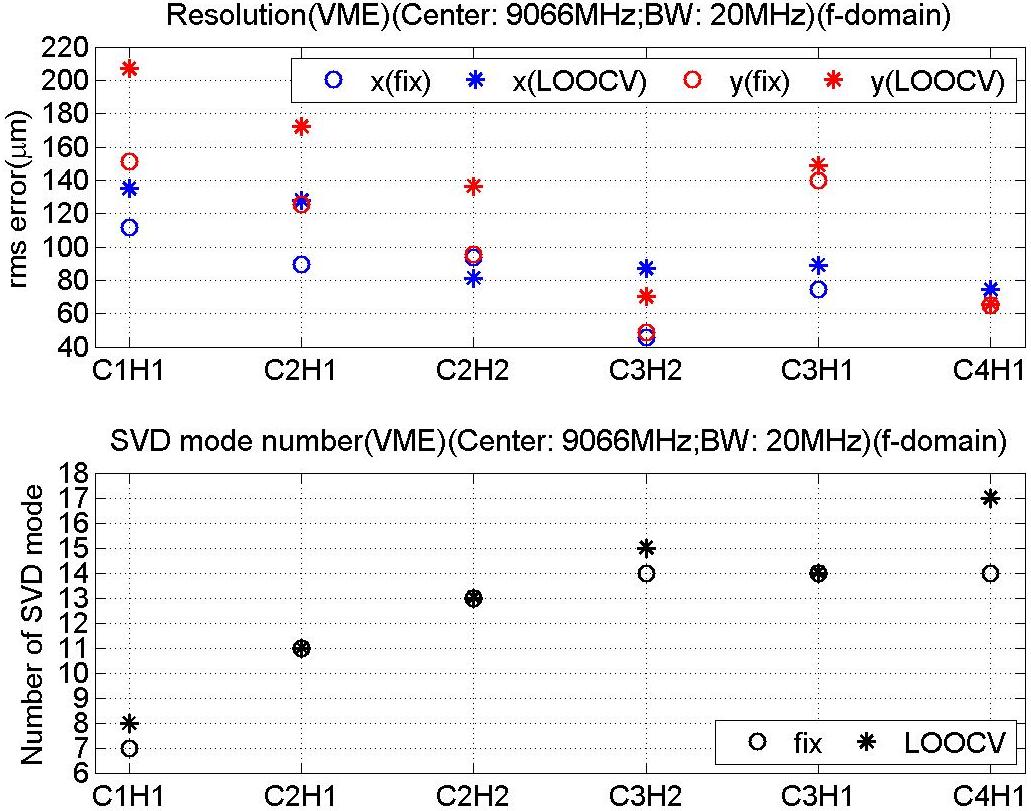}
\label{rms-8HOM-9066MHz-D503-VME-fft}
}
\caption{Position resolution and SVD mode number used for position determination for both time-domain and frequency-domain. The center frequency is 9066~MHz with a 20~MHz BPF.}
\label{rms-8HOM-9066MHz-D503-VME}
\end{figure}

The position resolution of these modes is shown in Fig.~\ref{rms-8HOM-9066MHz-D503-VME-wfm} for time-domain waveform and Fig.~\ref{rms-8HOM-9066MHz-D503-VME-fft} for frequency-domain f{}f{}t amplitude along with the number of SVD modes used in the respective regression. The number of SVD modes is dif{}ferent for individual couplers due to the dissimilar HOM signal extracted from each coupler. 

\begin{figure}[h]\center
\includegraphics[width=0.8\textwidth]{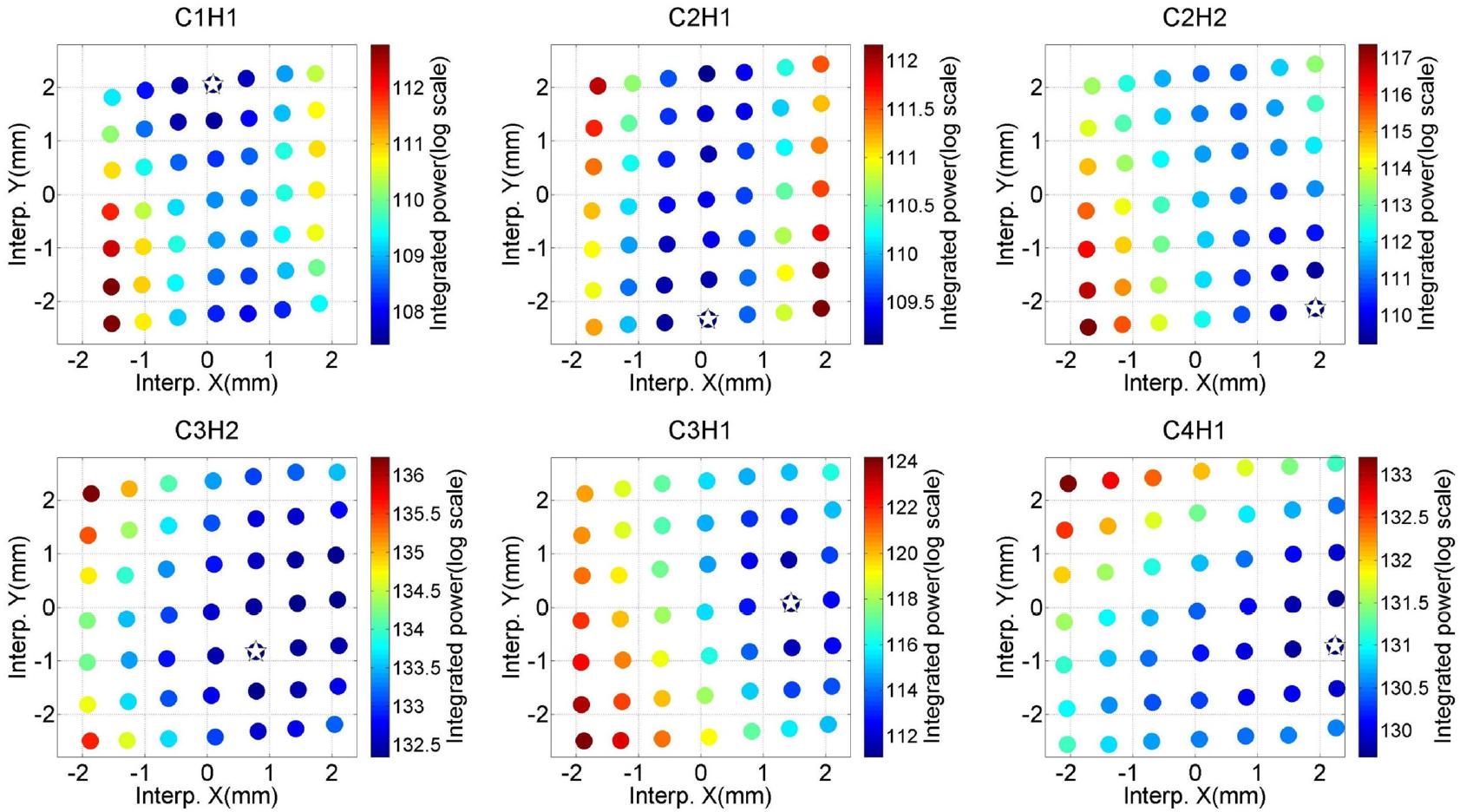}
\caption{Integrated power as a function of transverse position interpolated in each cavity. The log scale magnitude of power is denoted by dif{}ferent color. The minimum power is marked with white pentagon. The center frequency is 9066~MHz with a 20~MHz BPF.}
\label{power-well-B-9066MHz-D503}
\end{figure}

\begin{figure}[h]\center
\includegraphics[width=0.95\textwidth]{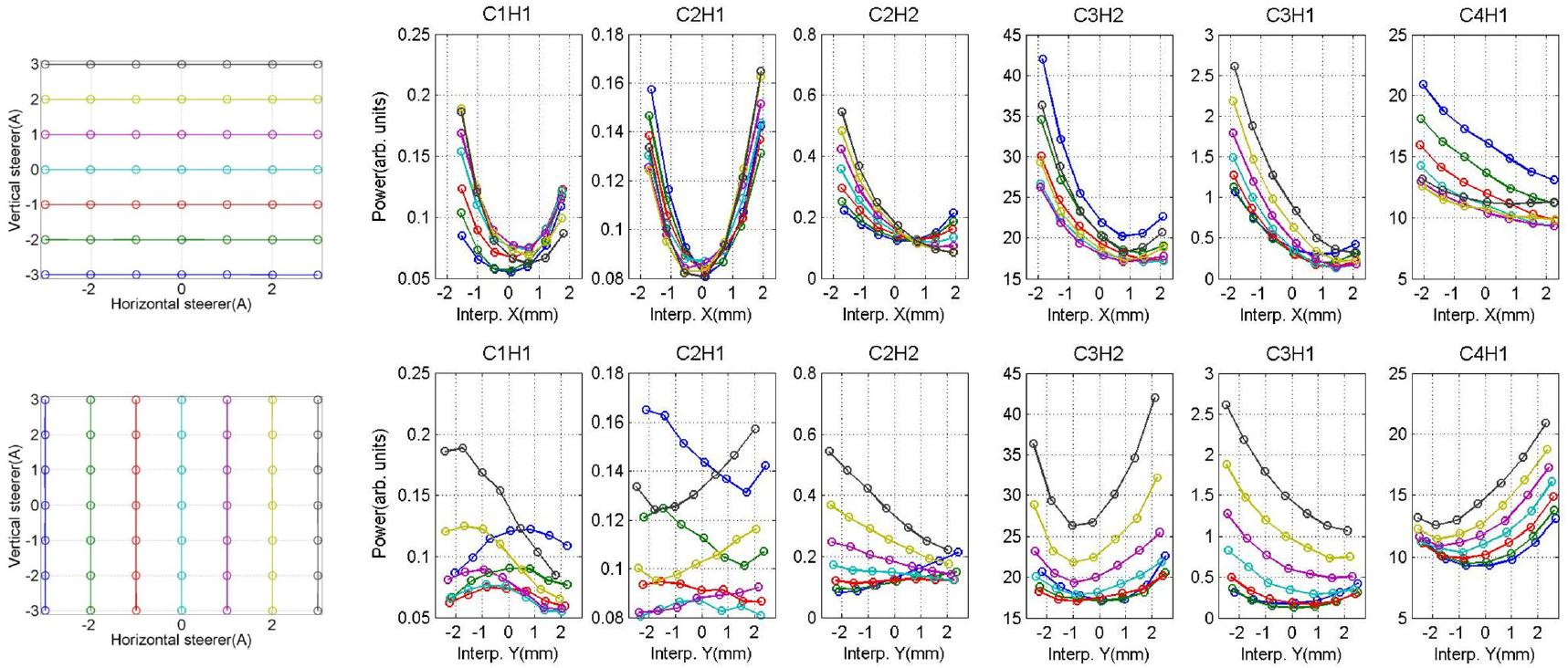}
\caption{Integrated power as a function of transverse position interpolated in each cavity. The center frequency is 9066~MHz with a 20~MHz BPF.}
\label{power-scatter-B-9066MHz-D503}
\end{figure}

\FloatBarrier
The integrated power over the frequency range shown in Fig.~\ref{wfm-fft-9066MHz-D503} (red spectrum) is calculated for each beam position. Fig.~\ref{power-well-B-9066MHz-D503} shows the integrated power distribution for each HOM coupler. The position, which has minimum integrated power, is marked with white pentagon. Neither the power distribution nor the power minimum is similar among couplers. This might be a proof that modes are trapped inside each cavity presented in the frequency range for power integration. It might also be an indication of dif{}ferent electrical axes for dif{}ferent dipole modes of individual cavities. This dif{}ference can be attributed to asymmetric structure due to couplers and fabrication tolerance of the cavities. Fig.~\ref{power-scatter-B-9066MHz-D503} decomposes the power distribution plot into horizontal and vertical moves. 

\FloatBarrier
\section{Estimation of the Resolution}
Fundamentally the measurement resolution is limited by the thermal noise, which is unavoidable at non-zero temperature. The smallest measurable thermal energy, $U_{th}$, is \cite{phy-1}
\begin{equation}
U_{th}=\frac{1}{2}k_BT,
\label{eq:thermal}
\end{equation}
where $T$ is the temperature in $K$ and $k_B$ is the Boltzmann constant ($8.6\times10^{-5}~eV\cdot K^{-1}$). Assuming a room temperature of 300~$K$, $U_{th}=0.0129~eV$. 

A beam with charge $q$ traversing a cavity loses an energy $U$ in a particular mode as \cite{phy-2}
\begin{equation}
U = \frac{\omega}{2} \cdot \left(\frac{R}{Q}\right) \cdot q^2,
\label{eq:mode-energy}
\end{equation}
where $R/Q$ (the shunt impedance divided by the quality factor) has the unit of $\Omega/cm^{-2}$, $\omega$ is the angular frequency of the mode. This energy radiates to the HOM coupler. 

For the third harmonic 3.9~GHz cavity, the strongest coupling mode in the second dipole band has a frequency of 5.4427~GHz and a $R/Q$ value of 20.877~$\Omega/cm^{-2}$ \cite{cst-1}, and its energy deposited by a 1~nC beam is
\begin{equation}
U [eV/cm^2] = \frac{2\pi \cdot 5.4427GHz}{2}\cdot 20.877\Omega/cm^{-2} \cdot \left(1nC\right)^2 \cdot 6.2\times10^{18} = 2.2\times10^{12} eV/cm^2.
\label{eq:mode-energy-ex1}
\end{equation}
This corresponds to a transverse beam of{}fset
\begin{equation}
r = \sqrt{\frac{U_{th}}{U}} = 0.8~nm.
\label{eq:mode-energy-ex2}
\end{equation}

The HOM signals are brought from the coupler inside FLASH tunnel to the HOM patch panel outside the tunnel by long cables. The measured cable loss is approximately 20~dB, the electronics attenuation is 26~dB (10~dB in the RF section and 16~dB in the IF section), this degrades the resolution by a factor of $f_{cable}$ as
\begin{equation}
r^{\ast} = f_{cable}\cdot r = 10^{\frac{20dB+26dB}{20}} \cdot r = 199.5\times 0.8nm = 160~nm.
\label{eq:cable-loss}
\end{equation}
The theoretical minimum resolution is therefore 160~nm, as compared to the observed resolution of 20-50~$\mu m$ (see Fig.~\ref{rms-8HOM-5437MHz-D208-VME-wfm}). This discrepancy may come from three main aspects.

F{}irst, HOM signals are all normalized to bunch charge before the position prediction. The bunch charge was recorded from the nearby toroid, which has a measured resolution of 3~pC. The measured bunch charge was $\sim$0.5~nC throughout our studies presented in this report, therefore it corresponds to a relative resolution of $\sim$6\%. At 1~mm beam of{}fset, this contributes $\sim$6~$\mu m$ to the position resolution. 

Second, readouts of two BPMs are used to interpolate the beam position into the cavity/module and then to calibrate the HOM signals. These two BPMs have a measured resolution of $\sim$20~$\mu m$, contributing to the resolution of the HOM-based system.

Third, the phase noise of the LO used to down-mix the HOM signal can also contribute to the resolution. This need to be studied in future measurements.

Therefore, it can be seen that the measured position resolution is dominated by the BPM resolution used for position interpolation. For the measurement of local position inside each cavity, BPMs can be used to calibrate HOM signals in C1 and C3, the position measured in these two cavities can then be used to make a prediction about the position in C2. In this case, the limitations of BPMs are eliminated from the position prediction in C2, and will consequently improve the resolution.

\chapter{Summary}
We have studied three modal options with the specially designed test electronics for the third harmonic 3.9~GHz cavities at FLASH. Localized modes are used to determine the individual beam position in each cavity, and the resolution is approximately 50--100~$\mu m$. The transverse beam position at the center of the ACC39 module can be determined by using the coupled modes, and the resolution is approximately 20--50~$\mu m$ due to the strong coupling of these modes to the beam. The resolution estimation with the test electronics for all modal options is shown in Table~\ref{table-res-sum}. Based on these results we decided to build HOM electronics for the second dipole band and the f{}ifth dipole band, so that we will have both high resolution measurements for the whole module, and localized measurements for individual cavity. A prototype HOM electronics is being built by Fermilab and planned to be tested in FLASH by the end of 2012.
\begin{table}[h]\center
\caption{Resolution estimation for various modal options with the test electronics.}
\label{table-res-sum}
\begin{tabular}{|c|c|c|c|c|}
\hline
& \multicolumn{2}{|c|}{\textbf{Localized Modes}} & \multicolumn{2}{c|}{\textbf{Coupled Modes}} \\
\hline
& Beam-pipe modes & 5$^{th}$ dipole band & 1$^{st}$ dipole band & 2$^{nd}$ dipole band \\
\hline
\textbf{x} ($\mu m$) & 40$-$100 & 40$-$60 & 20$-$30 & 20$-$30\\
\hline
\textbf{y} ($\mu m$) & 80$-$180 & 50$-$100 & 30$-$70 & 20$-$50\\
\hline
\end{tabular}
\end{table}

\begin{abstract}
We thank Dr.~Jacek~Sekutowicz for carefully reading this manuscript and many useful comments. This work received support from the European Commission under the FP7 Research Infrastructures grant agreement No.227579.
\end{abstract}

\bibliography{refs}

\begin{thebibliography}{10}

\bibitem{flash-1}
{W. Ackermann} {\em et~al.}, ``{Operation of a free-electron laser from the
  extreme ultraviolet to the water window},'' {\em Nature Photonics}, vol.~1,
  pp.~336--342, 2007.

\bibitem{tesla-1}
{J. Sekutowicz}, {\em {Multi-cell Superconducting Structures for High Energy
  e+e- Colliders and Free Electron Laser Linacs}}.
\newblock Warsaw, Poland: Warsaw University of Technology Publishing House,
  first~ed., 2008.

\bibitem{acc39-1}
{K.~Floettmann, T.~Limberg and Ph.~Piot}, ``{Generation of Ultrashort Electron
  Bunches by Cancellation of Nonlinear Distortions in the Longitudinal Phase
  Space},'' TESLA-FEL Report: TESLA-FEL 2001-06, 2001.

\bibitem{acc39-2}
{J. Sekutowicz, R. Wanzenberg, W.F.O. Mueller and T. Weiland}, ``{A Design of a
  3rd Harmonic Cavity for the TTF 2 Photoinjector},'' TESLA-FEL Report:
  TESLA-FEL 2002-05, 2002.

\bibitem{wake-1}
{K.L.F. Bane}, ``{Wake Field Ef{}fects in a Linear Collider},'' SLAC Note:
  SLAC-PUB-4169, 1986.

\bibitem{tesla-hombpm-1}
{S. Molloy} {\em et~al.}, ``High precision superconducting cavity diagnostics
  with higher order mode measurements,'' {\em Phys. Rev. ST Accel. Beams},
  vol.~9, p.~112801, 2006.

\bibitem{tesla-hombpm-2}
{G. Devanz} {\em et~al.}, ``{HOM Beam Coupling Measurements at the TESLA Test
  Facility (TTF)},'' in {\em Proceedings of EPAC2002}, (Paris, France),
  pp.~230--232, 2002.

\bibitem{tesla-hombpm-3}
{N. Baboi} {\em et~al.}, ``{Preliminary Study on HOM-Based Beam Alignment in
  the TESLA Test Facility},'' in {\em Proceedings of LINAC 2004},
  (L$\ddot{u}$beck, France), pp.~117--119, 2004.

\bibitem{mafia-0}
{MAFIA Release 4}.
\newblock CST AG, Darmstadt, Germany.

\bibitem{mafia}
{T. Khabibouline} {\em et~al.}, ``{Higher Order Modes of a 3rd Harmonic Cavity
  with an Increased End-cup Iris},'' TESLA-FEL Report: TESLA-FEL 2003-01, 2003.

\bibitem{hfss}
{ANSYS\textregistered HFSS}.
\newblock Release 11.2, ANSYS Inc., USA.

\bibitem{hfss-1}
{I.R.R. Shinton} {\em et~al.}, ``{Compendium of Eigenmodes in Third Harmonic
  Cavities for FLASH and the XFEL},'' DESY Report: DESY 12-053, 2012.

\bibitem{cst}
{CST Microwave Studio\textregistered}.
\newblock Ver. 2011, CST AG, Darmstadt, Germany.

\bibitem{cst-1}
{P.~Zhang, N.~Baboi and R.M.~Jones}, ``{Eigenmode simulations of third harmonic
  superconducting accelerating cavities for FLASH and the European XFEL},''
  DESY Report: DESY 12-101, 2012.

\bibitem{acc39-hombpm-2}
{I.R.R. Shinton} {\em et~al.}, ``{Higher Order Modes in Third Harmonic Cavities
  for XFEL/FLASH},'' in {\em Proceedings of IPAC'10}, (Kyoto, Japan),
  pp.~3007--3009, 2010.

\bibitem{acc39-hombpm-4}
{I.R.R. Shinton} {\em et~al.}, ``{Higher Order Modes in Third Harmonic Cavities
  at FLASH},'' in {\em Proceedings of Linear Accelerator Conference LINAC2010},
  (Tsukuba, Japan), pp.~785--787, 2010.

\bibitem{acc39-hombpm-10}
{T. Flisgen} {\em et~al.}, ``{A Concatenation Scheme for the Computation of
  Beam Excited Higher Order Mode Port Signals},'' in {\em Proceedings of
  IPAC2011}, (San Sebastian, Spain), pp.~2238--2240, 2011.

\bibitem{acc39-hombpm-9}
{I.R.R. Shinton} {\em et~al.}, ``{Higher Order Modes in Coupled Cavities of the
  FLASH Module ACC39},'' in {\em Proceedings of IPAC2011}, (San Sebastian,
  Spain), pp.~2301--2303, 2011.

\bibitem{acc39-hombpm-11}
{I.R.R. Shinton} {\em et~al.}, ``{Simulations of higher order modes in the
  ACC39 module of FLASH},'' in {\em IPAC2012}, (New Orleans, USA,), 2012.
\newblock TUPPR037.

\bibitem{acc39-hombpm-12}
{P.~Zhang, N.~Baboi and R.M.~Jones}, ``{Higher order mode spectra and the
  dependence of localized dipole modes on the transverse beam position in third
  harmonic superconducting cavities at FLASH},'' DESY Report: DESY 12-109,
  2012.

\bibitem{acc39-hombpm-3}
{P. Zhang} {\em et~al.}, ``{First Beam Spectra of SC Third Harmonic Cavity at
  FLASH},'' in {\em Proceedings of Linear Accelerator Conference LINAC2010},
  (Tsukuba, Japan), pp.~782--784, 2010.

\bibitem{acc39-hombpm-5}
{P. Zhang} {\em et~al.}, ``{Beam-based HOM Study in Third Harmonic SC Cavities
  for Beam Alignment at FLASH},'' in {\em Proceedings of DIPAC2011}, (Hamburg,
  Germany), pp.~77--79, 2011.

\bibitem{acc39-hombpm-6}
{H.-W. Glock} {\em et~al.}, ``{Diode Down-mixing of HOM Coupler Signals for
  Beam Position Determination in 1.3-GHz- and 3.9-GHz-Cavities at FLASH},'' in
  {\em Proceedings of DIPAC2011}, (Hamburg, Germany), pp.~101--103, 2011.

\bibitem{acc39-hombpm-8}
{P. Zhang} {\em et~al.}, ``{Study of Beam Diagnostics with Trapped Modes in
  Third Harmonic Superconducting Cavities at FLASH},'' in {\em Proceedings of
  IPAC2011}, (San Sebastian, Spain), pp.~2891--2893, 2011.

\bibitem{acc39-hombpm-7}
{N. Baboi} {\em et~al.}, ``{Higher Order Modes for Beam Diagnostics in Third
  Harmonic 3.9 GHz Accelerating Modules},'' in {\em Proceedings of SRF2011},
  (Chicago, USA), pp.~239--243, 2011.

\bibitem{uTCA}
{R.S.~Larsen}, ``{PICMG xTCA Standards Extensions for Physics: New Developments
  and Future Plans},'' SLAC Note: SLAC-PUB-14182, 2010.

\bibitem{epics}
{EPICS}.
\newblock http://www.aps.anl.gov/epics/.

\bibitem{doocs}
{DOOCS}.
\newblock http://tesla.desy.de/doocs/.

\bibitem{matlab}
{MATLAB\textregistered}.
\newblock Ver. R2011b, The MathWorks Inc., USA.

\bibitem{stat-1}
{R.L. Ott} and {M. Longnecker}, {\em {An Introduction to Statistical Methods
  and Data Analysis}}, ch.~11.7, p.~611.
\newblock Duxbury Press, sixth~ed., 2008.

\bibitem{stat-4}
{G.H. Golub and C.F. Van Loan}, {\em {Matrix Computations}}, ch.~2.3,
  pp.~16--20.
\newblock The John Hopkins University Press, second~ed., 1984.

\bibitem{stat-8}
{R. Kohavi}, ``{A Study of Cross-Validation and Bootstrap for Accuracy
  Estimation and Model Selection},'' in {\em Proceedings of the 14th
  International Joint Conference on Artif\mbox{}icial Intelligence}, vol.~2,
  (Montreal, Quebec, Canada), pp.~1137--1143, 1995.

\bibitem{phy-1}
{W.B.~Davenport and W.L.~Root}, {\em {An introduction to the theory of random
  signals and noise}}, ch.~9-4, p.~185.
\newblock IEEE Press, 1987.

\bibitem{phy-2}
{A.~Chao and M.~Tigner}, {\em {Handbook of Accelerator Physics and
  Engineering}}, ch.~3.2.7, p.~213.
\newblock Singapore: World Scientif{}ic, first~ed., 1999.

\end{thebibliography}
\bibliographystyle{ieeetr}
\appendix
\chapter{The Analog Box Circuit}\label{app:draw}

\begin{figure}[h]\center
\includegraphics[width=0.9\textwidth]{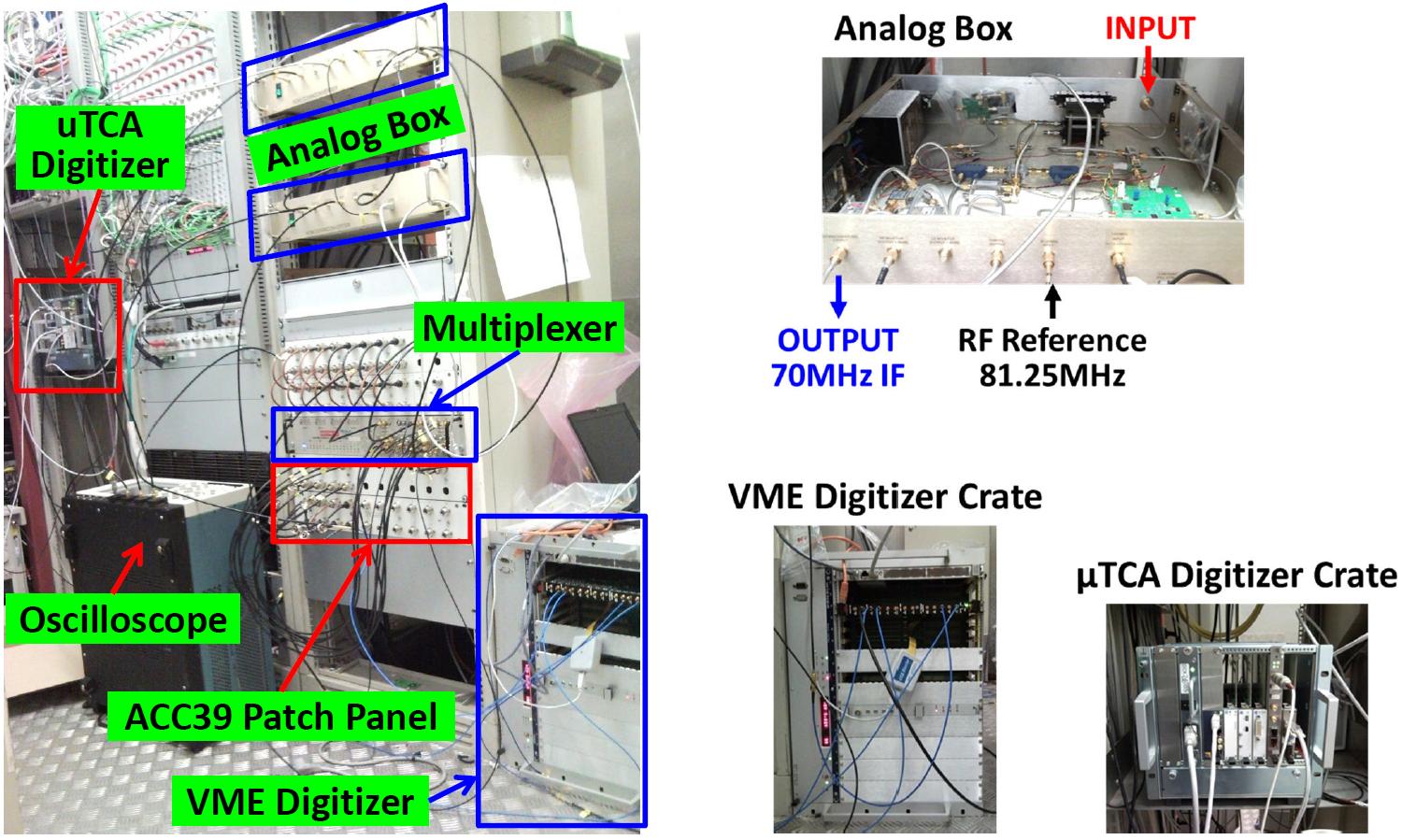}
\caption{Photo of the device setup in the FLASH injector barrack outside the tunnel, the analog box and two types of digitizer.}
\label{device-pic}
\end{figure}

\begin{landscape}
\begin{figure}[h]\center
\includegraphics[width=1.4\textwidth]{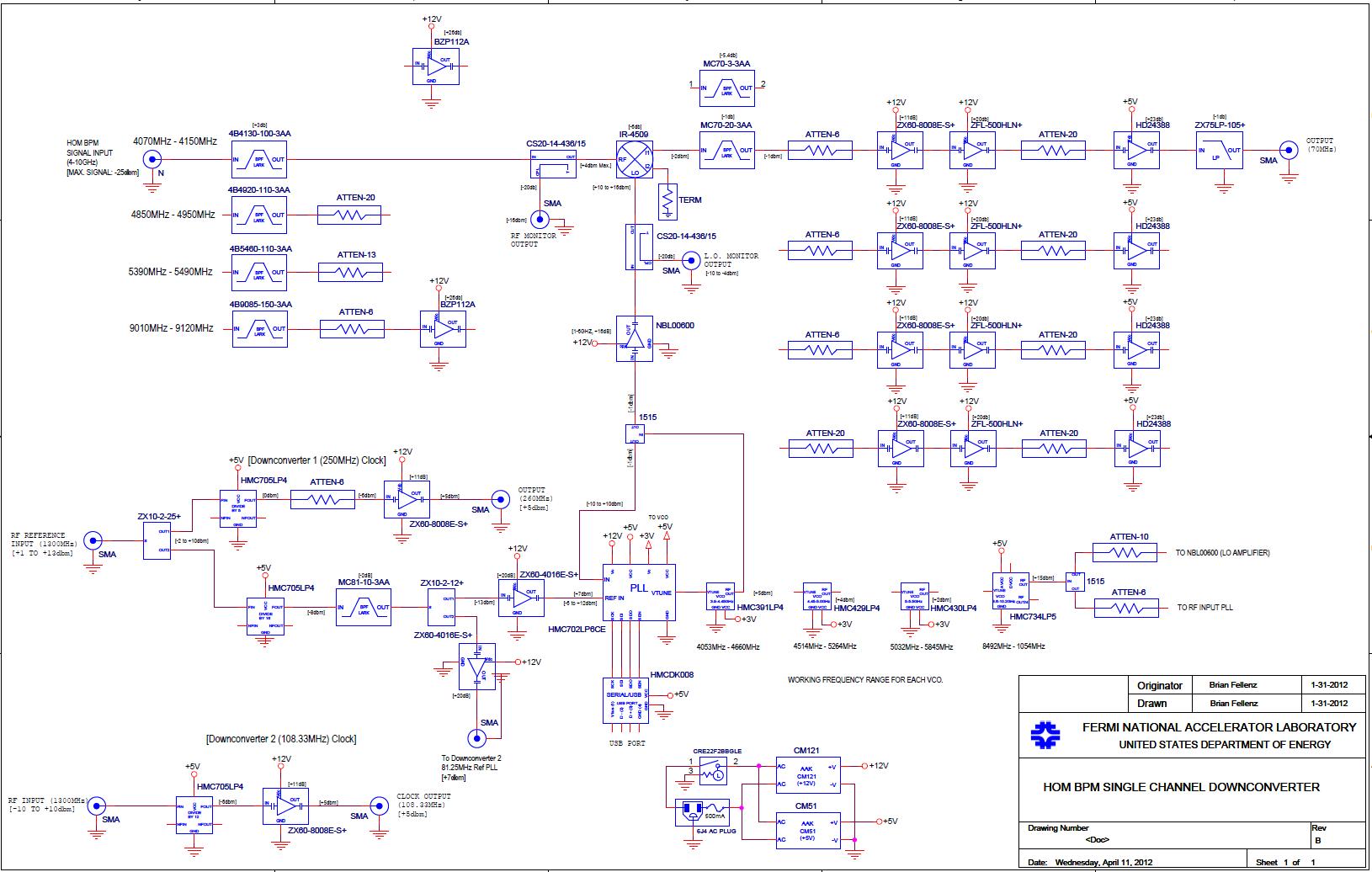}
\caption{Drawing of the analog box.}
\label{analog-box-drawing}
\end{figure}
\end{landscape}

\chapter{Settings in the Analog Box}\label{app:atten}
HOM signals vary amongst couplers and frequency bands. Therefore various amplif{}ication and attenuations were used for each frequency band. They are listed in Table~\ref{table-atten}.
\begin{table}[h]\center
\caption{Settings in the analog box for each frequency band. ``$+$'' means amplif{}ication and ``$-$'' means attenuation.}
\label{table-atten}
\begin{tabular}{c|c|c|c|c|c}
\hline
Band & Center frequency & Bandwidth & RF section & IF section& Total \\
\hline
\multirow{2}{*}{Beam-pipe modes} & 4082~MHz & 20~MHz & \multirow{2}{*}{$+$0~dB} & \multirow{2}{*}{$+$33~dB} & \multirow{2}{*}{$+$33~dB}\\
& 4118~MHz & 20~MHz & & & \\
\hline
\multirow{4}{*}{1$^{st}$ dipole band} & 4859~MHz & 20~MHz & \multirow{4}{*}{$-$13~dB} & \multirow{4}{*}{$+$32~dB} & \multirow{4}{*}{$+$19~dB}\\
& 4904~MHz & 20~MHz & & & \\
& 4940~MHz & 20~MHz & & & \\
& 4900~MHz & 100~MHz & & & \\
\hline
\multirow{4}{*}{2$^{nd}$ dipole band} & 5437~MHz & 20~MHz & \multirow{4}{*}{$-$10~dB} & \multirow{4}{*}{$+$36~dB} & \multirow{4}{*}{$+$26~dB}\\
& 5464~MHz & 20~MHz & & & \\
& 5482~MHz & 20~MHz & & & \\
& 5450~MHz & 100~MHz & & & \\
\hline
\multirow{2}{*}{5$^{th}$ dipole band} & 9048~MHz & 20~MHz & \multirow{2}{*}{$-$6~dB} & \multirow{2}{*}{$+$52~dB} & \multirow{2}{*}{$+$46~dB}\\
& 9066~MHz & 20~MHz & & & \\
\hline
\end{tabular}
\end{table}

\chapter{Tables: Position Resolutions}\label{app:res}
Resolutions in predicting beam positions for validation samples are listed in this appendix. The method applied is \textit{singular value decomposition} (SVD). The number of SVD modes used for regression is determined based on a stable and good position resolution. These are listed in Appendix~\ref{app:nsvd}. Both the f{}ixed sample split and a \textit{leave-one-out cross-validation} (LOOCV) are shown. The reading jumps of the BPM-A readouts have been corrected in $y$ plane for all modal options.

\section{The Localized Beam-pipe modes}\label{app:res-bp}
\begin{figure}[h]\center
\includegraphics[width=0.95\textwidth]{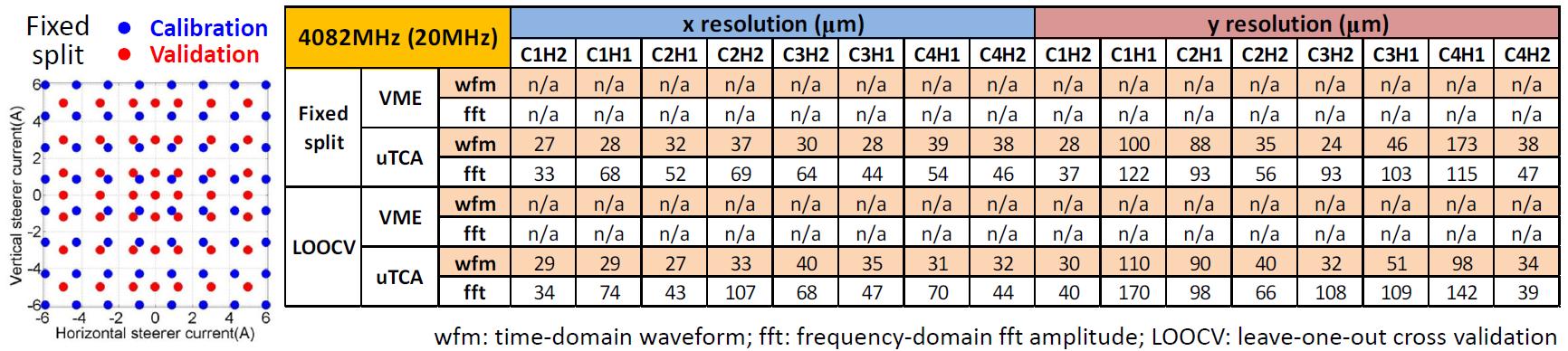}
\caption{Position resolutions of beam-pipe modes centered at 4082~MHz with a 20~MHz bandwidth.}
\label{res-BP-4082MHz}
\end{figure}

\begin{figure}[h]\center
\includegraphics[width=0.95\textwidth]{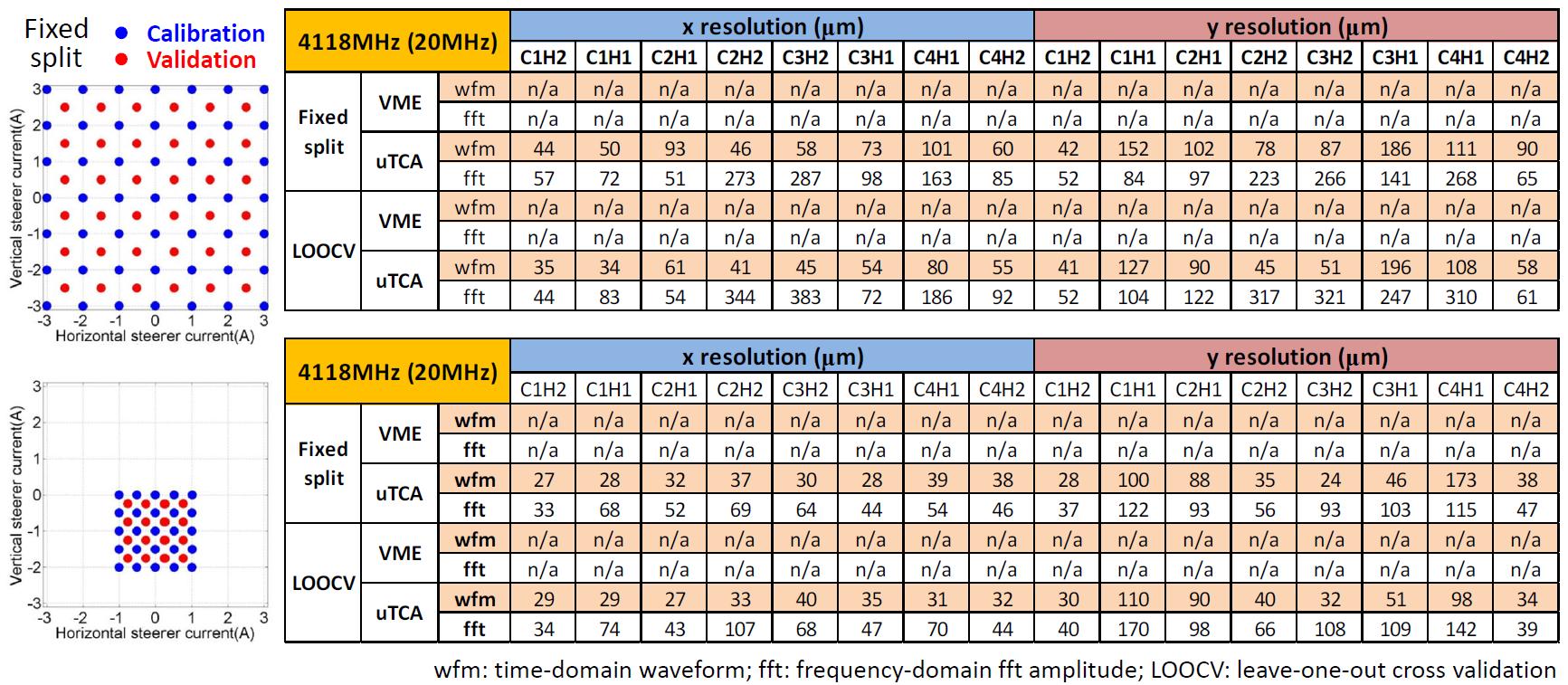}
\caption{Position resolutions of beam-pipe modes centered at 4118~MHz with a 20~MHz bandwidth.}
\label{res-BP-4082MHz}
\end{figure}

\FloatBarrier
\section{Coupled Cavity Modes - The First Dipole Band}\label{app:res-D1}
\begin{figure}[h]\center
\includegraphics[width=0.95\textwidth]{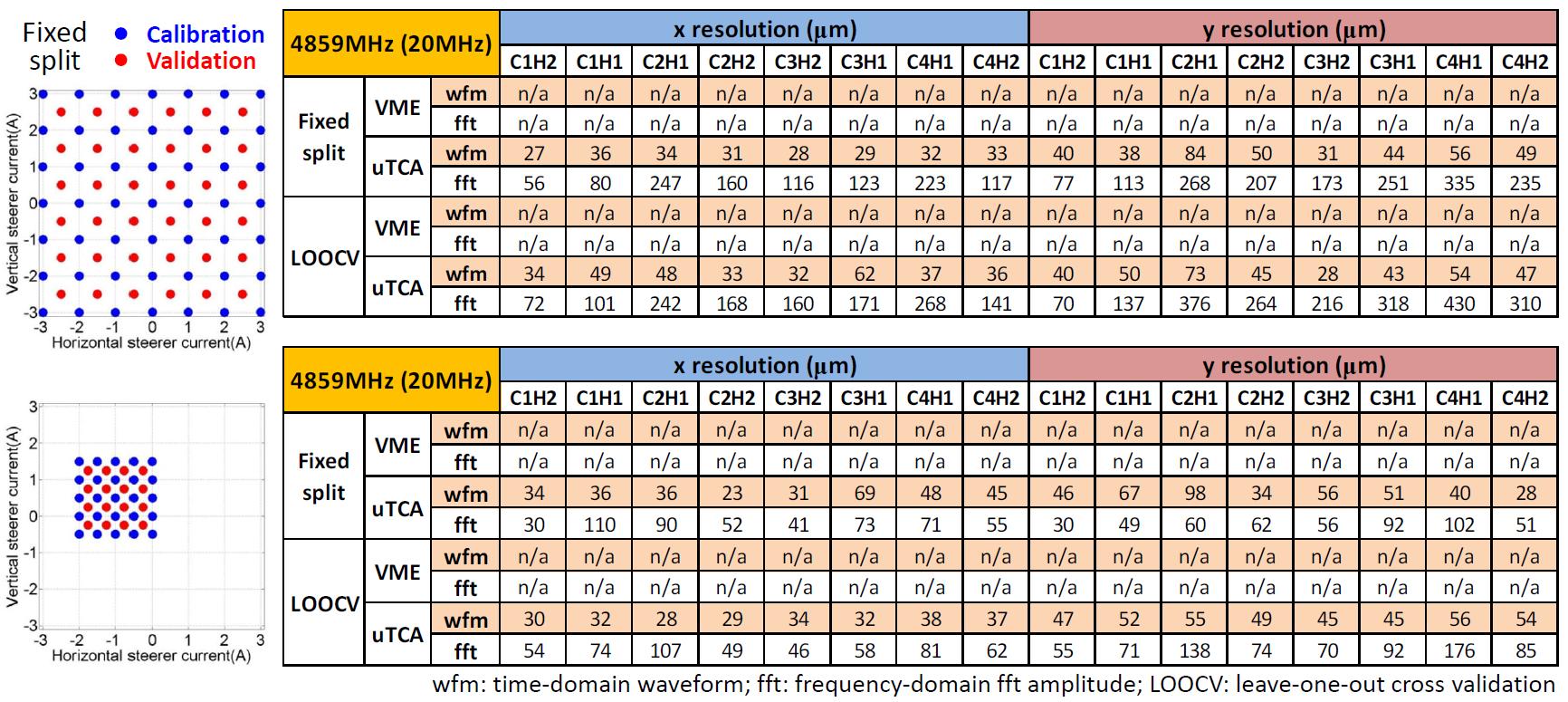}
\caption{Position resolutions of cavity modes centered at 4859~MHz with a 20~MHz bandwidth.}
\label{res-D1-4859MHz}
\end{figure}

\begin{figure}[h]\center
\includegraphics[width=0.95\textwidth]{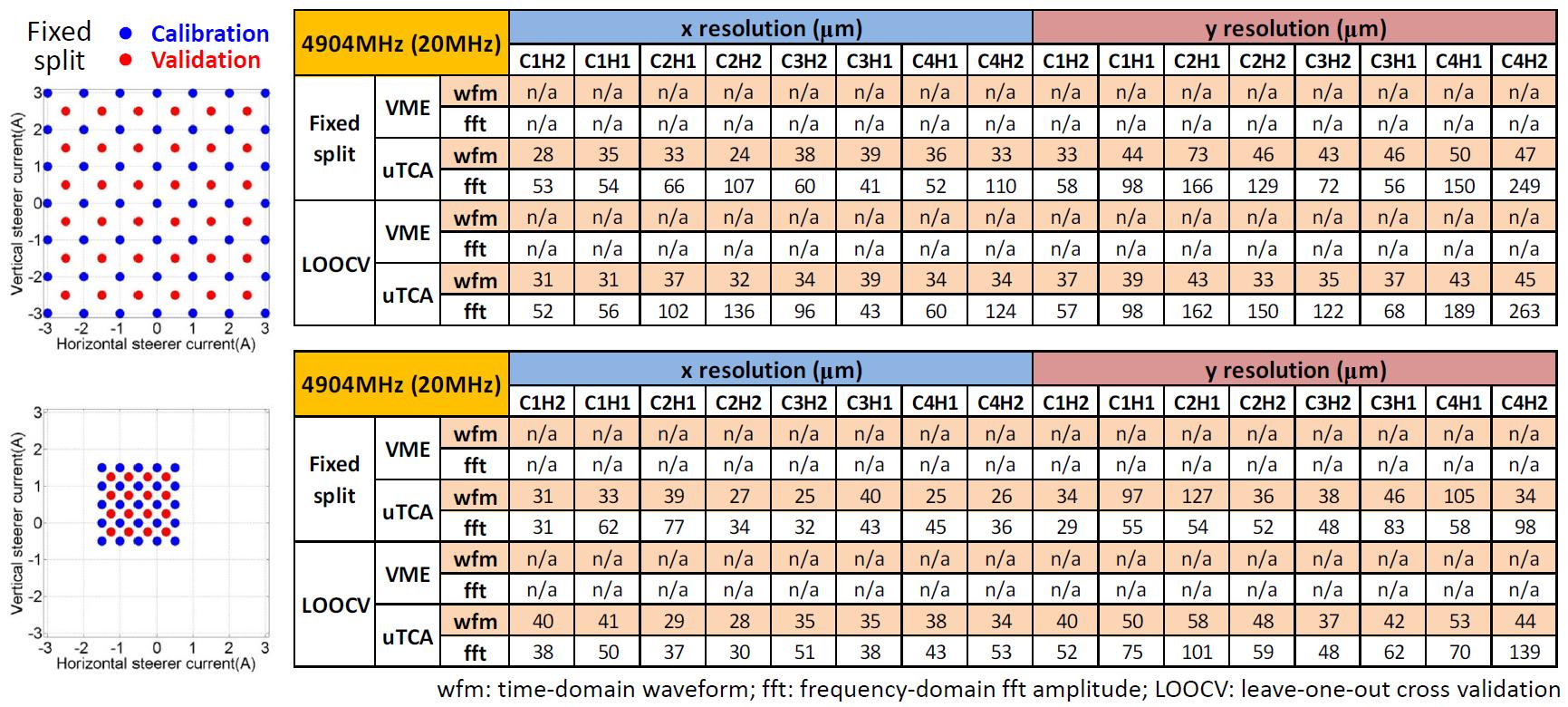}
\caption{Position resolutions of cavity modes centered at 4904~MHz with a 20~MHz bandwidth.}
\label{res-D1-4904MHz}
\end{figure}

\begin{figure}[h]\center
\includegraphics[width=0.95\textwidth]{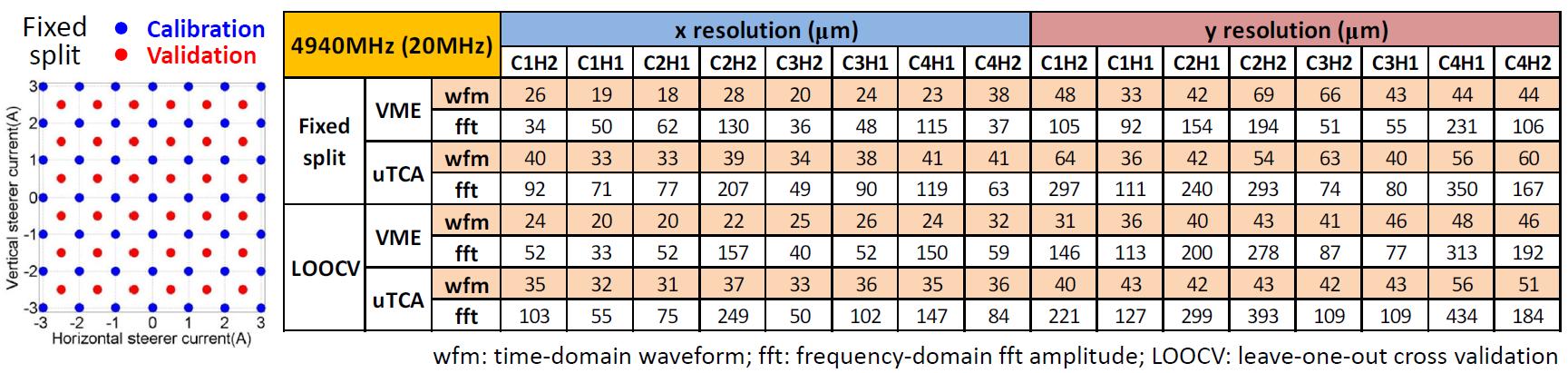}
\caption{Position resolutions of cavity modes centered at 4940~MHz with a 20~MHz bandwidth.}
\label{res-D1-4940MHz}
\end{figure}

\begin{figure}[h]\center
\includegraphics[width=0.95\textwidth]{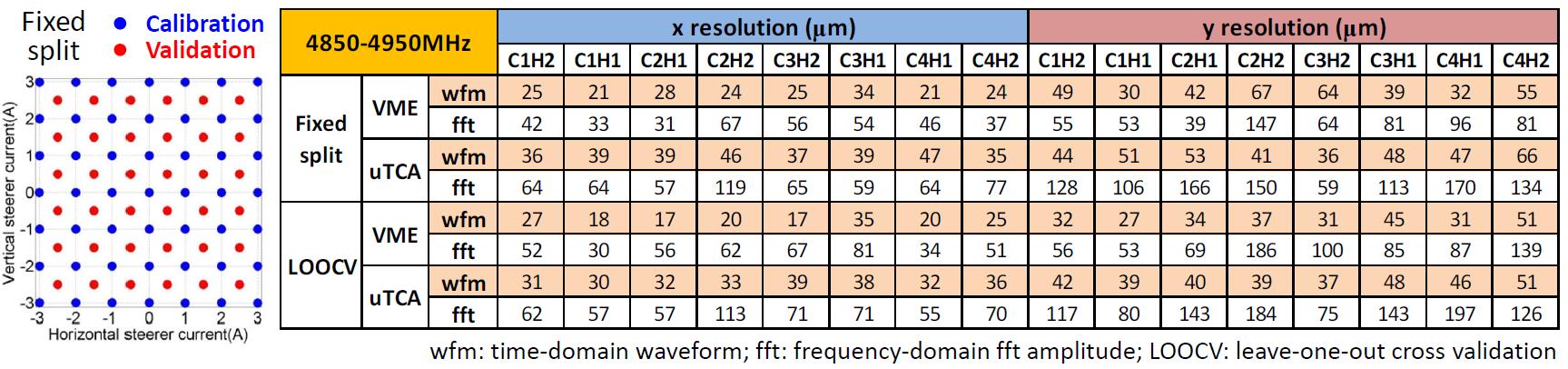}
\caption{Position resolutions of cavity modes from 4850~MHz to 4950~MHz.}
\label{res-D1-4850-4950MHz}
\end{figure}

\FloatBarrier
\section{Coupled Cavity Modes - The Second Dipole Band}\label{app:res-D2}
\begin{figure}[h]\center
\includegraphics[width=0.95\textwidth]{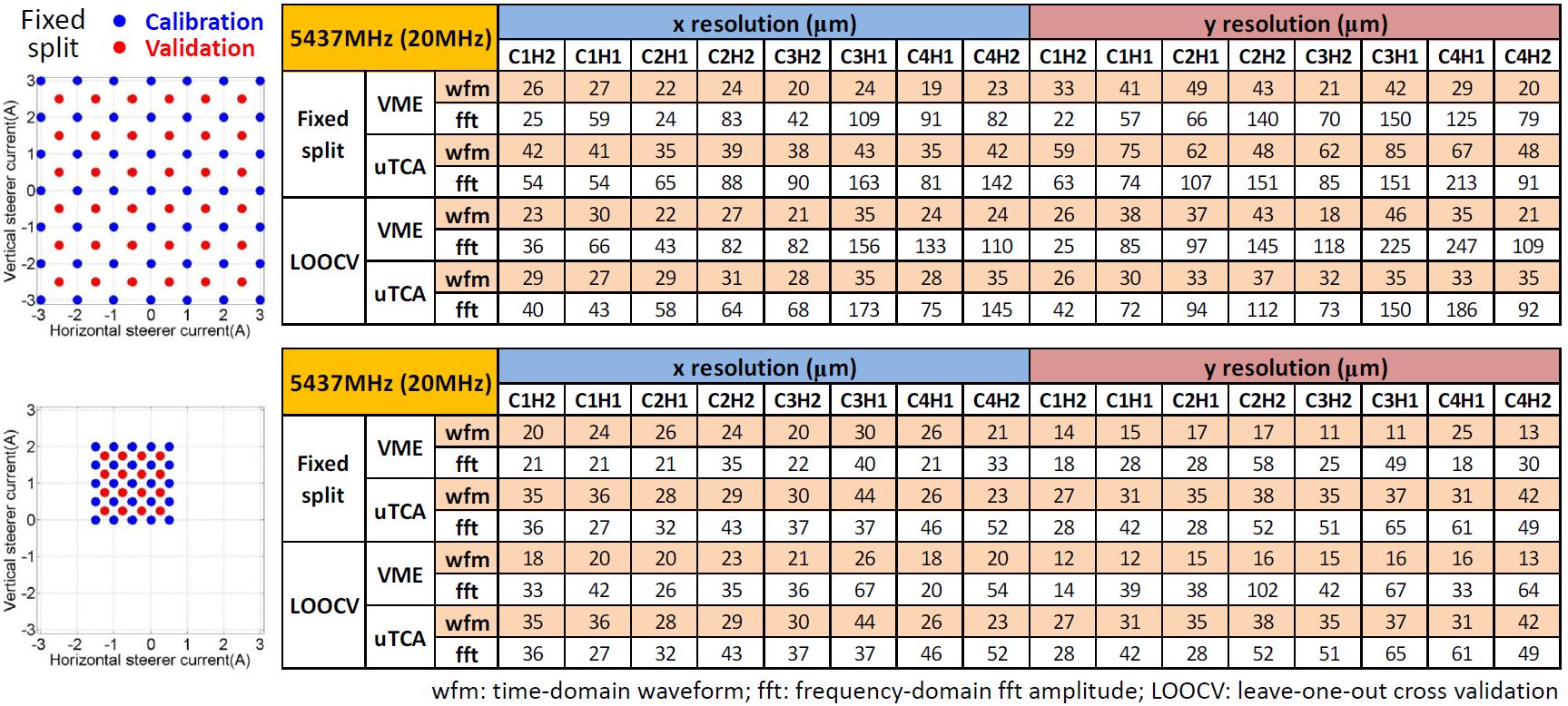}
\caption{Position resolutions of cavity modes centered at 5437~MHz with a 20~MHz bandwidth.}
\label{res-D2-5437MHz}
\end{figure}

\begin{figure}[h]\center
\includegraphics[width=0.95\textwidth]{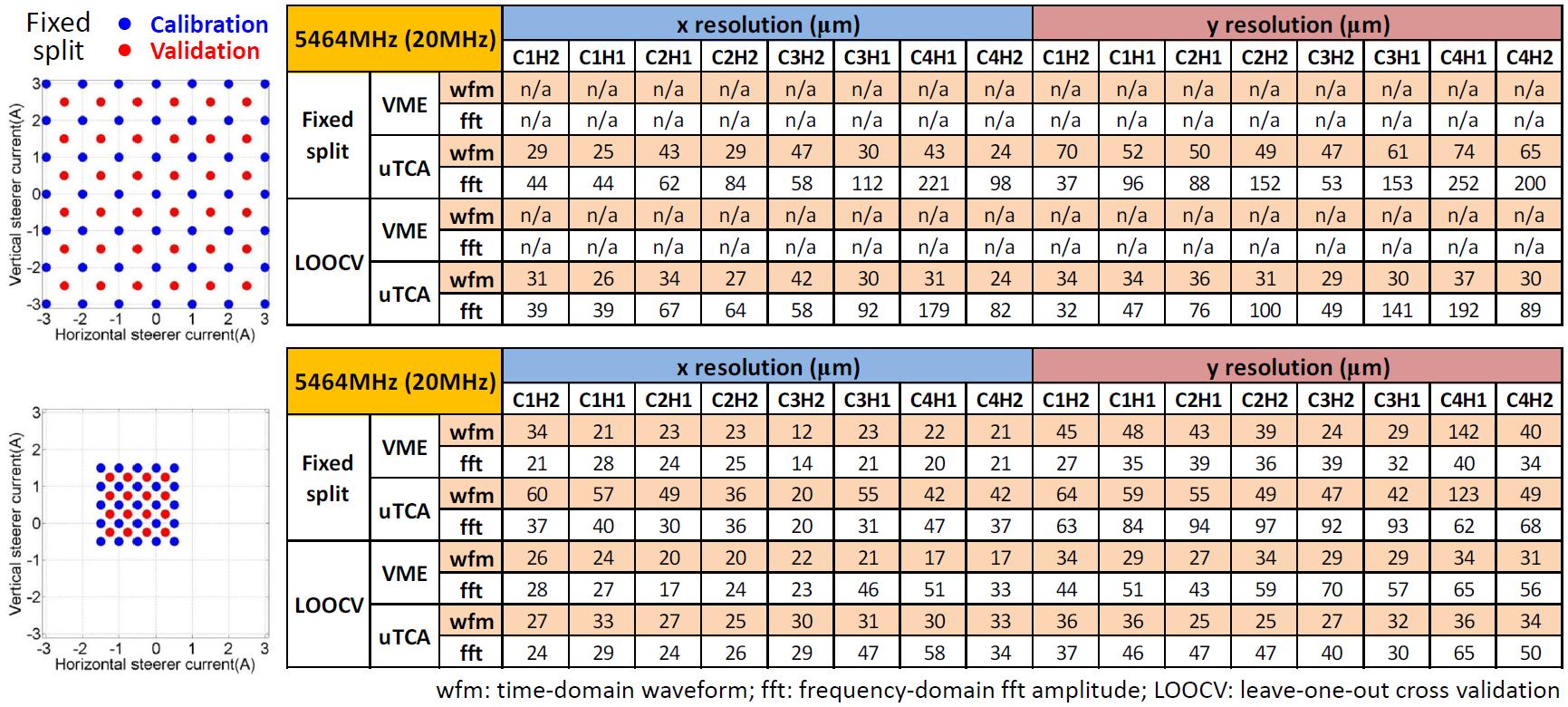}
\caption{Position resolutions of cavity modes centered at 5464~MHz with a 20~MHz bandwidth.}
\label{res-D2-5464MHz}
\end{figure}

\begin{figure}[h]\center
\includegraphics[width=0.95\textwidth]{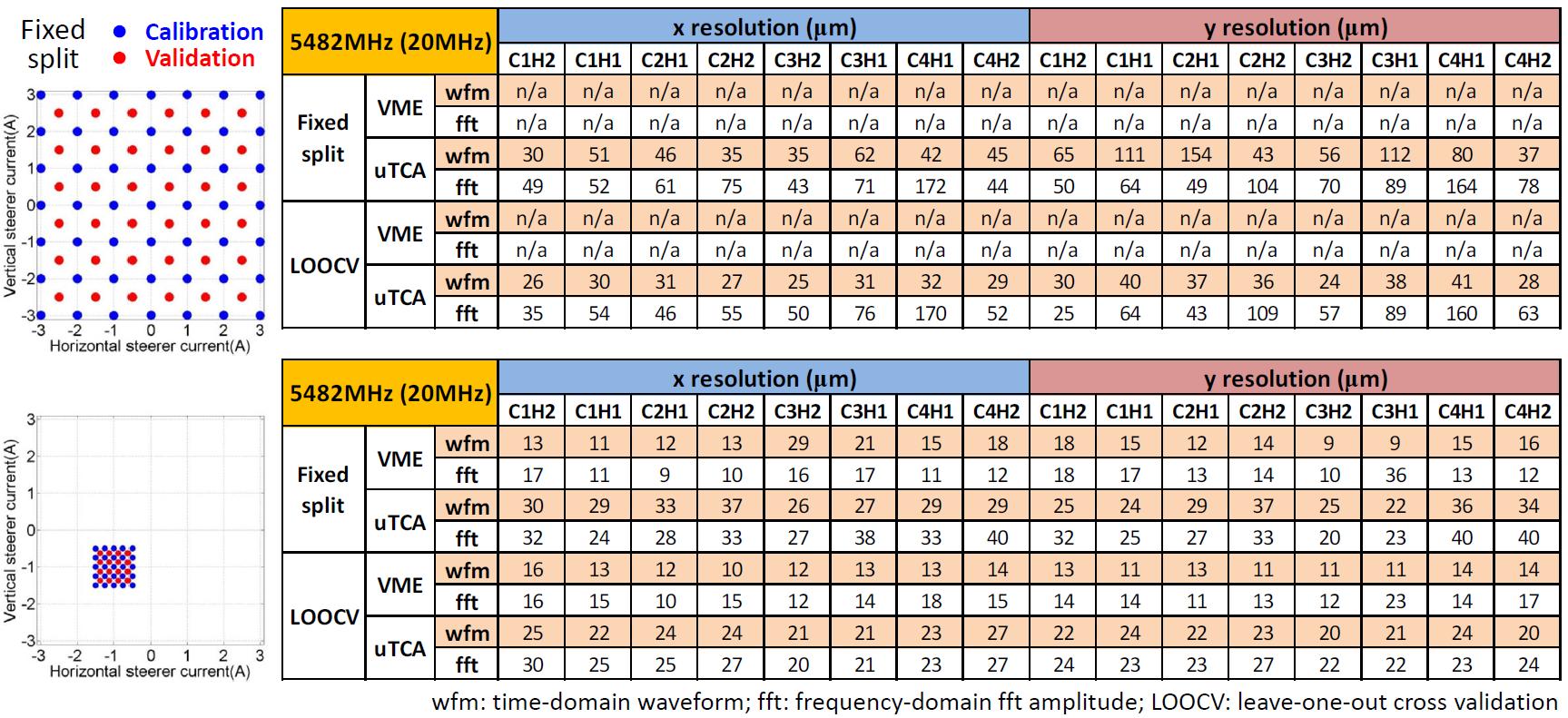}
\caption{Position resolutions of cavity modes centered at 5482~MHz with a 20~MHz bandwidth.}
\label{res-D2-5482MHz}
\end{figure}

\begin{figure}[h]\center
\includegraphics[width=0.95\textwidth]{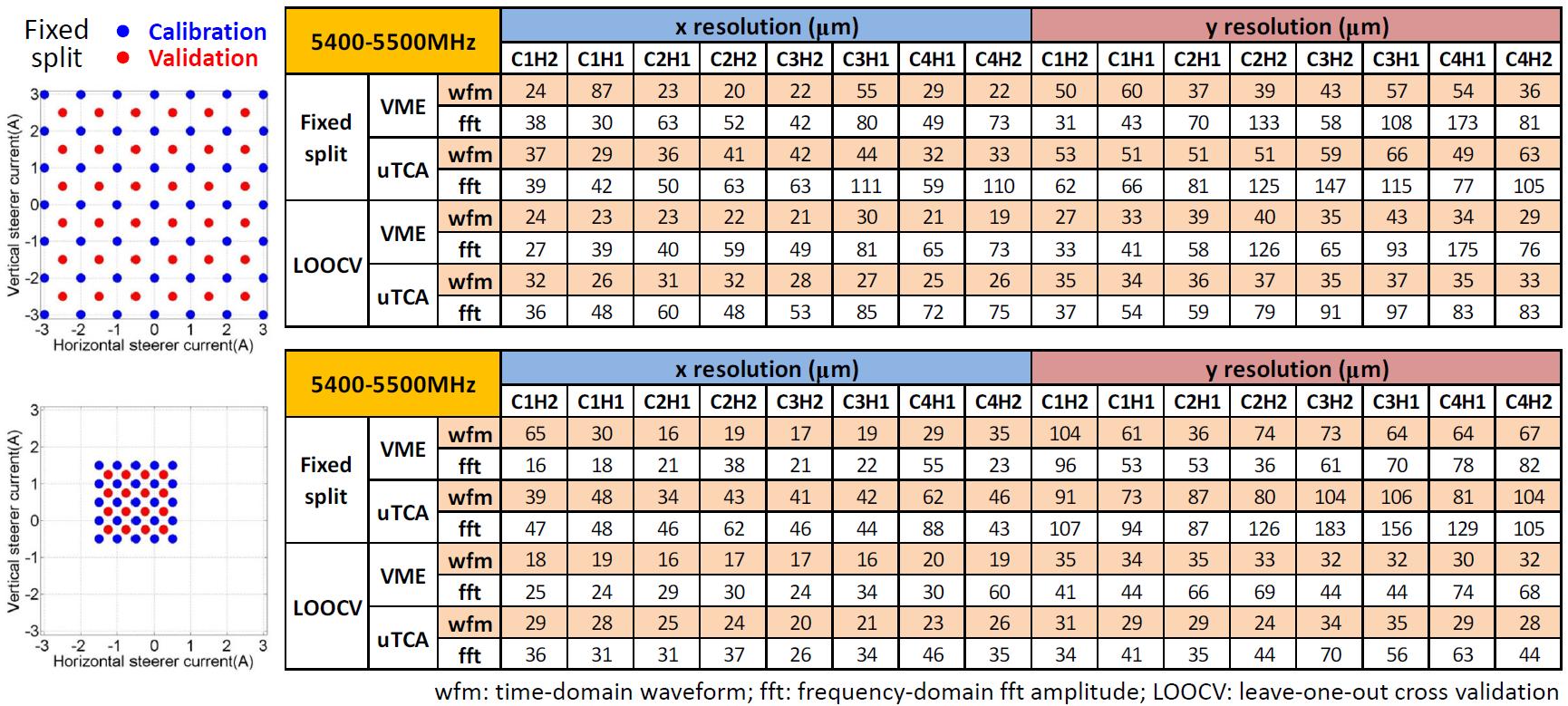}
\caption{Position resolutions of cavity modes from 5400~MHz to 5500~MHz.}
\label{res-D2-5400-5500MHz}
\end{figure}

\FloatBarrier
\section{Trapped Cavity Modes - The Fif{}th Dipole Band}\label{app:res-D5}
\begin{figure}[h]\center
\includegraphics[width=0.95\textwidth]{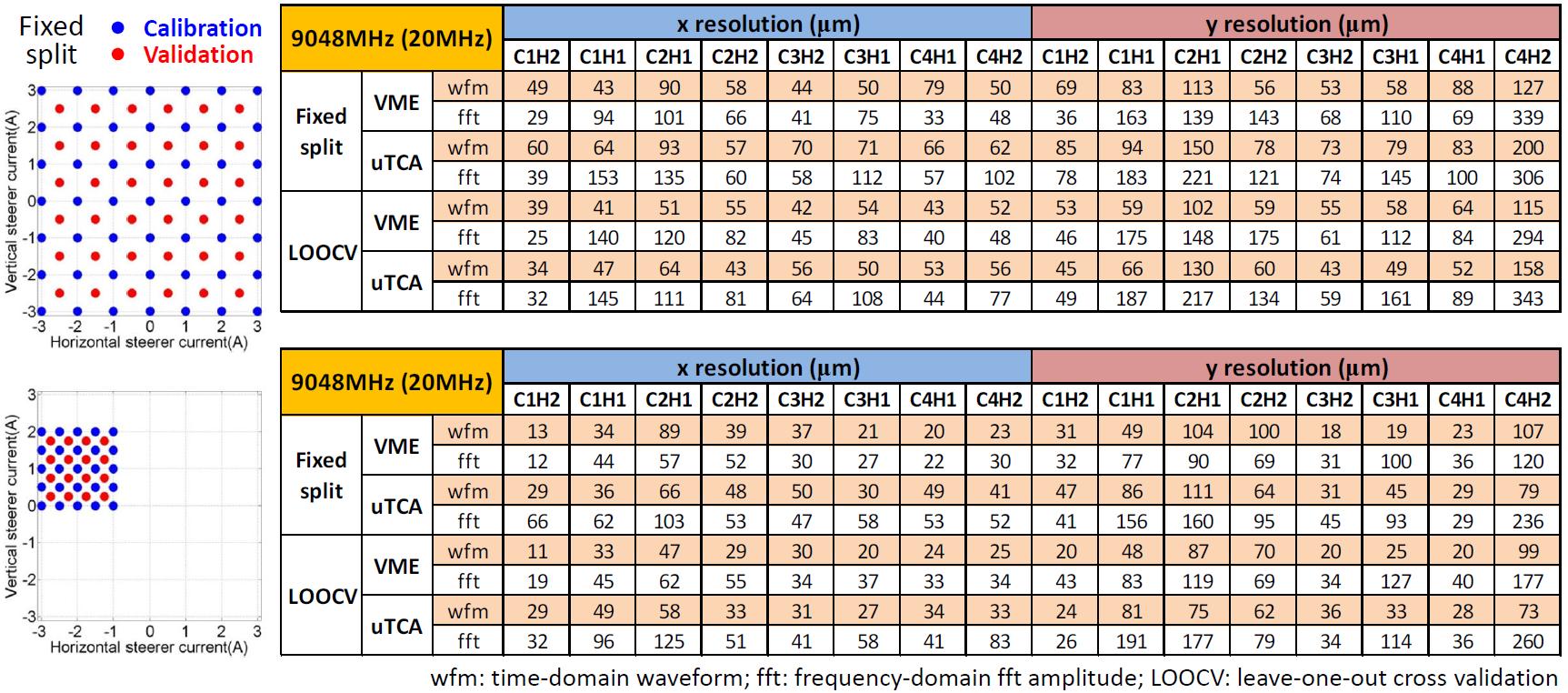}
\caption{Position resolutions of cavity modes centered at 9048~MHz with a 20~MHz bandwidth.}
\label{res-D5-9048MHz}
\end{figure}

\begin{figure}[h]\center
\includegraphics[width=0.95\textwidth]{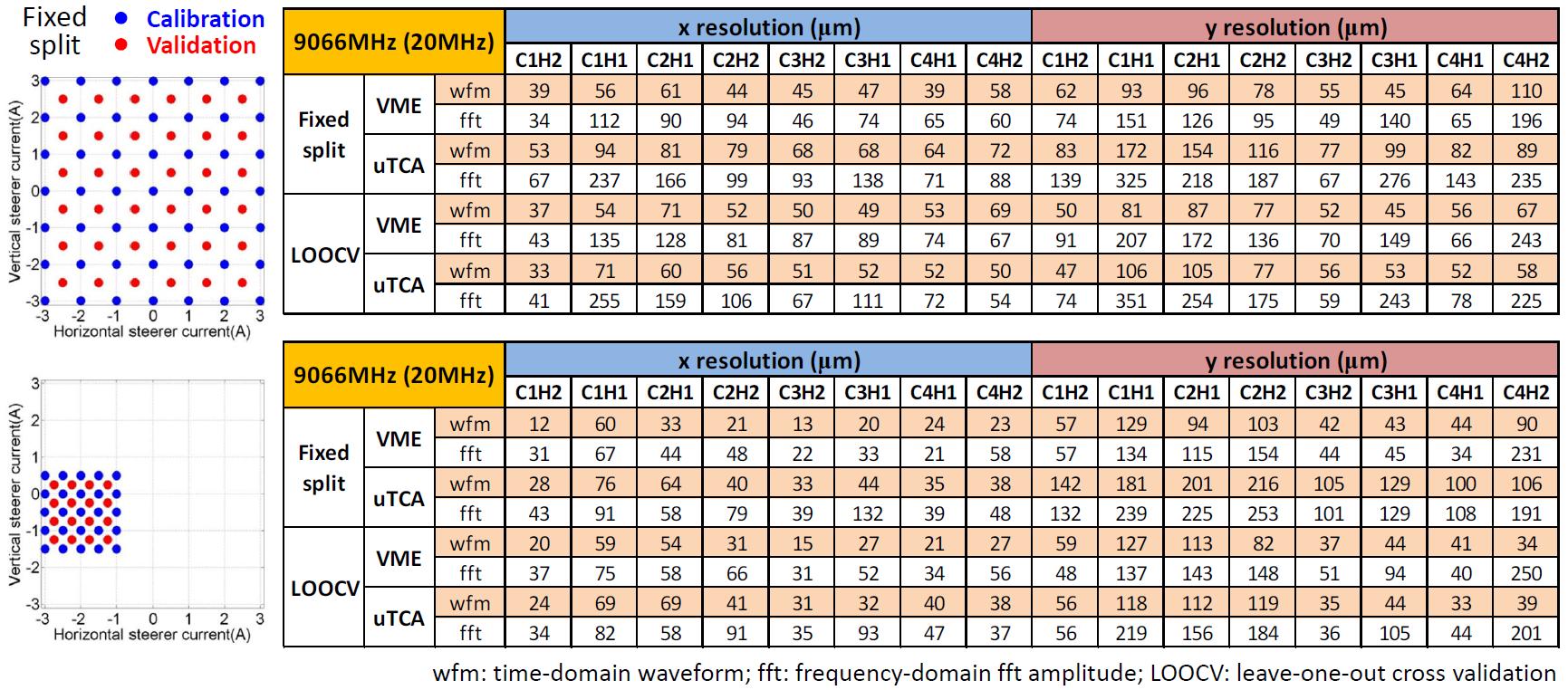}
\caption{Position resolutions of cavity modes centered at 9066~MHz with a 20~MHz bandwidth.}
\label{res-D5-9066MHz}
\end{figure}

\chapter{Tables: Number of SVD Modes}\label{app:nsvd}
This appendix shows the number of SVD modes used in position determination shown in Appendix~\ref{app:res}. The number of SVD modes used for regression is determined based on a stable and good position resolution. 

\section{The Localized Beam-pipe modes}\label{app:nsvd-bp}
\begin{figure}[h]\center
\includegraphics[width=0.95\textwidth]{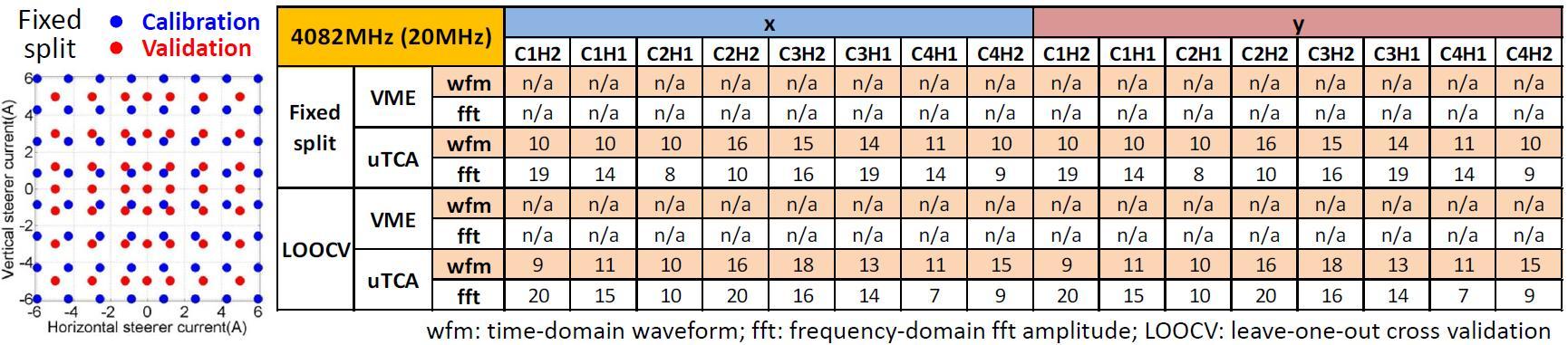}
\caption{Number of SVD modes used for position determination of beam-pipe modes centered at 4082~MHz with a 20~MHz bandwidth.}
\label{nsvd-BP-4082MHz}
\end{figure}

\begin{figure}[h]\center
\includegraphics[width=0.95\textwidth]{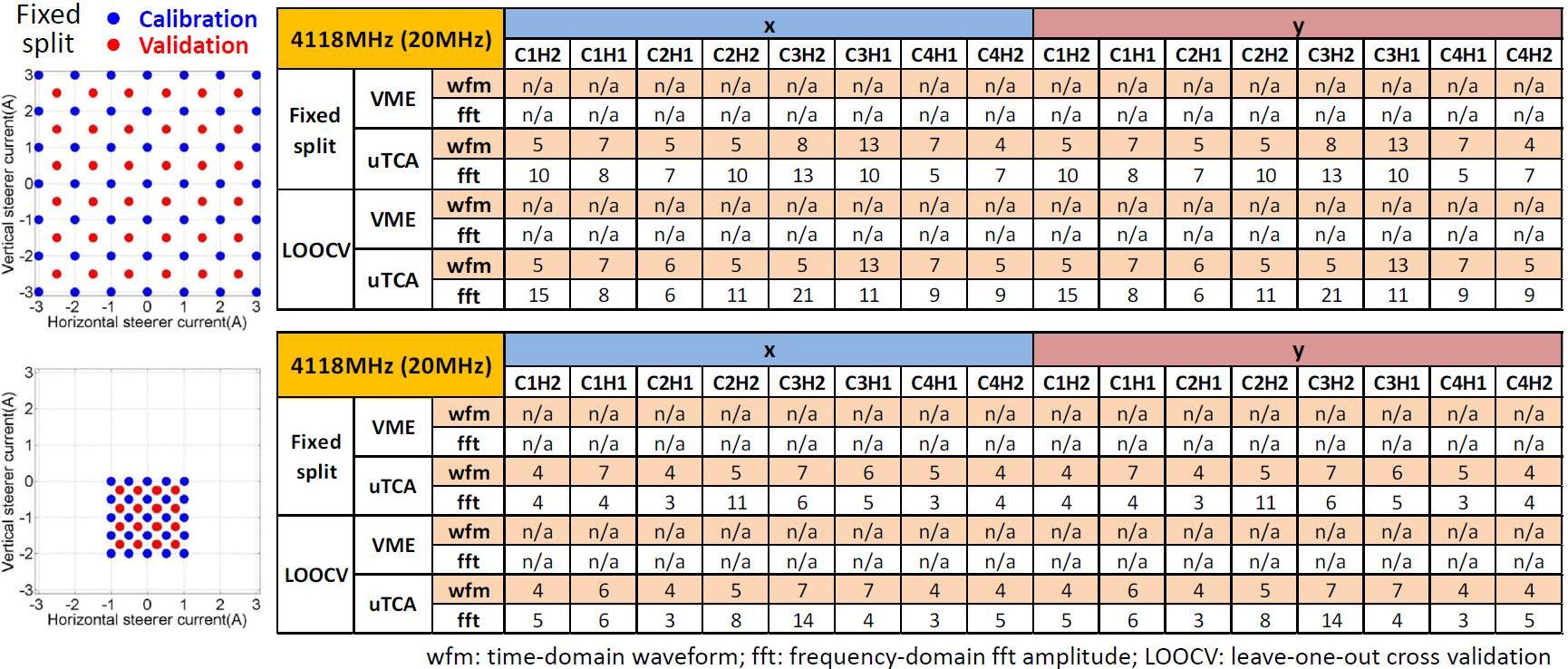}
\caption{Number of SVD modes used for position determination of beam-pipe modes centered at 4118~MHz with a 20~MHz bandwidth.}
\label{nsvd-BP-4082MHz}
\end{figure}

\FloatBarrier
\section{Coupled Cavity Modes - The First Dipole Band}\label{app:nsvd-D1}
\begin{figure}[h]\center
\includegraphics[width=0.95\textwidth]{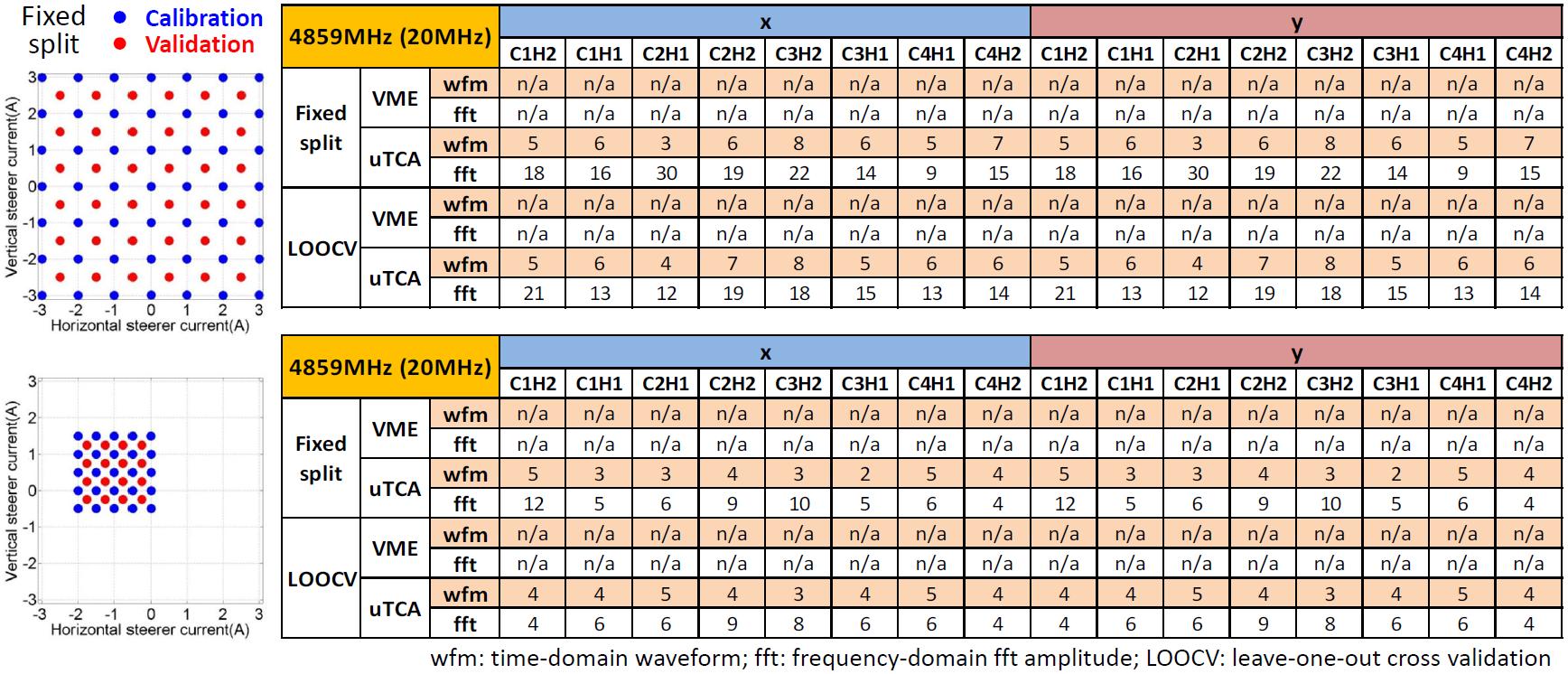}
\caption{Number of SVD modes used for position determination of cavity modes centered at 4859~MHz with a 20~MHz bandwidth.}
\label{nsvd-D1-4859MHz}
\end{figure}

\begin{figure}[h]\center
\includegraphics[width=0.95\textwidth]{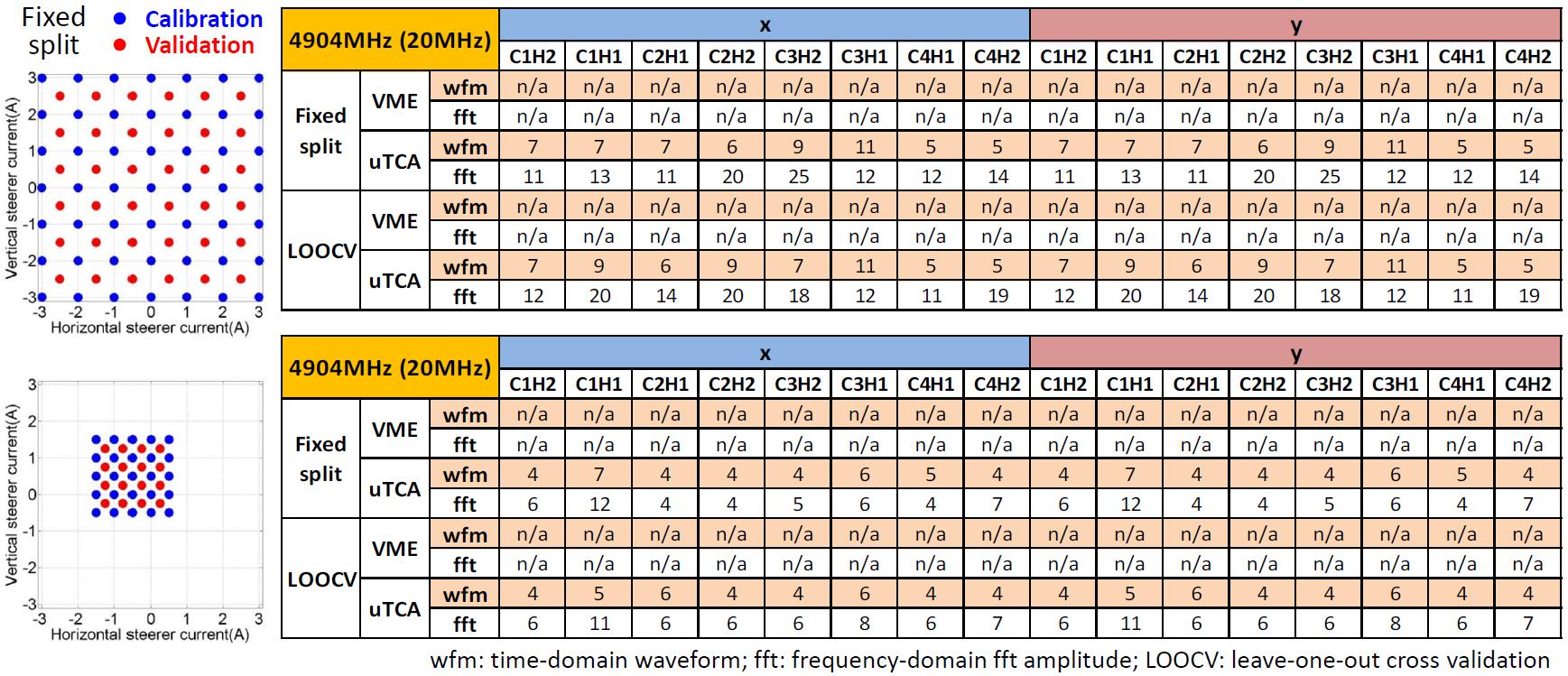}
\caption{Number of SVD modes used for position determination of cavity modes centered at 4904~MHz with a 20~MHz bandwidth.}
\label{nsvd-D1-4904MHz}
\end{figure}

\begin{figure}[h]\center
\includegraphics[width=0.95\textwidth]{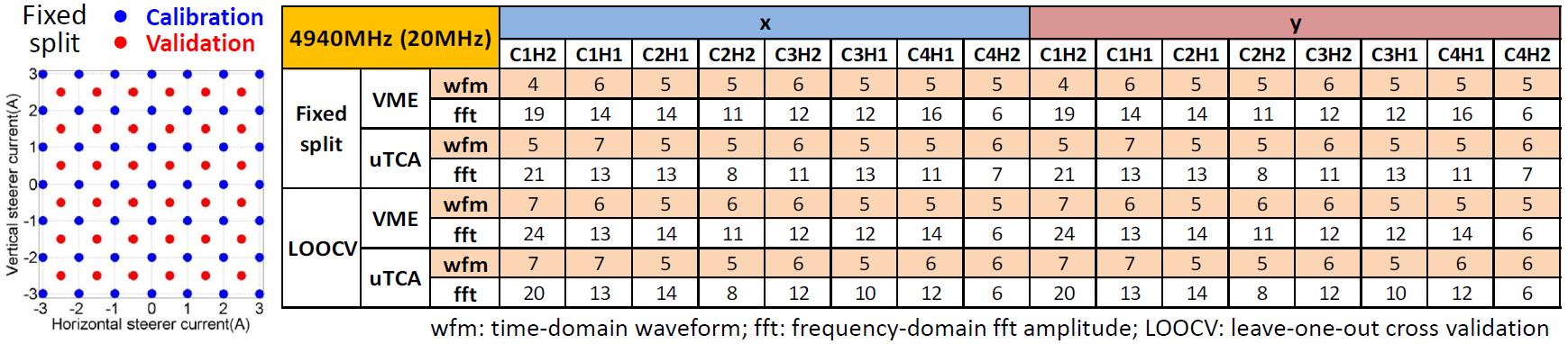}
\caption{Number of SVD modes used for position determination of cavity modes centered at 4940~MHz with a 20~MHz bandwidth.}
\label{nsvd-D1-4940MHz}
\end{figure}

\begin{figure}[h]\center
\includegraphics[width=0.95\textwidth]{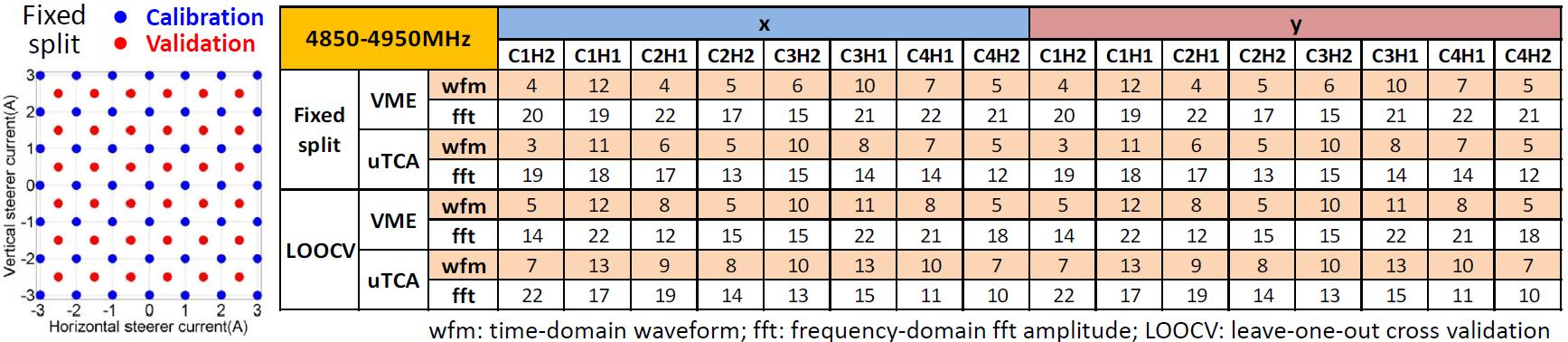}
\caption{Number of SVD modes used for position determination of cavity modes from 4850~MHz to 4950~MHz.}
\label{nsvd-D1-4850-4950MHz}
\end{figure}

\FloatBarrier
\section{Coupled Cavity Modes - The Second Dipole Band}\label{app:nsvd-D2}
\begin{figure}[h]\center
\includegraphics[width=0.95\textwidth]{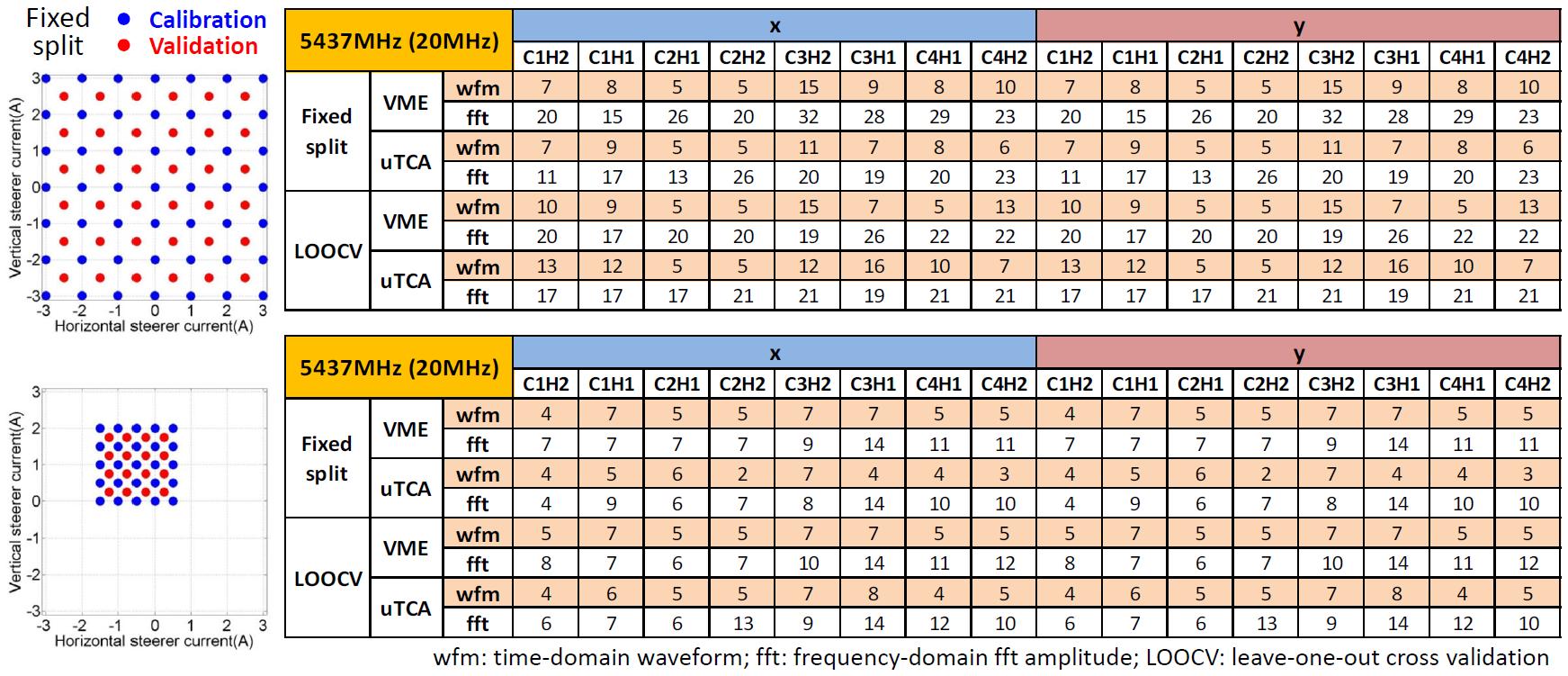}
\caption{Number of SVD modes used for position determination of cavity modes centered at 5437~MHz with a 20~MHz bandwidth.}
\label{nsvd-D2-5437MHz}
\end{figure}

\begin{figure}[h]\center
\includegraphics[width=0.95\textwidth]{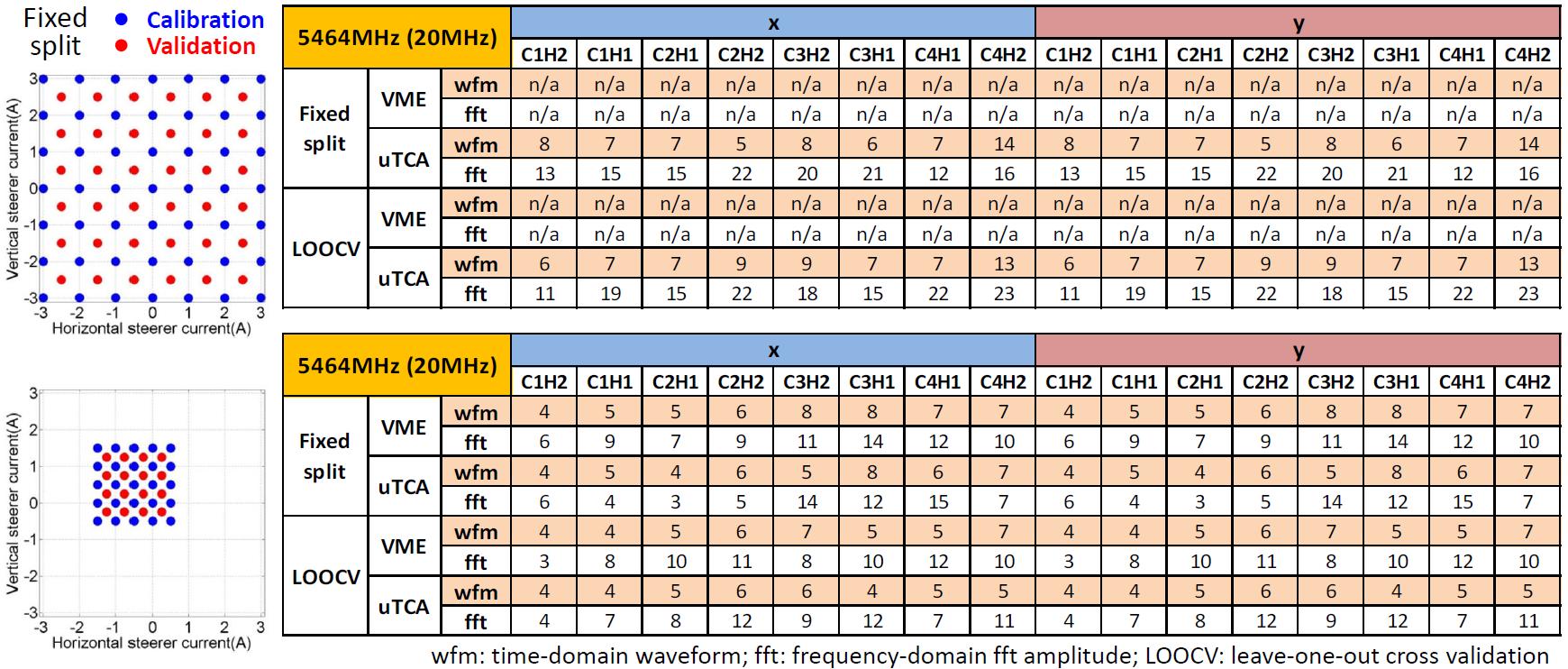}
\caption{Number of SVD modes used for position determination of cavity modes centered at 5464~MHz with a 20~MHz bandwidth.}
\label{nsvd-D2-5464MHz}
\end{figure}

\begin{figure}[h]\center
\includegraphics[width=0.95\textwidth]{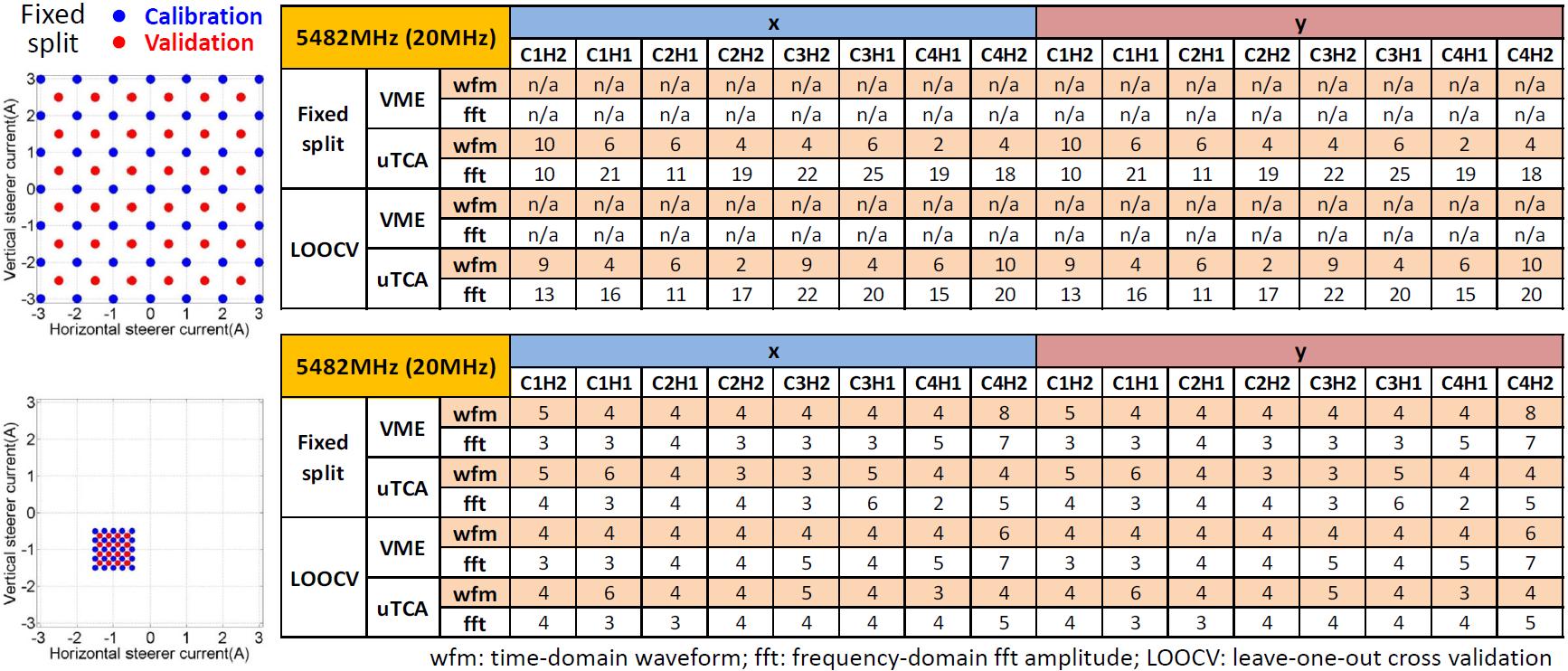}
\caption{Number of SVD modes used for position determination of cavity modes centered at 5482~MHz with a 20~MHz bandwidth.}
\label{nsvd-D2-5482MHz}
\end{figure}

\begin{figure}[h]\center
\includegraphics[width=0.95\textwidth]{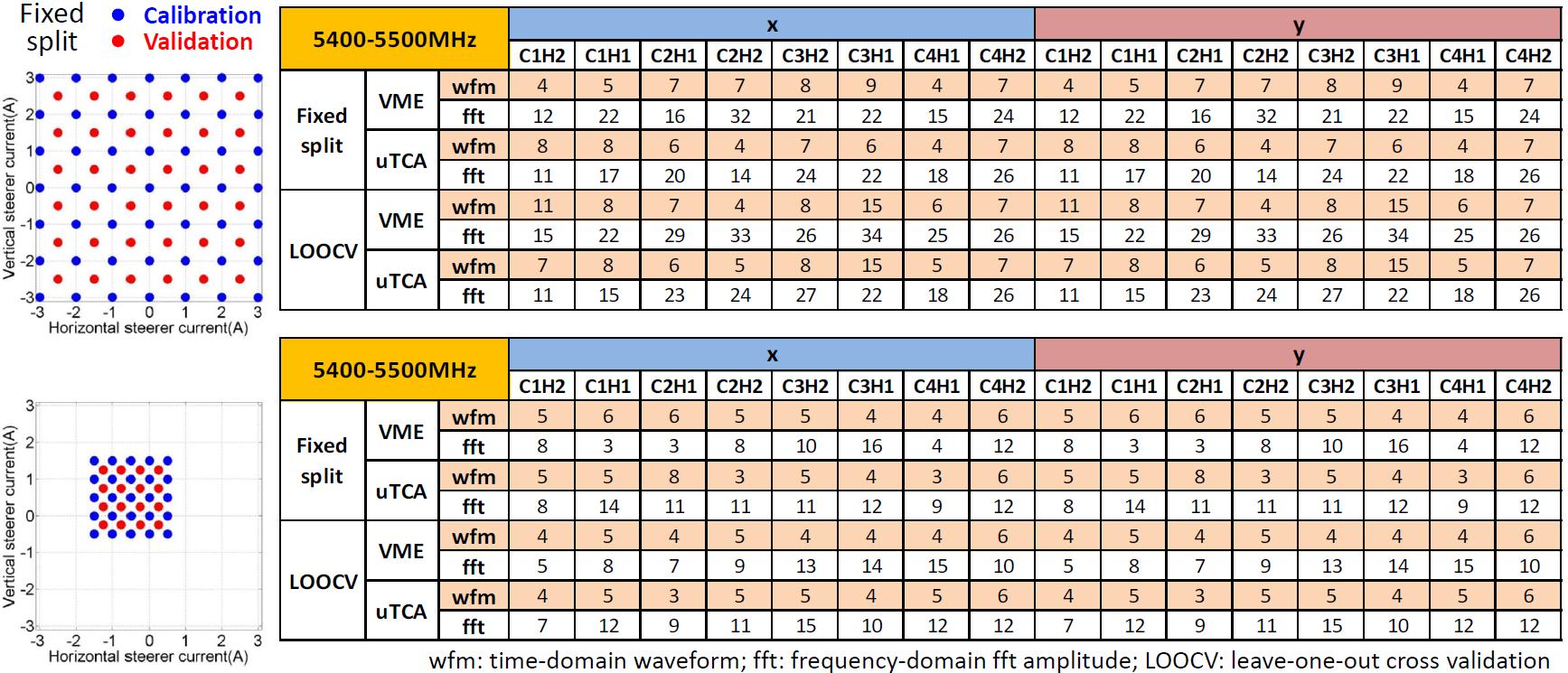}
\caption{Number of SVD modes used for position determination of cavity modes from 5400~MHz to 5500~MHz.}
\label{nsvd-D2-5400-5500MHz}
\end{figure}

\FloatBarrier
\section{Trapped Cavity Modes - The Fif{}th Dipole Band}\label{app:nsvd-D5}
\begin{figure}[h]\center
\includegraphics[width=0.95\textwidth]{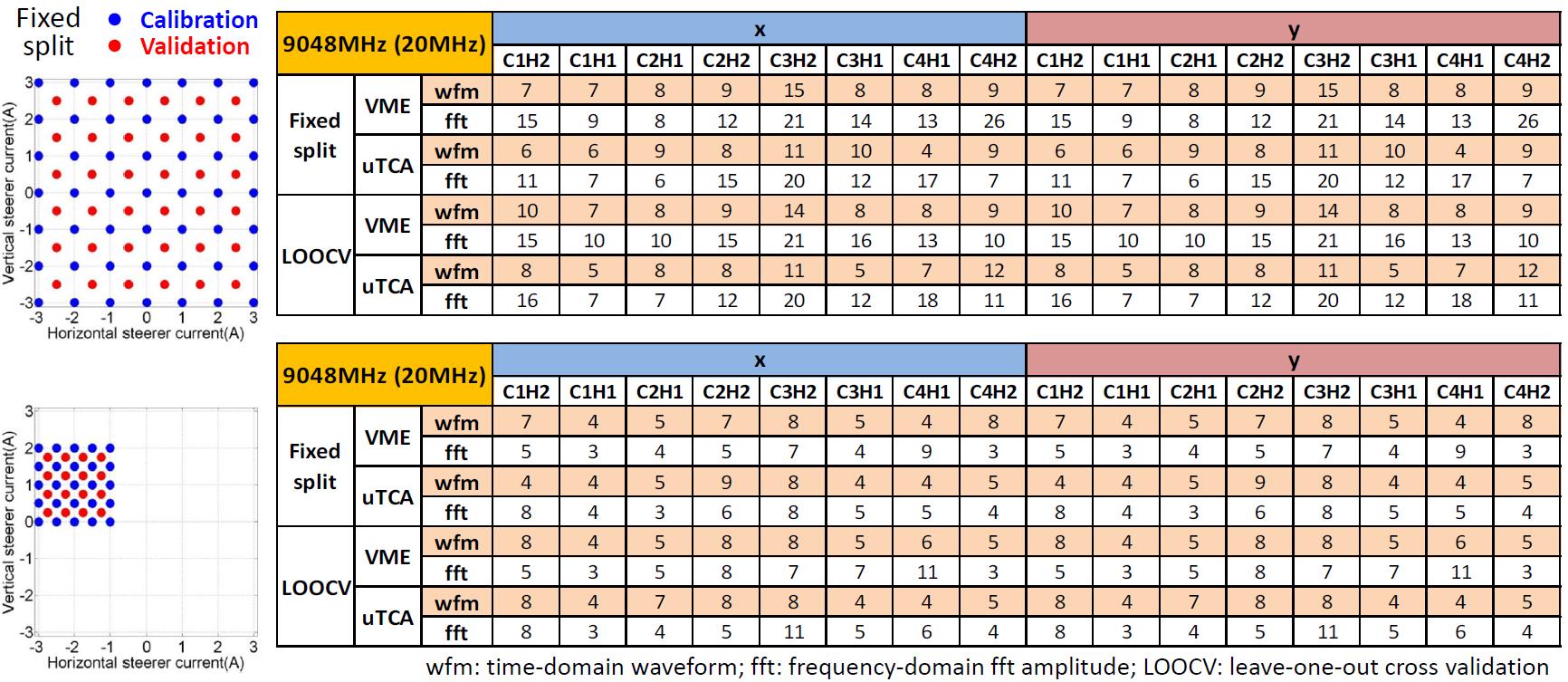}
\caption{Number of SVD modes used for position determination of cavity modes centered at 9048~MHz with a 20~MHz bandwidth.}
\label{nsvd-D5-9048MHz}
\end{figure}

\begin{figure}[h]\center
\includegraphics[width=0.95\textwidth]{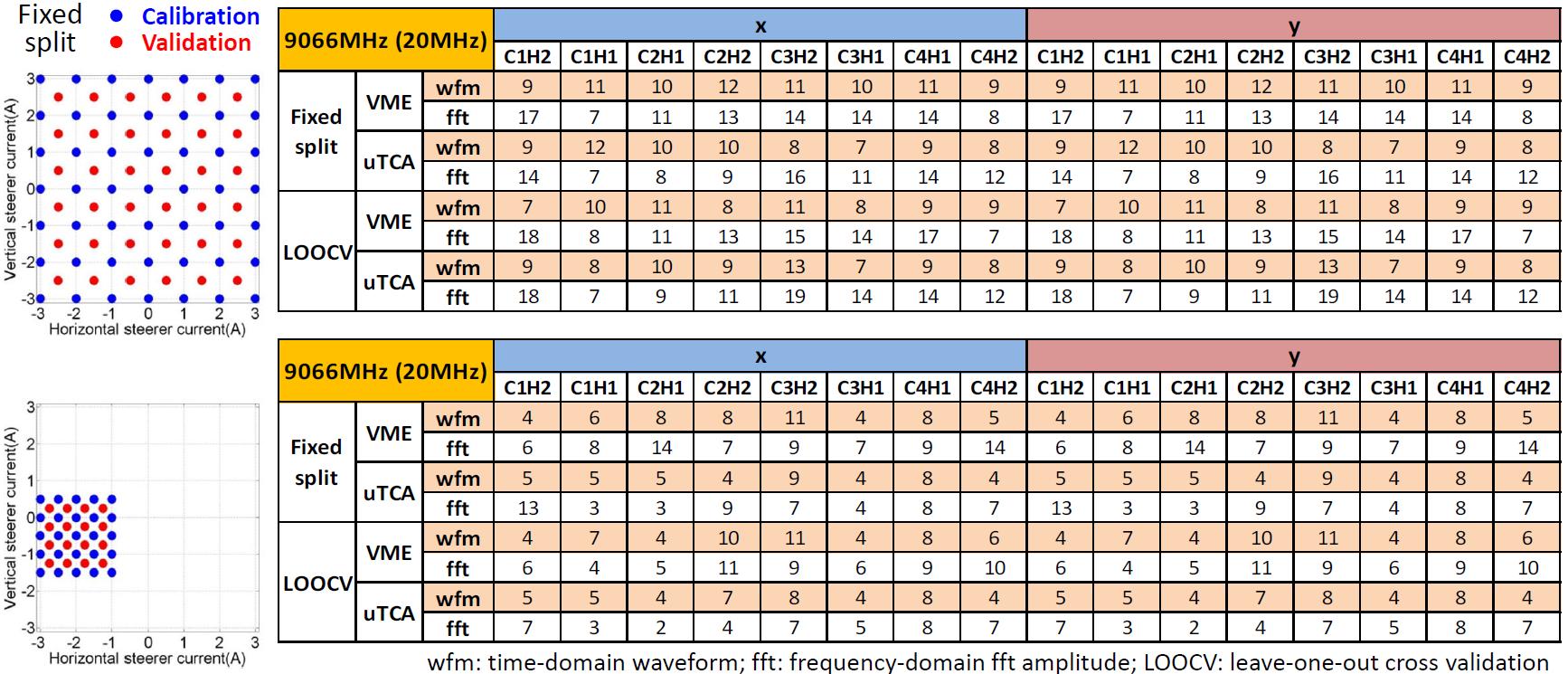}
\caption{Number of SVD modes used for position determination of cavity modes centered at 9066~MHz with a 20~MHz bandwidth.}
\label{nsvd-D5-9066MHz}
\end{figure}

\end{document}